\documentclass[useAMS,usenatbib,twocolumn]{mn2e}

\usepackage{graphicx,amsmath,amssymb,setspace,psfrag,color}
\usepackage[table]{xcolor}
\usepackage[breaklinks,colorlinks,citecolor=blue,linkcolor=blue,urlcolor=blue]{hyperref}

\newcommand\aj{{AJ}}%
\newcommand\araa{{ARA\&A}}%
\newcommand\apj{{ApJ}}%
\newcommand\apjl{{ApJ}}%
\newcommand\apjs{{ApJS}}%
\newcommand\aap{{A\&A}}%
\newcommand\aaps{{A\&AS}}%
\newcommand\mnras{{MNRAS}}%
\newcommand\pasp{{PASP}}%
\newcommand\pasj{{PASJ}}%
\newcommand\ssr{{Space~Sci.~Rev.}}%
\newcommand\nat{{Nature}}%
\newcommand\iaucirc{{IAU~Circ.}}%
\newcommand\na{{NewA}}%
\newcommand\nar{{NewAR}}%
\newcommand\scchg{ScChG}
\newcommand{\todo}{\ifmmode {\Huge \bullet} \else {\Huge$\bullet$}\fi}
\newcommand{\E        }[1]{\ifmmode 10^{#1} \else $10^{#1}$\fi}
\newcommand{\til}{\ifmmode \sim \else $\sim$\fi}
\renewcommand{\~} {\ifmmode \sim \else $\sim$\fi}

\newcommand\ltsim{\raisebox{-.5ex}{$\;\stackrel{<}{\sim}\;$}}
\newcommand\gtsim{\raisebox{-.5ex}{$\;\stackrel{>}{\sim}\;$}}

\def\Msun{\ifmmode M_{\odot} \else $M_{\odot}$\fi}
\def\msun{\ifmmode M_{\odot} \else $M_{\odot}$\fi}
\def\Lsun{\ifmmode L_{\odot} \else $L_{\odot}$\fi}
\def\mpyr{\ifmmode \Msun\,{\rm yr}^{-1} \else $\Msun\,{\rm yr}^{-1}$ \fi}

\def\U{\textit{u$^\prime$}}
\def\G{\textit{g$^\prime$}}
\def\R{\textit{r$^\prime$}}
\def\I{\textit{i$^\prime$}}

\def\vespa{{\scriptsize VESPA}} 
\def\snid{{\scriptsize SNID}}
\def\salt{{\scriptsize SALT2}}
\def\sex{\normalsize SE\scriptsize XTRACTOR\normalsize}

\def\Ngal{739\,584}
\def\NIIgal{215\,114}
\def\RvolII{$0.621^{+0.197}_{-0.154}~({\rm stat})~^{+0.024}_{-0.063}~({\rm sys}) \times 10^{-4}$ yr$^{-1}$ Mpc$^{-3}$}
\def\RvolCC{$1.04^{+0.33}_{-0.26}~({\rm stat})~^{+0.04}_{-0.11}~({\rm sys}) \times 10^{-4}$ yr$^{-1}$ Mpc$^{-3}$}
\def\SNuMIaall{$0.10\pm0.01~({\rm stat})~\pm0.01~({\rm sys})\times 10^{-12}~{\rm M_\odot^{-1}~yr^{-1}}$}

\def\SNuMII{$0.52^{+0.16}_{-0.13}~({\rm stat})~^{+0.02}_{-0.05}~({\rm sys})\times 10^{-12}~{\rm M_\odot^{-1}~yr^{-1}}$}

\title[Explaining the SN Rate-Mass Dependency]{A unified explanation for the supernova rate-galaxy mass dependency based on supernovae detected in Sloan galaxy spectra}

\author[Graur, Bianco, \& Modjaz]
{Or~Graur,$^{1,2}$\thanks{E-mail: \href{mailto:orgraur@nyu.edu}{orgraur@nyu.edu}}
 Federica B. Bianco,$^1$
 and Maryam Modjaz$^1$
\\
$^1$CCPP, New York University, 4 Washington Place, New York, NY 10003, USA \\
$^2$Department of Astrophysics, American Museum of Natural History, Central Park West and 79th Street, New York, NY 10024-5192, USA \\
}

\begin{document}

\maketitle


\setstretch{1}

\begin{abstract}
\noindent
Using a method to discover and classify supernovae (SNe) in galaxy spectra, we detect 91 Type Ia SNe (SNe Ia) and 16 Type II SNe (SNe II) among $\sim740\,000$ galaxies of all types and $\sim215\,000$ star-forming galaxies without active galactic nuclei, respectively, in Data Release 9 of the Sloan Digital Sky Survey. Of these SNe, 15 SNe Ia and 8 SNe II are new discoveries reported here for the first time. We use our SN samples to measure SN rates per unit mass as a function of galaxy stellar mass, star-formation rate (SFR), and specific SFR (sSFR), as derived by the MPA-JHU Galspec pipeline. We show that correlations between SN Ia and SN II rates per unit mass and galaxy stellar mass, SFR, and sSFR can be explained by a combination of the respective SN delay-time distributions (the distributions of times that elapse between the formation of a stellar population and all ensuing SNe), the ages of the surveyed galaxies, the redshifts at which they are observed, and their star-formation histories. This model was first suggested by Kistler et al. for the SN Ia rate-mass correlation, but is expanded here to SNe II and to correlations with galaxy SFR and sSFR. Finally, we measure a volumetric SN II rate at redshift 0.075 of $R_{\rm II,V} = $~\RvolII. Assuming that SNe IIP and IIL account for 60 per cent of all core-collapse (CC) SNe, the CC SN rate is $R_{\rm CC,V} = $~\RvolCC.

\end{abstract}



\section{INTRODUCTION}
\label{sec:intro}

Supernovae (SNe) play many roles in the Universe, from accelerating cosmic rays (e.g., \citealt{2009Sci...325..719H,2012SSRv..173..369H}), through assisting in galaxy evolution (e.g., \citealt*{2013MNRAS.429.1922C}), to seeding the interstellar medium with heavy elements (e.g., \citealt{2008NewA...13..606B,2011MNRAS.415..353W,Graur2011}). Type Ia SNe (SNe Ia) have been used to measure extragalactic distances, leading to the discovery that the expansion of the Universe is accelerating \citep{1998AJ....116.1009R,1999ApJ...517..565P}. Other SN types, such as core-collapse (CC) SNe IIP and superluminous SNe (see review by \citealt{2012Sci...337..927G}), may also be useful as cosmological distance probes (e.g., \citealt{2002ApJ...566L..63H,2006ApJ...645..841N,2009ApJ...694.1067P,2014ApJ...796...87I}).

To fully understand how SNe explode and affect their surroundings, we must identify which types of progenitor stars end up exploding as different types of SNe. Pre-explosion images of some SNe IIP and IIn have linked these SN subtypes to red and blue supergiants, respectively (\citealt{smartt2009mnras}; see review by \citealt{Smartt2009review}). The natures of the progenitors of all other SN types remain contested (see \citealt{2013MNRAS.436..774E} for SNe Ib/c and a review by \citealt*{2014ARA&A..52..107M} for SNe Ia). Of the different methods to constrain SN progenitors (e.g., \citealt*{2007ApJ...662..472B,2008ARA&A..46..433W,2011ApJ...741...20B,2011MNRAS.412.1522S,2012ApJ...749L..11B,Schaefer2012,2013MNRAS.436..222M,2013ApJ...762L...5S,2014MNRAS.442L..28G,2014MNRAS.442.1079J}), we focus here on measuring SN rates and searching for possible correlations between them and various galaxy properties. 

\citet[hereafter L11]{li2011rates} used a local sample of SNe discovered by the Lick Observatory Supernova Search (LOSS; \citealt{2011MNRAS.412.1419L,li2011LF}) to measure mass-normalized rates and discovered that the SN rate per unit mass decreased with increasing galaxy stellar mass. This `rate-mass' correlation was a surprising result, as naively one would expect that once the SN rates were normalized by the stellar masses of the observed galaxies, they would be independent of them. \citet{2006ApJ...648..868S} showed a similar rate-mass correlation for SNe Ia when they observed that although the SN Ia rates per galaxy in their sample increased with increasing stellar mass, the slope of the increase in star-forming galaxies was $<1$, so that the mass-normalized rates in these galaxies decreased with increasing stellar mass. Interestingly, the passive galaxies in the \citet{2006ApJ...648..868S} sample did not show this correlation; with a slope of $1.10\pm0.12$, the mass-normalized SN Ia rates were independent of galaxy stellar mass. 

In an effort to explain the rate-mass correlation, L11 compared the slopes of the power laws they fit to their rates ($-0.55\pm0.10$ for CC SNe and $-0.50\pm-0.10$ for SNe Ia) to that of the correlation between specific star-formation rate (sSFR) and galaxy stellar mass, as measured by \citet{2007ApJS..173..315S}. The latter found a slope of $-0.36$ for star-forming galaxies and a shallower slope of $-0.16$ for passive galaxies. L11 averaged the data in figure 7 of \citet{2007ApJS..173..315S} to result in an average slope of $-0.55\pm0.09$. They note that this average slope is similar to the slope of the CC SN rate-mass correlation and understand this similarity to indicate that both sSFR and the mass-normalized CC SN rates are consistent indicators of star-formation activity. In this work, we re-examine this assertion and show that although the L11 explanation does not hold up on its own, it is part of a more comprehensive explanation, which is valid for both CC SNe and SNe Ia.

In \citet[hereafter GM13]{GraurMaoz2013}, we developed a method to detect SNe hidden in galaxy spectra. We optimized our method to detect SNe Ia and detected 90 SNe Ia among $\sim700\,000$ galaxy spectra in the seventh data release (DR7; \citealt{SDSS-DR7}) of the Sloan Digital Sky Survey (SDSS; \citealt{2000AJ....120.1579Y}). We used this sample to measure mass-normalized SN Ia rates and confirmed the L11 rate-mass correlation for SNe Ia at a median redshift of $\sim0.1$. Follwing \citet{Kistler2011}, we showed that this rate-mass correlation could be explained by a combination of a power-law delay-time distribution (DTD; the distribution of times that elapse between the birth of a stellar population and all subsequent SNe; see \citealt{2012NewAR..56..122W,2013FrPhy...8..116H} for reviews) with an index of $-1$ and the correlation reported by \citet{2005MNRAS.362...41G} between the ages and stellar masses of galaxies, colloquially known as galaxy `downsizing' (i.e., older galaxies are, on average, more massive than younger galaxies).

In this paper, we optimize our SN detection method to discover SNe II (specifically, SNe IIP and IIL) and apply it to the ninth SDSS data release (DR9; \citealt{2012ApJS..203...21A}). In Section~\ref{sec:galaxies}, we describe our galaxy sample and how it differs from the GM13 galaxy sample. We briefly describe the changes we have made to our SN detection and classification method in Section~\ref{sec:method}. In Section~\ref{sec:sample}, we report the discovery of 91 SNe Ia among $\sim740\,000$ spectra of galaxies of all types and 16 SNe II among $\sim215\,000$ spectra of star-forming galaxies that do not host active galactic nuclei (AGNs). Of these SNe, 15 SNe Ia and 8 SNe II are reported here for the first time. We use these SN samples in Section~\ref{sec:results} to measure mass-normalized SN Ia and SN II rates and find correlations between these rates and galaxy stellar mass, SFR, and sSFR. In this Section, we also convert the mass-normalized SN II rate, averaged over all masses, into a volumetric SN II rate at a median redshift of $0.075$, from which we estimate a volumetric CC SN rate. We have already performed such a measurement of the SN Ia volumetric rate in GM13; we do not repeat it here, as the result would not change appreciably. We discuss our results in Section~\ref{sec:discuss}, where we make several predictions based on our preferred explanation for the correlations. Our conclusions are summarized in Section~\ref{sec:summary}. 

Throughout this paper, we assume a $\Lambda$-cold-dark-matter cosmological model with parameters $\Omega_{\Lambda} = 0.7$, $\Omega_m = 0.3$, and $H_0 = 70$~km~s$^{-1}$~Mpc$^{-1}$. Magnitudes are on the \citet{1983ApJ...266..713O} AB system, unless noted otherwise.


\section{GALAXY SAMPLE}
\label{sec:galaxies}

In GM13, our galaxy sample was the subsample of galaxies in SDSS DR7 that had star-formation histories (SFHs) computed with the VErsatile SPectral Analysis code (\vespa;\footnote{\url{http://www-wfau.roe.ac.uk/vespa/}} \citealt{Tojeiro2007,Tojeiro2009}). In this work, we selected our galaxy sample from SDSS DR9, which includes updated reductions of the galaxy spectra from previous Data Releases. The DR9 galaxy sample comprises $\sim860\,000$ unique galaxy spectra from the SDSS-I and SDSS-II surveys, as well as $\sim490\,000$ unique galaxies from the SDSS-III Baryon Oscillation Spectroscopic Survey (BOSS; \citealt{Dawson2012BOSS}). Here, our galaxy selection criteria are different than in GM13 not only because the selection is based on SDSS DR9 data instead of DR7, but also because we rely on galaxy properties derived by the MPA-JHU Galspec pipeline\footnote{\url{http://www.sdss3.org/dr9/algorithms/galaxy\_mpa\_jhu.php}} \citep{2003MNRAS.341...33K,2004MNRAS.351.1151B,2004ApJ...613..898T}, which was run on galaxy spectra from the earlier SDSS-I and SDSS-II surveys (i.e., up to and including DR7). Of the $\sim860\,000$ galaxy spectra that were run through the Galspec pipeline, we have selected those spectra that conformed to the following criteria.
\begin{enumerate}
 \item The spectra must have at least 500 good pixels, defined as those pixels for which the MASK bit flag is set to either `0' (good) or `40000000' (emission line). 
 \item The spectra were run through the Galspec pipeline, as determined by requiring that the PLATE, MJD, and FIBER header keywords in the galSpecInfo table be set to values different than $-1$ and that the RELIABLE keyword be different than 0.
 \item The stellar mass within the fibre aperture, as measured by Galspec, must be smaller than the total stellar mass of the galaxy.
 \item The redshifts of the galaxies, $z$, be greater than zero, their uncertainties $\Delta(z)\le 0.015$, and without any error flags.
 \item All SFR and sSFR values measured by Galspec be different than either 0 or $-9999$.
\end{enumerate}
\noindent After applying these criteria, we were left with 768\,756 galaxies.

Of the galaxies in this sample, 69\,973 galaxies were targeted for spectroscopy more than once. So that we do not count any SNe more than once, we require that each SN appear in only one spectrum (in other words, we require that our observation epochs be independent). Due to the shape of their light curves, SNe Ia are expected to be visible in our sample for $<60$ d. However, as most CC SNe are SNe IIP (e.g., \citealt{li2011LF}), we are most likely to catch them during their plateau phase, so that most SNe IIP we observe will be visible in our sample for $\sim100$ d. Hence, we chose to exclude any observations spaced less than 120 d apart. After this step, we were left with 741\,440 galaxies, of which 29\,160 had multiple observations, accounting for 58\,576 spectra. Of these galaxies, 28\,904 had one more observation, and 256 had two more.

Some of the objects classified by the SDSS imaging pipeline as galaxies and later targeted for spectroscopy may be misclassified H\,{\small II} regions within galaxies or fragments of large, spatially-resolved galaxies. As Galspec would compute low stellar masses for such galaxy `fragments,' they are a source of contamination for our sample. To remove any such objects, we have used the source identification program \sex\footnote{\url{http://www.astromatic.net/software/sextractor}} \citep{sextractor} to detect our galaxies in the SDSS \R-band images and measure their centroids. We then calculated the distance, in arcsec, between these centroids and those measured by the SDSS pipeline. We have chosen to exclude any objects with a distance between centroids of $>3$ arcsec, i.e., larger than the diameter of the SDSS fibre aperture. We have calculated that of the `giant' galaxies in our sample, defined here as galaxies with stellar masses larger than that of the Large Magellanic Cloud ($3\times10^9$ ${\rm M_\odot}$; \citealt{2002AJ....124.2639V}), only $\sim0.2$ per cent might be galaxy fragments, so we do not remove them, as they will have a negligible effect on the SN rates calculated in Section~\ref{sec:results}. Of the 52\,041 `dwarf galaxies' in our sample ($\le 3\times10^9$ ${\rm M_\odot}$), 1\,856 (3.6 per cent of the subsample) are suspected of being galaxy fragments. As this constitutes a non-negligible fraction of the dwarf-galaxy subsample, it is removed, leaving a final dwarf-galaxy subsample of 50\,185 galaxies. As discussed in Section~\ref{sec:sample}, we also apply this test to all SN host galaxies in our sample. Our final galaxy sample numbers \Ngal\ galaxies

\begin{figure}
 \begin{center}
  \includegraphics[width=0.475\textwidth]{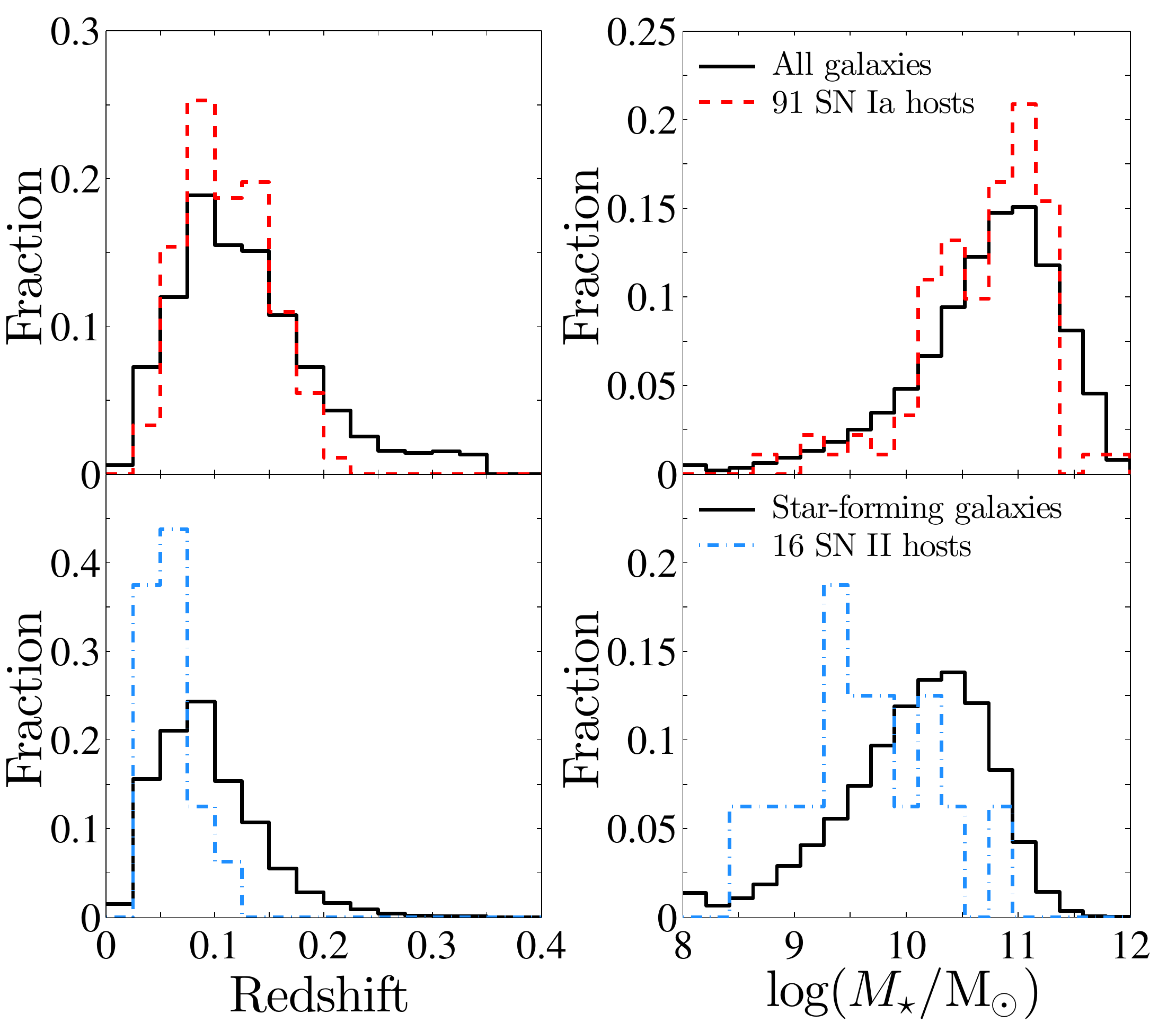}
  \caption{Redshift (left) and total stellar mass, $M_\star$ (right), distributions for all galaxies in our sample (upper panels, solid curve) and for star-forming galaxies that do not host AGNs (lower panels, solid curve). The upper and lower panels also display the distributions of the SN~Ia (red dashed curve) and SN~II (blue dot-dashed curve) host galaxies, respectively.}
  \label{fig:gals} 
 \end{center}
\end{figure}

\begin{figure}
 \begin{center}
  \includegraphics[width=0.475\textwidth]{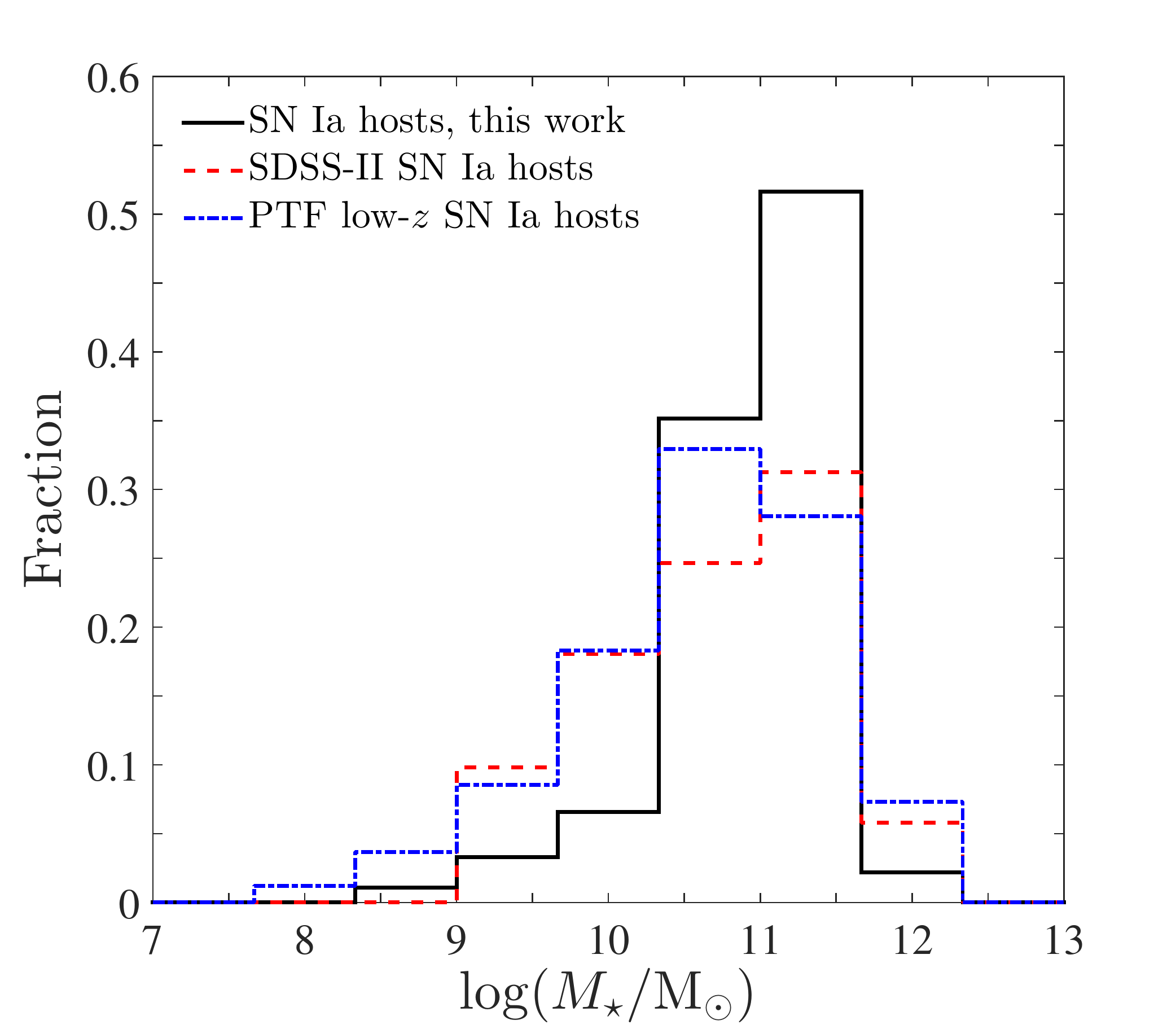}
  \caption{A comparison between the distribution of stellar masses of the SN Ia host galaxies in our sample (solid black) with those from the SDSS-II SN survey (dashed red; 499 spectroscopically-classified SNe Ia; \citealt{2014arXiv1401.3317S}) and the low-redshift PTF sample (dot-dashed blue; 82 SNe Ia; \citealt{2014MNRAS.438.1391P}). Our SN Ia sample is biased towards host galaxies with higher stellar masses than those in the SDSS and PTF samples.}
  \label{fig:ptf_sdss}
 \end{center}
\end{figure}

In Fig.~\ref{fig:gals}, we show the redshift and stellar mass distributions of the galaxies and SN-hosting galaxies in our sample. For our SN Ia sample, we display the appropriate distributions for all 739\,584 galaxies in our sample in the upper panels. However, for reasons that will be elaborated in Section~\ref{sec:sample}, we limit our SN II sample to those SNe detected among the \NIIgal\ star-forming galaxies that display no sign of an AGN. The redshift and stellar mass distributions for these galaxies are shown in the lower panels of Fig.~\ref{fig:gals}. 

Fig.~\ref{fig:ptf_sdss} presents a comparison of the stellar mass distribution of our SN Ia host galaxies with those from two other SN Ia samples in redshift ranges comparable to that of our own: 499 host galaxies of spectroscopically classified SNe Ia from the SDSS-II SN survey \citep{2014arXiv1401.3317S} and 82 host galaxies of low-redshift ($z<0.087$) SNe Ia from the Palomar Transient Factory (PTF; \citealt{2009PASP..121.1334R,2014MNRAS.438.1391P}). A two-sample Kolmogorov-Smirnov test shows that the SDSS and PTF samples are consistent with each other at a 5 per cent significance level. However, both of the latter are inconsistent with our sample at the same significance level. Relative to these samples, our SN Ia host galaxy sample is biased towards galaxies with high stellar masses ($5\times10^{10} \le M_\star \le 5\times10^{11}~{\rm M_\odot}$). This bias is due to the target selection strategy employed by the SDSS spectroscopic survey.

Although our final galaxy sample is larger by 31\,792 galaxies (4.5 per cent) than the GM13 sample, 85 per cent of the galaxies in the current sample also appear in the GM13 sample. Yet, the two samples differ in three key aspects:
\begin{enumerate}
 \item The spectra of the galaxies in the new sample have been re-reduced with the more advanced version of the SDSS reduction pipeline implemented in DR9. As we describe in Appendix~A, this has a direct effect on the number of SNe detected in this work.
 \item Although 70 per cent of the dwarf galaxies in the current sample are also included in the GM13 galaxy sample, the remaining 30 per cent host half of the SNe II reported in Section~\ref{sec:sample}, as well as four SNe Ia.
 \item The properties of the current galaxy sample have been measured with a different pipeline than the one used in GM13 (i.e., Galspec as opposed to \vespa). This will allow us to test any systematic uncertainty on the SN rates caused by these pipelines.
\end{enumerate}


\section{Supernova discovery and classification}
\label{sec:method}

For a full description of the method we use to detect and classify SNe in galaxy spectra, we refer the reader to section 3 of GM13. We make no changes to our method when searching for SNe Ia, but make the following changes when searching for CC SNe. 

To optimize the detection method to the discovery of SNe~Ia, in GM13 we introduced a figure of merit, $\chi^2_\lambda$, composed of the reduced $\chi^2$ value, $\chi^2_r$, divided by the wavelength range covered by the transient template. This figure of merit was used to select the best-fitting transient template to the suspected SN signal detected in the SDSS spectrum. Here, we use the same $\chi^2_\lambda$ figure of merit when searching for SNe~Ia, but revert to the standard $\chi^2_r$ figure of merit when searching for CC~SNe, as $\chi^2_\lambda$ penalizes CC~SNe, which have fewer templates in the Supernova Identification code (\snid;\footnote{\url{http://people.lam.fr/blondin.stephane/software/snid/}} \citealt{Blondin2007}), with narrower wavelength coverage, relative to the SN~Ia templates. As in GM13, we use the second version of the \snid\ spectral library. A future version of our detection code will also include the SN Ia library published by \citet{Silverman2012SNDB} and the SN Ib/c library described in \citet{2014AJ....147...99M} and \citet{2014arXiv1405.1437L}.

In GM13 we included SNe~IIP, IIL, IIn, and peculiar SNe~II (i.e., most resembling SN1987A) in our SN II category. Here, we remove the SN IIn and SN1987A templates from our classification library and require that any suspected SNe II be classified as either a SN~IIP or IIL by both classification stages. This minimizes confusion between SNe~IIP/L and SNe~IIn, which share a prominent H$\alpha$ feature. 

Finally, as in any SN survey, we may miss SNe in our data for various reasons, such as data with low signal-to-noise (S/N) ratios; the SN being too faint, relative to its galaxy, to be detected; or failing to meet any of our classification criteria. We quantify these, and other systematic effects, by planting artificial SN spectra in our galaxy spectra (as made available by the SDSS pipeline, i.e., after reduction and extraction) at random and counting what fraction of these SNe is recovered by our method. As we have made no changes to our detection method when searching for SNe~Ia, we continue to use the detection efficiency vs. SN~Ia \R-band magnitude curves we measured in GM13. In order to measure similar curves for the SNe II detected here, we plant 15\,000 artificial SNe~IIP and SNe~IIL in a random subsample of our galaxies. As SNe~II are known to explode only in star-forming galaxies, we choose only those galaxies that have a specific star-formation rate of ${\rm log(sSFR/yr^{-1})}\ge-12$ \citep{2006ApJ...648..868S}, which leaves 523\,143 galaxies, or $\sim70.7$ per cent of the sample. The artificial SNe~II are planted according to the luminosity functions and population fractions presented in tables 6 and 7 of \citet{li2011LF}, namely 58.9 and 41.1 per cent SNe~IIP and SNe~IIL, respectively. 

We perform two simulations of the detection efficiency, each time randomly selecting the absolute magnitudes of the artificial SNe~II, relative to maximum light in the $B$ band, from different luminosity functions (LFs). In one simulation, we use the empirical LFs measured by \citet{li2011LF}, which are uncorrected for extinction by dust in the host galaxies of the SNe. The means and standard deviations of these LFs for SNe IIP and IIL are: $M_{B,{\rm IIP,wd}}=-15.66$, $\sigma_{\rm IIP,wd}=1.23$, $M_{B,{\rm IIL,wd}}=-17.44$, and $\sigma_{\rm IIL,wd}=0.64$ mags (Vega). For the second simulation, we use a version of these LFs which has been corrected for host-galaxy dust extinction \citep{Graur2014,2014AJ....148...13R}, with means and standard deviations of $M_{B,{\rm IIP,nd}}=-16.56$, $\sigma_{\rm IIP,nd}=0.80$, $M_{B,{\rm IIL,nd}}=-17.66$, and $\sigma_{\rm IIL,nd}=0.42$. When using the latter set of LFs, we redden the artificial SN spectra according to the \citet*{1989ApJ...345..245C} reddening law and $A_V$ values chosen at random from the positive side of a Gaussian distribution centred on $A_V=0$, with a standard deviation of $\sigma_{A_V}=0.93$ \citep{neill2006}. Although the mean of the SN~II \R-band magnitude distribution that results from the dust-corrected LFs is 0.1 mag larger (i.e., fainter) than the corresponding distribution that results from the uncorrected \citet{li2011LF} LFs, a two-sample Kolmogorov-Smirnov test cannot reject the null hypothesis that both sets of \R-band magnitudes originate in the same distribution, at a one per cent significance level.

\begin{figure}
 \begin{center}
  \includegraphics[width=0.475\textwidth]{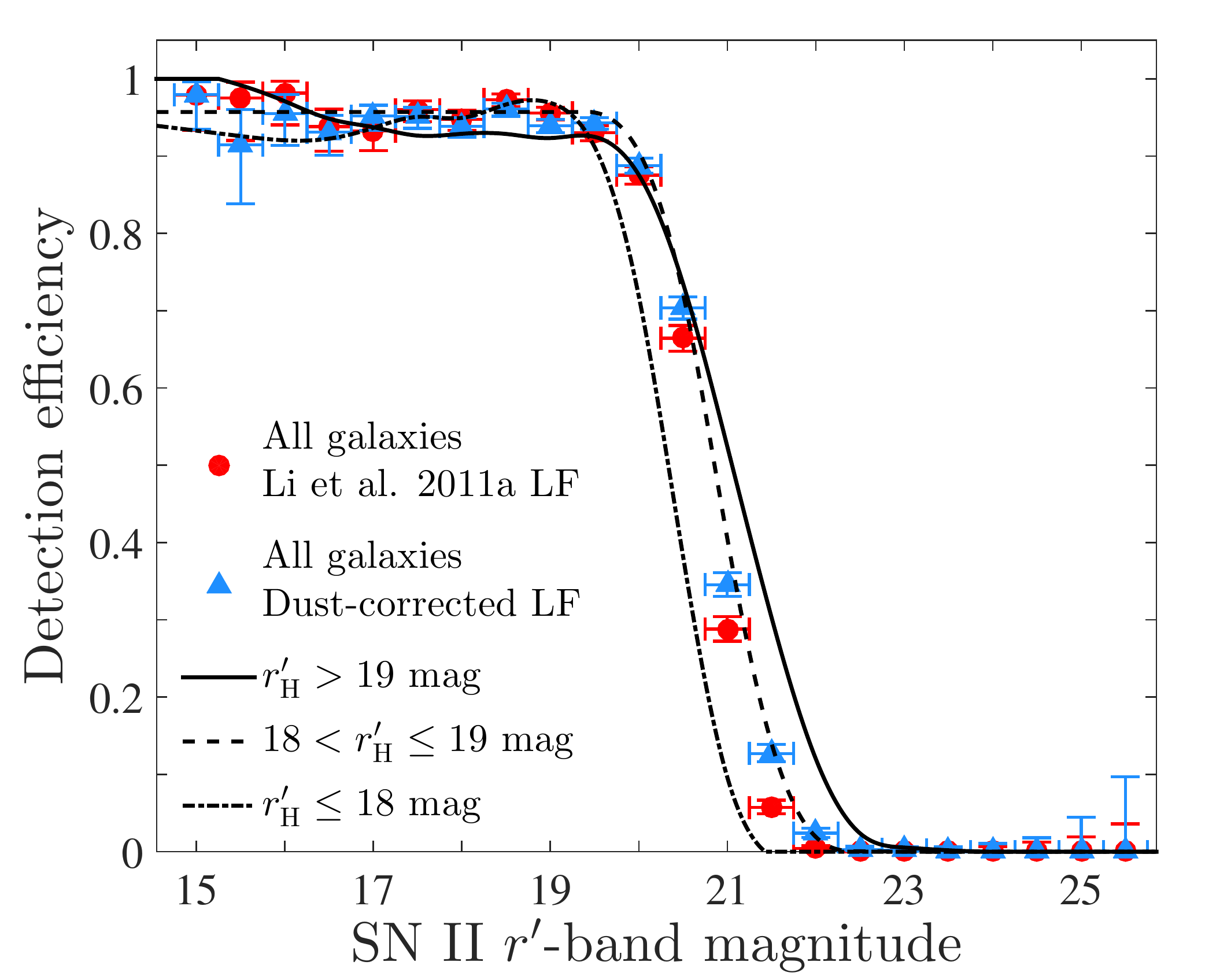}
  \caption{SN~II detection efficiency as a function of SN~II \R-band magnitude. Filled red circles denote the fraction of artificial SNe~II detected in all galaxies in 0.5-mag-wide bins, for artificial SNe II planted using the empiricial \citet{li2011LF} LFs. The filled blue triangles show the same type of measurement for artificial SNe II planted using the dust-corrected LFs used in \citet{Graur2014} and \citet{2014AJ....148...13R}. Error bars indicate 1$\sigma$ binomial uncertainties. Curves are cubic-spline fits to the detection-efficiency measurements of artificial SNe~II planted using the dust-corrected LFs in host galaxies with $\R_{\rm H}$-band magnitudes in different ranges, as marked. The SN~II detection efficiency declines to 50 per cent at 20.8 mag when using either of the above LFs.}
  \label{fig:fakes_eff}
 \end{center}
\end{figure}

In Fig.~\ref{fig:fakes_eff}, we show our SN~II detection efficiency as a function of the \R-band apparent magnitude of the artificial SNe when using the two different sets of LFs described above. We reach 50 per cent detection efficiency at $\R=20.8$ mag when using either of the LF sets. This value is 0.4 mag fainter than the GM13 50-per-cent efficiency mark for SNe~Ia, at $\R=20.4$ mag. We attribute this difference to the prominent H$\alpha$ feature in SN~II spectra, which remains detectable in the galaxy spectra even for SNe~II, which are generally fainter than SNe~Ia by 1--2 mags. Because the probability of detecting a SN using our method depends on the brightness of the host galaxy, we divide the SN~II detection efficiency measurements into three subsets according to the \R-band magnitudes of their host galaxies, $\R_{\rm H}$: $\R_{\rm H}>19$, $18<\R_{\rm H}\le19$, and $\R_{\rm H}\le18$. We have fit each of these subsets with cubic splines, as shown in Fig.~\ref{fig:fakes_eff}.

Based on the results of the detection efficiency simulation, we find that there is a negligible probability that our SN~II sample would be contaminated by either misclassified SNe~Ia or SNe~Ib/c. Of the artificial SNe II recovered in this simulation, 99.36 per cent were correctly classified as SNe II when using the \citet{li2011LF} LF. When using the dust-corrected LF, the fraction of SNe classified correctly was 99.20 per cent. When using either the \citet{li2011LF} or the dust-corrected (in parentheses) LFs, only 0.10 (0.05) per cent of the SNe wre classified as SNe~Ia, 0.01 (0.01) per cent as SNe~Ia/Ic, 0 (0.06) per cent as SNe~Ib/c, 0 (0) per cent as SNe~Ic/Ia, and 0.53 (0.69) per cent as AGNs. From a visual inspection of the 48 and 63 SNe~II misclassified as AGNs in each simulation, we find that in most cases the SN was planted in an AGN-hosting galaxy. We take this possibility into account when searching for real SNe by running our detection method on the galaxy sample twice: once without taking AGNs into account, and once by adding an AGN spectral template to the mix of galaxy eigenspectra and transient templates in the data-fitting stages.

\section{Supernova sample}
\label{sec:sample}

Here, we describe how we arrive at our final sample of 91 SNe Ia and 16 SNe II, as well as differences between this and the GM13 samples. Throughout this work, we will refer to the SNe in our sample by the numbers of the SDSS plate, Modified Julian Date (MJD), and fibre in which they were detected. For example, the SN~II referred to as 0437-51876-322 was detected in a galaxy named SDSS J075813.33+440108.1, which was observed on MJD 51876 (Nov. 28 2000) with the 322nd fibre on the 437th SDSS plate. Without light curves, we cannot say with confidence which of our 16 SNe II are SNe IIP and which are SNe IIL (see section 4 of GM13), and so refer to them only as SNe II. 

We have detected a total of 109 SNe among the $\sim740\,000$ galaxy spectra in our sample, of which 17 are SNe~II and 92 are SNe~Ia. Our detection and classification code found 117, 86, and 292 SN Ia, Ia/Ic, and II candidates, respectively. As some spectra may have been misclassified as containing SN signals due to random noise spikes or bad emission-line fits when the galaxy template was constructed, we visually inspected the SN Ia and Ia/Ic candidates and removed 31 and 81 spectra, respectively. Most of the SN II candidates were classified as SNe II solely by a broad H$\alpha$ feature, which could also be attributed to an AGN in the galaxy. To break the degeneracy between SNe II and AGNs, we used the \citet*[hereafter BPT]{1981PASP...93....5B} diagram from \citet{2003MNRAS.346.1055K} to limit our galaxy sample to \NIIgal\ star-forming galaxies that show no sign of AGNs in their spectra. This limited our SN II sample to 17 SNe II. The majority of the SN/AGN candidates removed by this cut contained only a broad H$\alpha$ feature; none of them contained P-Cygni profiles or other features that would have conclusively classified them as SNe II. A visual inspection of the resulting SN II sample revealed no faulty spectra (i.e., showing no evidence of random noise spikes or other non-SN features) or signs of contaminating AGNs. 

A visual check of the SN host galaxies revealed that one SN~Ia, namely 1745-53061-056, did not explode in the centre of its galaxy but was erroneously targeted as a galaxy fragment and evaded our fragment exclusion test from Section~\ref{sec:galaxies}. This SN~Ia was previously identified and reported as SN2004ck by \citet{2004IAUC.8359....1C}. We have removed it from our sample.

\begin{table*}
 \caption{SNe discovered in this work.}\label{table:SNe}
 \begin{tabular}{l c c c c c c c c c c c c c}
  \hline
  \hline
  \multicolumn{1}{c}{SDSS Name} & Plate-MJD-fibre & Date & $z$ & Age$_1$ & Age$_2$ & $s$ & $\R_{\rm SN}$ & $\R_{\rm H}$ & $\chi^2_{\rm gal}$ & $\chi^2_{\rm SN}$ & Type \\
  \multicolumn{1}{c}{(1)} & (2) & (3) & (4) & (5) & (6) & (7) & (8) & (9) & (10) & (11) & (12) \\
  \hline
   J153856.21+474546.4 & 1167-52738-214 & 09/04/03 & 0.070 & -9 & -11 & 0.91 & 20.57 & 19.11 & 1.4 & 1.0 & Ia \\
   J112148.00+125250.6 & 1605-53062-528 & 27/02/04 & 0.101 & 0 & -2 & 1.01 & 20.22 & 18.63 & 1.7 & 1.1 & Ia \\
   J095842.45+200817.2 & 2371-53762-404 & 27/01/06 & 0.039 & 123 & 123 & $\cdots$ & 20.48 & 19.66 & 1.1 & 1.0 & Ia \\
   J141852.38+005318.6 & 0304-51609-436 & 06/03/00 & 0.129 & -1 & -7 & 0.90 & 19.47 & 17.99 & 1.9 & 1.2 & Ia \\
   J232650.82--095632.8 & 0646-52523-183 & 06/09/02 & 0.052 & -5 & -5 & 1.28 & 19.99 & 17.28 & 1.4 & 1.3 & Ia \\
   J010939.83+000346.9 & 0694-52209-152 & 27/10/01 & 0.078 & 21 & 59 & 0.89 & 20.59 & 17.42 & 1.5 & 1.3 & Ia \\
   J092620.08+502157.2 & 0767-52252-123 & 09/12/01 & 0.060 & 3 & -1 & $\cdots$ & 20.38 & 17.83 & 1.2 & 1.2 & Ia \\
   J114447.93+041652.3 & 0838-52378-021 & 14/04/02 & 0.104 & 2 & -1 & 0.98 & 19.63 & 18.30 & 1.5 & 1.1 & Ia \\
   J032108.86+411510.9 & 1665-52976-155 & 03/12/03 & 0.016 & 21 & 15 & 1.05 & 20.54 & 16.70 & 1.7 & 1.5 & Ia \\
   J102934.76+131637.3 & 1747-53075-177 & 11/03/04 & 0.090 & 41 & 33 & 1.02 & 20.53 & 17.79 & 1.2 & 1.0 & Ia \\
   J140309.73+060754.3 & 1808-54176-107 & 17/03/07 & 0.117 & 3 & -4 & 1.14 & 19.26 & 18.01 & 2.2 & 1.2 & Ia \\
   J135439.29+280952.9 & 2118-53820-468 & 26/03/06 & 0.073 & 12 & 9 & 0.85 & 20.02 & 17.45 & 1.1 & 1.0 & Ia \\
   J142608.24+152501.9 & 2746-54232-635 & 12/05/07 & 0.053 & 2 & -5 & 0.78 & 18.58 & 17.11 & 3.7 & 1.5 & Ia \\
   J162333.74+252420.7 & 1574-53476-461 & 16/04/05 & 0.190 & -0 & 8 & $\cdots$ & 19.86 & 18.53 & 1.2 & 1.1 & Ia/Ic\\
   J095313.01+305122.4 & 1946-53432-030 & 03/03/05 & 0.045 & 3 & 22 & $\cdots$ & 19.24 & 17.91 & 2.1 & 1.2 & Ia/Ic \\
   J075813.33+440108.1 & 0437-51876-322 & 28/11/00 & 0.047 & 24 & 0 & $\cdots$ & 19.98 & 17.46 & 1.6 & 1.3 & II \\
   J073753.33+315331.0 & 0541-51959-057 & 19/02/01 & 0.098 & 4 & -3 & $\cdots$ & 20.65 & 18.44 & 1.6 & 1.4 & II \\
   J114447.10+535501.4 & 1015-52734-019 & 05/04/03 & 0.062 & 19 & 44 & $\cdots$ & 20.02 & 18.94 & 1.9 & 1.4 & II \\
   J163305.64+350600.9 & 1339-52767-327 & 08/05/03 & 0.035 & -4 & -4 & $\cdots$ & 19.40 & 17.57 & 1.6 & 1.2 & II \\
   J170626.69+232409.9 & 1689-53177-325 & 21/06/04 & 0.063 & 250 & 310 & $\cdots$ & 21.95 & 17.37 & 1.5 & 1.2 & II \\
   J074820.66+471214.2 & 1737-53055-369 & 20/02/04 & 0.062 & -7 & -6 & $\cdots$ & 19.05 & 18.05 & 1.7 & 1.4 & II \\
   J133057.65+365921.2 & 2102-54115-072 & 15/01/07 & 0.058 & 64 & 96 & $\cdots$ & 20.61 & 19.41 & 1.7 & 1.0 & II \\
   J131503.77+223522.7 & 2651-54507-488 & 11/02/08 & 0.023 & 250 & 51 & $\cdots$ & 21.20 & 17.36 & 1.6 & 1.2 & II \\
  \hline
  \multicolumn{12}{l}{{\it Note.} A complete version of this table, detailing all SNe in our sample, is available in the electronic version of this paper -- see} \\
  \multicolumn{12}{l}{Supporting Information.} \\
  \multicolumn{12}{l}{(1) -- SDSS name, composed of right ascension and declination (J2000).} \\
  \multicolumn{12}{l}{(2) -- SDSS DR9 Plate, MJD, and fibre in which the SN was discovered.} \\
  \multicolumn{12}{l}{(3) -- Date on which the SN was captured, in dd/mm/yy.} \\
  \multicolumn{12}{l}{(4) -- SN host-galaxy redshift.} \\
  \multicolumn{12}{l}{(5) and (6) -- SN age, in days, as measured by SVD and \snid, respectively (with an uncertainty of $\pm6$ d).} \\ 
  \multicolumn{12}{l}{(7) -- SN stretch, as measured with the \salt\ templates. All stretches have an uncertainty of $^{+0.10}_{-0.14}$.} \\
  \multicolumn{12}{l}{(8) and (9) -- SN and host-galaxy \R-band magnitudes.} \\
  \multicolumn{12}{l}{(10) and (11) -- Reduced $\chi^2$ value of galaxy and galaxy+transient fits.} \\
  \multicolumn{12}{l}{(12) -- SN type.}
 \end{tabular}
\end{table*}

We have checked whether any of the SNe in our sample may have exploded outside the area covered by the spectral aperture by comparing the observed \R-band magnitudes of the SNe with those one would expect given their measured ages, Galactic extinction along the line of sight to their host galaxies, host galaxy extinction (simulated using the \citealt{neill2006} exinction model mentioned in Section~\ref{sec:method}), and in the case of the SNe Ia in our sample -- stretch values as well. We found that one SN II, namely 2138-53757-256, was $\sim5.6$ magnitudes brighter than expected, and so might be an example of a superluminous SN. As it is too luminous to be considered a normal SN II, we remove it from our sample, leaving 16 SNe II. Four of the SNe Ia in our sample appeared abnormally faint: 1452-53112-120, 0646-52523-183, 0767-52252-123, and 1665-52976-155, which were 2.3, 2.2, 2.0, and 4.4 magnitudes too faint, compared to the mean SN Ia magnitude at that phase. Five SNe Ia were classified by our method as Ia/Ic, which means they may be misclassified SNe Ic. We keep these nine SNe Ia in our sample, but treat them as a systematic uncertainty in the derivation of the SN~Ia rates, as detailed below in Section~\ref{subsec:systematics}.

Our final sample comprises 91 SNe Ia and 16 SNe II. The residual spectra and best-fitting SN templates of the SNe detected in this work and not shared with the GM13 sample are available in the electronic version of the paper -- see Supporting Information. The SN spectra of the complete sample, along with those from the GM13 sample, have been added to the Weizmann Interactive Supernova data REPository (WISeREP;\footnote{\url{http://wiserep.weizmann.ac.il/}} \citealt{Yaron2012WISeREP}). We caution that the SN spectra from this work and from GM13 should be used carefully, as there are known systematics affecting the shapes of their (pseudo)-continuua, and thus their luminosities; see section 3.3 of GM13, where we describe how the flow of flux between the galaxy eigenspectra and transient spectra used in the fits causes the resulting residual spectra to appear, on average, systematically brighter in the \G\ band and fainter in the \R\ and \I\ bands.

Of the SNe in our final sample, 15 SNe Ia and 8 SNe II are reported here for the first time. These SNe are listed in Table~\ref{table:SNe}. For a description of how we measure the SN properties listed in the table, we refer the reader to section 4 of GM13. An expanded version of Table~\ref{table:SNe}, detailing all of the SNe in our final sample, is available in the electronic version of the paper -- see Supporting Information. Appendix A describes in detail differences between the GM13 SN sample and our current one, as well as which SNe detected here were previously reported in other works.


\section{Supernova Rate Analysis}
\label{sec:results}

In this Section, we use the SN samples from Section~\ref{sec:sample} and visbility times we calculate in Section~\ref{subsec:vistime} to measure SN Ia and SN II rates per unit mass as a function of stellar mass, SFR, and sSFR in Sections~\ref{subsec:rates_Ia} and \ref{subsec:rate_mass}, respectively. In Section~\ref{subsec:rate_vol}, we convert the mass-normalized SN II rate into a volumetric rate at a median redshift of 0.075. Finally, in Section~\ref{subsec:systematics}, we discuss several sources of systematic uncertainty that affect the rates measured here.

\subsection{Visibility time}
\label{subsec:vistime}

To measure SN rates, we must first calculate the visibility time for each of the galaxies in our sample. This is the period of time during which the SN emission was bright enough to be detected by our method. Section 5.1 in GM13 describes how we measure the visibility times for SNe Ia. We follow the same procedure here for SNe Ia and make the following changes in order to measure visibility times for SNe II:
\begin{enumerate}
 \item The SN II light curves are constructed using the \citet*{Gilliland1999} SN~IIP and SN~IIL templates.\footnote{\url{https://c3.lbl.gov/nugent/nugent\_templates.html}}
 \item We assign the SN II light curves absolute $B$-band magnitudes at maximum light, which we draw from the dust-corrected LFs discussed in Section~\ref{sec:method}.
 \item For SNe Ia, we use the detection efficiency curves from GM13, while for SNe II we use the curves shown in Fig.~\ref{fig:fakes_eff}.
\end{enumerate}

In order to stretch the SN Ia light curves (see equation 1 in GM13), we take $\alpha=1.52$ \citep{2006A&A...447...31A}. This value is consistent with the `Combined \citet{2011ApJS..192....1C}' value of $\alpha=1.434\pm0.093$ presented in \citet{2014A&A...568A..22B}. In order to compare the rates measured here with those from GM13, we continue to use the \citet{2006A&A...447...31A} value.

\subsection{The Type Ia supernova rate-mass correlation}
\label{subsec:rates_Ia}

When measuring SN rates, we must normalize them by some parameter that reflects the stellar population that was surveyed during the experiment. Here, we take the stellar mass measured from the light inside the SDSS fibre aperture. This choice of normalization allows us to remove the bias caused by the fibre aperture covering different fractions of the light from each of the galaxies in our sample, depending on their angular size. We measure mass-normalized SN~Ia rates as a function of total galaxy stellar mass, SFR, and sSFR. In each case, we measure the rates in four bins, where the width of the bins is chosen so that each bin contains roughly the same number of SNe~Ia. In each bin $i$, we measure the mass-normalized rates $R_{{\rm Ia,M},i}$ by dividing the number of SNe~Ia in that bin, $N_{{\rm Ia},i}$, by the sum of the visbility times of the $n$ galaxies in that bin, $t_{v,j}$, multiplied by the stellar mass of each galaxy within the fibre aperture, $M_{\star_f,j}$:
\begin{equation}\label{eq:rates_Ia}
 R_{{\rm Ia,M},i} = \frac{N_{{\rm Ia},i}}{\sum\limits_{j=1}^n t_{v,j} M_{\star_f,j}}.
\end{equation}

\begin{figure}
 \centering
 \includegraphics[width=0.475\textwidth]{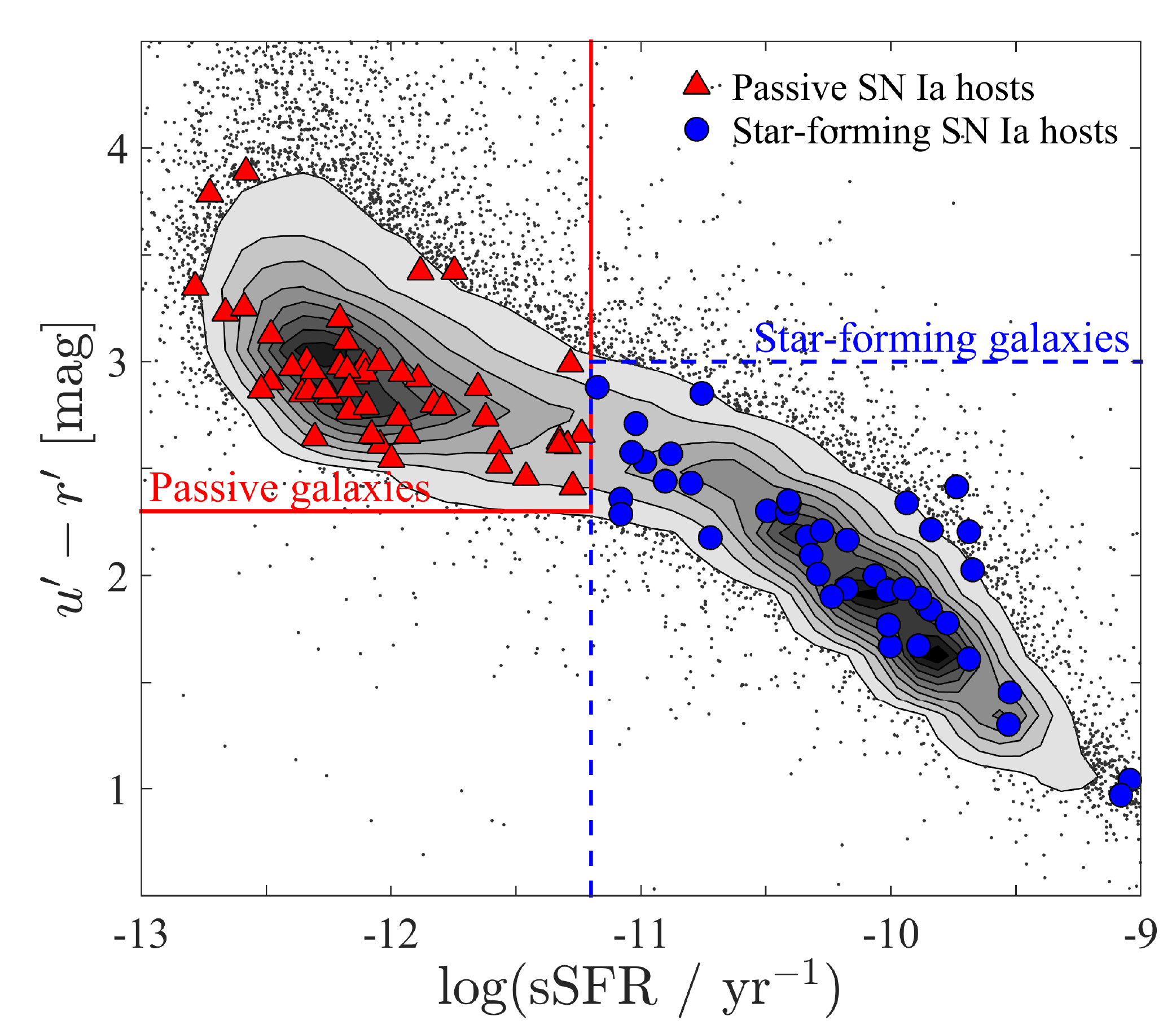}
 \caption{Galaxy colour vs. sSFR. The contours denote the density of the galaxies in our sample in the parameter space spanned by their $\U-\R$ colour (derived from model magnitudes) and sSFR, in decrements of 10 per cent. The solid red and dashed blue curves denote our cuts on this parameter space in order to divide the galaxy sample into passive and star-forming galaxies, respectively. The passive and star-forming SN Ia host galaxies are shown as red triangles and blue circles, respectively.}
 \label{fig:urssfr}
\end{figure}

We measure the SN~Ia rates in all the galaxies in our sample, as well as in subsamples of passive and star-forming galaxies, chosen according to the sSFR of the galaxy. In GM13, we followed \citet{2006ApJ...648..868S}, who used the sSFR of the galaxies in their sample to define passive $[{\rm log(sSFR/yr^{-1})}<-12]$, star-forming $[-12\le{\rm log(sSFR/yr^{-1})}<-9.5]$, and highly star-forming galaxies $[{\rm log(sSFR/yr^{-1})}\ge-9.5]$. However, the galaxies in our sample show a distinct separation in the $\U-\R$ vs. sSFR space, as shown in Fig.~\ref{fig:urssfr}, according to which we define passive galaxies as having ${\rm log(sSFR/yr^{-1})}<-11.2$ and star-forming galaxies as having ${\rm log(sSFR/yr^{-1})}\ge-11.2$. This definition happens to split our galaxy sample roughly evenly between passive and star-forming galaxies. The SN~Ia host galaxies are then divided into 49 and 42 passive and star-forming galaxies, respectively. A visual inspection of the SN~Ia host galaxies classified in this manner confirms this classification, i.e., all but one of the host galaxies classifed as passive have no emission lines due to ongoing star formation (some galaxies exhibit emission lines consistent with the galaxy harboring an AGN, as defined by the \citealt{2003MNRAS.346.1055K} BPT diagram). One of the 49 host galaxies classified as passive is classified by Galspec (based on the BPT diagram; \citealt{2004MNRAS.351.1151B}) as `low S/N star-forming' and so may represent the one contaminant of our passive host galaxy sample. Of the 42 host galaxies classified as star-forming by our criterion, 34 are classified by Galspec as `star forming' or `low S/N star-forming,' two are classified as `composite' (i.e., they fall inbetween the purely star-forming and purely AGN-hosting galaxies in figure 1 of \citealt{2004MNRAS.351.1151B}) and six are classified as AGNs. In Section~\ref{subsub:ssfr_criterion}, we discuss how this new criterion may affect the visibility times of the galaxies in the sSFR range $-12<{\rm log(sSFR/yr^{-1})}<-11.2$, and subsequently the SN Ia rates. 

\begin{table*}
 \center
 \caption{Mass-normalized SN rates vs. various galaxy properties.}\label{table:rates}
 \begin{tabular}{ccccccccc}
  \hline
  \hline
  Stellar mass$^a$ & SN rate$^b$ & SFR & SN rate & sSFR & SN rate & $N_{\rm SN}^c$ \\
  $(10^{10}~{\rm M_\odot})$ & ($10^{-12}$ yr$^{-1}$ M$_\odot^{-1}$) & (M$_\odot$~yr$^{-1}$) & ($10^{-12}$ yr$^{-1}$ M$_\odot^{-1}$) & ($10^{-12}$~yr$^{-1}$) & ($10^{-12}$ yr$^{-1}$ M$_\odot^{-1}$) & \\
  \hline
  \multicolumn{7}{c}{SNe Ia in all galaxies} \\
  $0.8^{+0.8}_{-0.6}$ & $0.271^{+0.069,+0.034}_{-0.056,-0.047}$ & $0.037^{+0.017}_{-0.017}$ & $0.099^{+0.025,+0.012}_{-0.020,-0.011}$ & $0.44^{+0.19}_{-0.17}$ & $0.080^{+0.021,+0.010}_{-0.017,-0.017}$ & 23 \\
  $3.5^{+1.2}_{-1.0}$ & $0.102^{+0.026,+0.014}_{-0.021,-0.009}$ & $0.10^{+0.05}_{-0.03}$    & $0.093^{+0.024,+0.020}_{-0.019,-0.017}$ & $1.5^{+1.7}_{-0.6}$    & $0.077^{+0.020,+0.025}_{-0.016,-0.007}$ & 22 \\
  $7.0^{+1.4}_{-1.2}$ & $0.104^{+0.026,+0.016}_{-0.021,-0.000}$ & $0.6^{+0.6}_{-0.3}$       & $0.090^{+0.023,+0.021}_{-0.019,-0.012}$ & $23^{+22}_{-14}$       & $0.104^{+0.026,+0.016}_{-0.021,-0.009}$ & 23 \\ 
  $16^{+14}_{-6}$     & $0.058^{+0.015,+0.013}_{-0.012,-0.010}$ & $3.4^{+3.9}_{-1.3}$       & $0.118^{+0.030,+0.018}_{-0.024,-0.010}$ & $140^{+150}_{-60}$     & $0.190^{+0.049,+0.026}_{-0.039,-0.002}$ & 23 \\
  $\mathbf{5^{+12}_{-4}}$ & $\mathbf{0.10\pm0.01\pm0.01}$ & $\mathbf{0.33^{+2.34}_{-0.27}}$ & $\mathbf{0.10\pm0.01\pm0.01}$ & $\mathbf{6^{+120}_{-6}}$ & $\mathbf{0.10\pm0.01\pm0.01}$ & $\mathbf{91}$ \\
  \multicolumn{7}{c}{SNe Ia in star-forming galaxies} \\
  $0.6^{+0.7}_{-0.4}$ & $0.30^{+0.10,+0.03}_{-0.08,-0.06}$      & $0.54^{+0.43}_{-0.38}$ & $0.14^{+0.05,+0.02}_{-0.04,-0.02}$      & $18^{+13}_{-9}$    & $0.086^{+0.030,+0.012}_{-0.023,-0.006}$ & 14 \\
  $2.7^{+1.0}_{-0.8}$ & $0.15^{+0.05,+0.02}_{-0.04,-0.02}$      & $1.8^{+0.5}_{-0.4}$    & $0.15^{+0.05,+0.02}_{-0.04,-0.02}$      & $70^{+30}_{-20}$   & $0.129^{+0.044,+0.017}_{-0.034,-0.009}$ & 14 \\ 
  $7.7^{+6.3}_{-2.6}$ & $0.076^{+0.026,+0.012}_{-0.020,-0.000}$ & $4.6^{+4.6}_{-1.5}$    & $0.106^{+0.037,+0.016}_{-0.028,-0.008}$ & $200^{+180}_{-60}$ & $0.265^{+0.091,+0.035}_{-0.070,-0.057}$ & 14 \\
  $\mathbf{2.3^{+5.4}_{-1.9}}$ & $\mathbf{0.129^{+0.023,+0.018}_{-0.020,-0.015}}$ & $1.6^{+3.1}_{-1.1}$ & $\mathbf{0.129^{+0.023,+0.018}_{-0.020,-0.015}}$ & $\mathbf{70^{+140}_{-60}}$ & $\mathbf{0.129^{+0.023,+0.018}_{-0.020,-0.015}}$ & $\mathbf{42}$ \\
  \multicolumn{7}{c}{SNe Ia in passive galaxies} \\
  $3.6^{+1.8}_{-1.8}$ & $0.091^{+0.029,+0.016}_{-0.023,-0.006}$ & $0.032^{+0.012}_{-0.013}$ & $0.091^{+0.029,+0.009}_{-0.023,-0.012}$ & $0.39^{+0.14}_{-0.15}$ & $0.068^{+0.022,+0.008}_{-0.017,-0.013}$ & 16 \\
  $7.8^{+1.2}_{-1.1}$ & $0.126^{+0.040,+0.025}_{-0.031,-0.001}$ & $0.069^{+0.017}_{-0.014}$ & $0.095^{+0.030,+0.016}_{-0.024,-0.007}$ & $0.81^{+0.21}_{-0.15}$ & $0.107^{+0.034,+0.018}_{-0.027,-0.014}$ & 16 \\
  $18^{+15}_{-7}$     & $0.056^{+0.017,+0.014}_{-0.013,-0.014}$ & $0.20^{+0.29}_{-0.08}$    & $0.065^{+0.020,+0.022}_{-0.016,-0.012}$ & $2.2^{+2.0}_{-0.9}$    & $0.078^{+0.024,+0.034}_{-0.019,-0.009}$ & 17 \\
  $\mathbf{10^{+16}_{-6}}$ & $\mathbf{0.081^{+0.013,+0.018}_{-0.012,-0.009}}$ & $\mathbf{0.08^{+0.18}_{-0.05}}$ & $\mathbf{0.081^{+0.013,+0.018}_{-0.012,-0.009}}$ & $\mathbf{0.8^{+1.7}_{-0.4}}$ & $\mathbf{0.081^{+0.013,+0.018}_{-0.012,-0.009}}$ & $\mathbf{49}$ \\
  \multicolumn{7}{c}{SNe II in star-forming galaxies without AGNs} \\
  $0.08^{+0.06}_{-0.05}$    & $5.5^{+3.7,+1.2}_{-2.4,-0.7}$           & $0.21^{+0.14}_{-0.13}$ & $1.82^{+1.23,+0.34}_{-0.79,-0.21}$      & $69^{+34}_{-35}$    & $0.25^{+0.17,+0.01}_{-0.11,-0.03}$ & 5 \\
  $0.35^{+0.15}_{-0.13}$    & $1.8^{+1.1,+0.3}_{-0.7,-0.2}$           & $0.65^{+0.17}_{-0.15}$ & $1.27^{+0.76,+0.16}_{-0.50,-0.11}$      & $170^{+60}_{-40}$   & $0.66^{+0.40,+0.03}_{-0.26,-0.05}$ & 6 \\
  $2.1^{+3.2}_{-1.2}$       & $0.187^{+0.126,+0.004}_{-0.081,-0.019}$ & $2.6^{+4.0}_{-1.3}$    & $0.213^{+0.144,+0.003}_{-0.092,-0.022}$ & $400^{+420}_{-100}$ & $2.62^{+1.77,+0.05}_{-1.13,-0.34}$ & 5 \\
  $\mathbf{1.0^{+3.0}_{-1.0}}$ & $\mathbf{0.52^{+0.16,+0.02}_{-0.13,-0.05}}$ & $\mathbf{1.5^{+3.3}_{-1.1}}$ & $\mathbf{0.52^{+0.16,+0.02}_{-0.13,-0.05}}$ & $\mathbf{130^{+150}_{-70}}$ & $\mathbf{0.52^{+0.16,+0.02}_{-0.13,-0.05}}$ & $\mathbf{16}$ \\
  \hline 
  \multicolumn{7}{l}{\textit{Note.} In the last line of each section, we present the SN rate averaged over all masses and redshifts of the galaxies in} \\
  \multicolumn{7}{l}{that specific subsample.} \\
  \multicolumn{7}{l}{$^a$ Median and 16th and 84th percentiles of the distribution of stellar mass, SFR, or sSFR of the galaxies in that bin.} \\
  \multicolumn{7}{l}{$^b$ Rate uncertainties are Poisson uncertainties on the number of SNe in each mass bin. Systematic uncertainties, from} \\ 
  \multicolumn{7}{l}{using different LFs and from SNe that were possibly misclassified or that exploded outside of the area covered by the} \\
  \multicolumn{7}{l}{fibre aperture, are separated by commas.} \\
  \multicolumn{7}{l}{$^c$ Number of SNe used for the measurements in that bin.}
 \end{tabular}
\end{table*}

\begin{figure}
 \centering
 \includegraphics[width=0.475\textwidth]{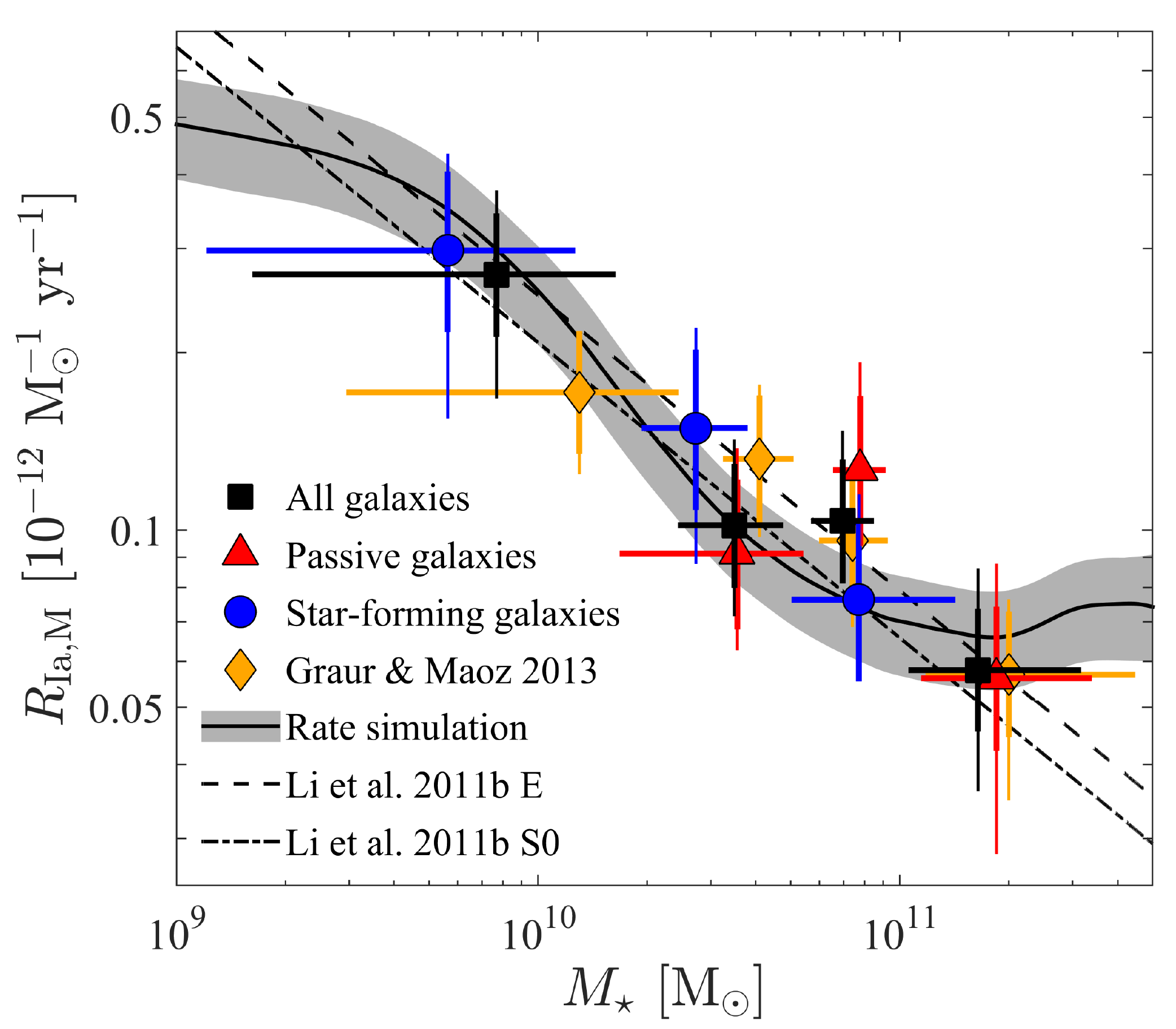}
 \caption{SN~Ia rates per unit mass, as a function of total galaxy stellar mass. The mass-normalized SN~Ia rates from this work are shown as black squares for all galaxies, red triangles for passive galaxies, and blue circles for star-forming galaxies. The GM13 rates are shown as yellow diamonds. Thick vertical error bars are based on the Poisson uncertainty on the number of SNe~Ia in the specific mass bin, thin vertical error bars show the added systematic uncertainties, and the horizontal error bars denote the range within which 68.3 per cent of the galaxies fall within each mass bin. The solid curve shows the best-fitting SN~Ia rate, simulated as a combination of a power-law DTD with an index of $-1$, and the \citet{2005MNRAS.362...41G} age-mass relation (i.e., galaxy `downsizing'). The shaded area is the confidence region resulting from the 68.3 per cent statistical uncertainty of the DTD amplitude $\Psi_1$, the only free parameter in the fit. The dashed and dot-dashed curves are the L11 power-law fits to their mass-normalized SN~Ia rates in local elliptical and S0 galaxies, respectively.}
 \label{fig:SNuM_Ia_mass}
\end{figure}

We report our SN~Ia rates per unit mass in Table~\ref{table:rates}, where the last line in each category notes the relevant SN rate measured in the full sample, averaged over all masses and redshifts. In Fig.~\ref{fig:SNuM_Ia_mass}, we show the SN~Ia rates per unit mass as a function of galaxy stellar mass measured here for all galaxies, as well as in star-forming and passive galaxies. For comparison, we also show the GM13 measurements in all galaxies. Although we have used stellar masses, SFR, and sSFR values derived by a different pipeline than those used in GM13, and roughly a quarter of the SNe~Ia in this work are new detections, the rates from this work and from GM13 are consistent (so that the choice of pipeline had no systematic effect on the rates) and show that the total rates decrease with increasing mass, as originally reported by L11. However, as observed by \citet{2006ApJ...648..868S}, our mass-normalized SN Ia rates in passive galaxies are consistent with being independent of galaxy stellar mass. 

\begin{figure*}
 \centering
 \begin{tabular}{cc}
  \includegraphics[width=0.475\textwidth]{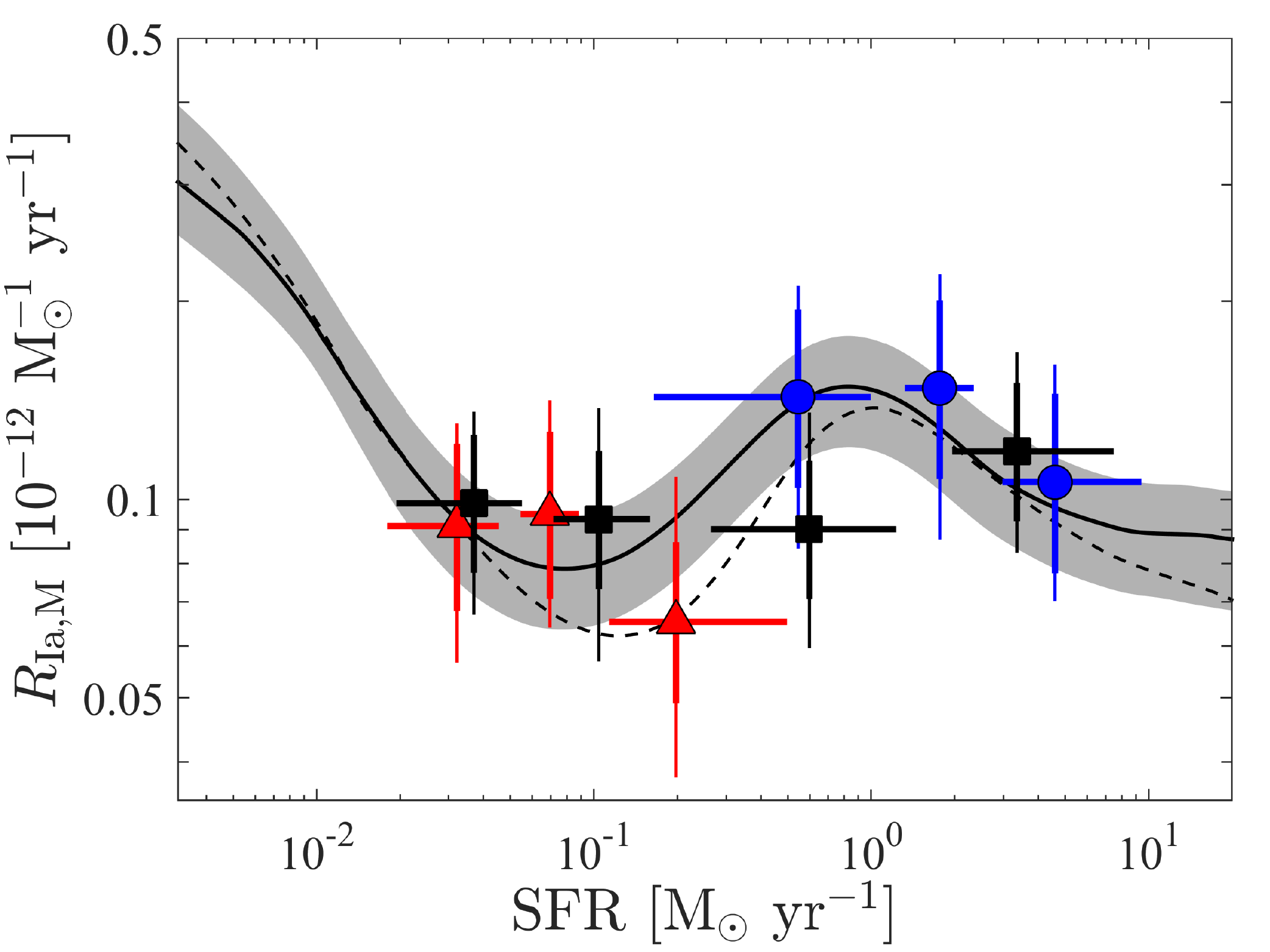} & \includegraphics[width=0.475\textwidth]{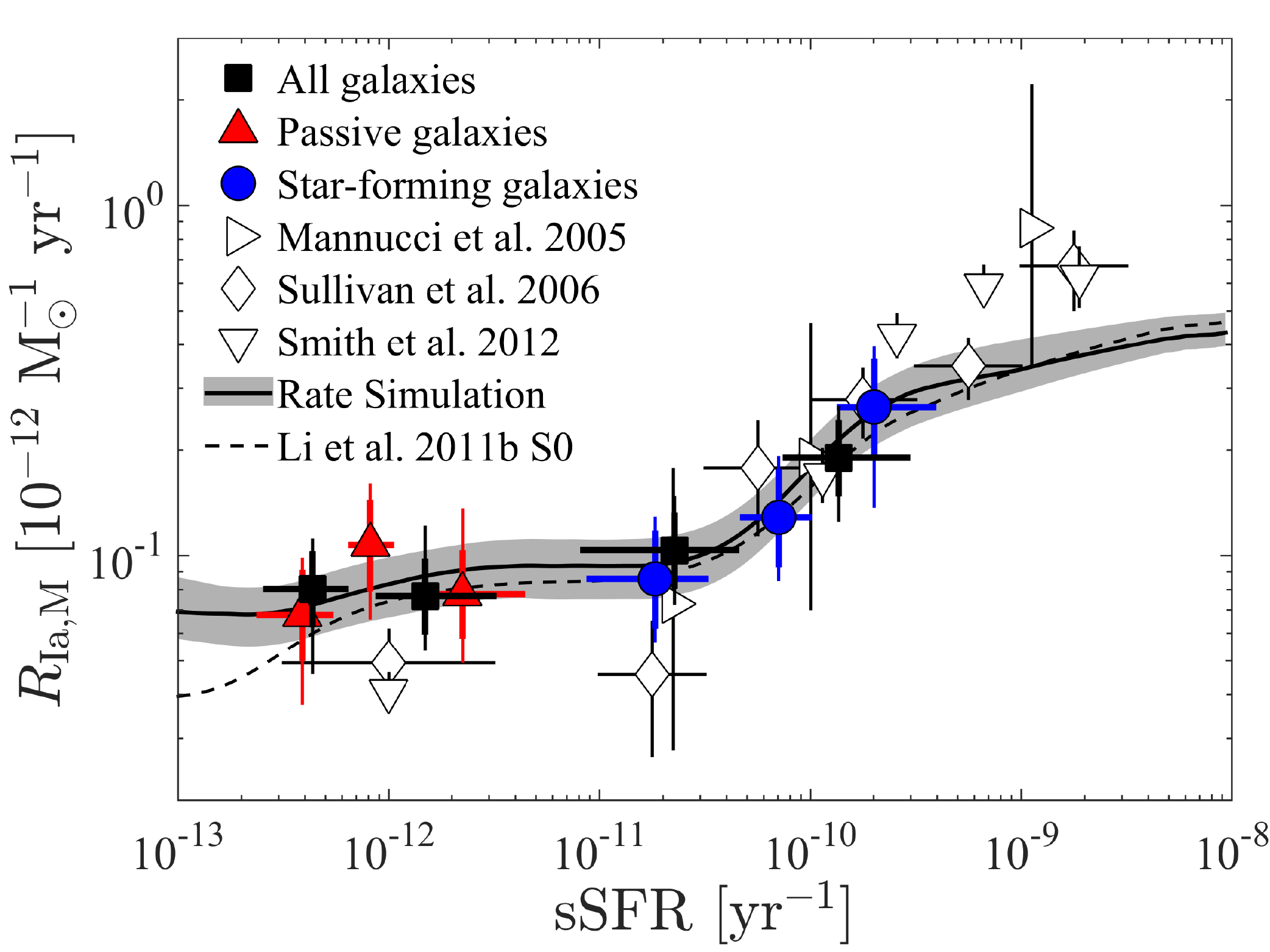} \\
  \includegraphics[width=0.475\textwidth]{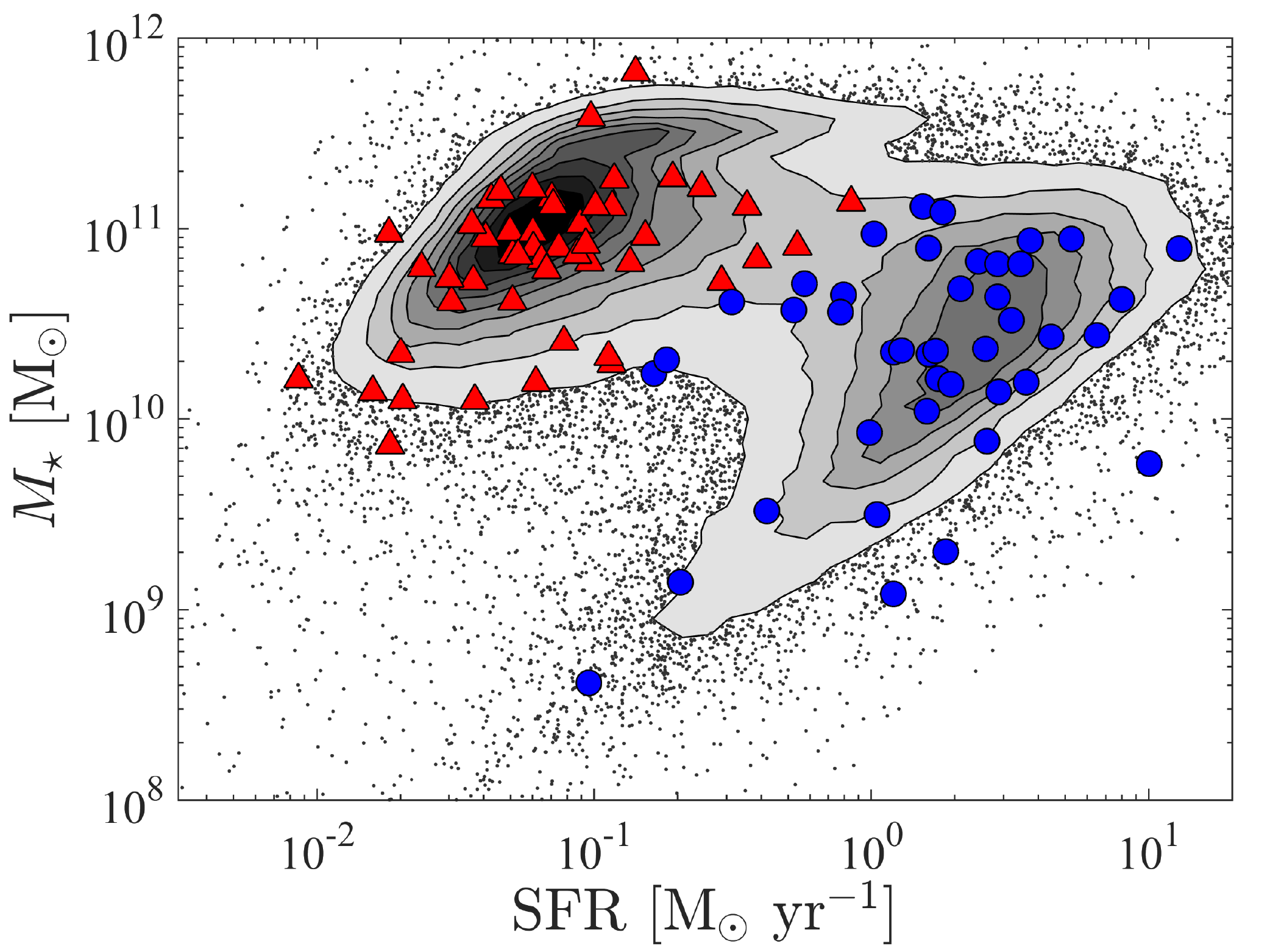} & \includegraphics[width=0.475\textwidth]{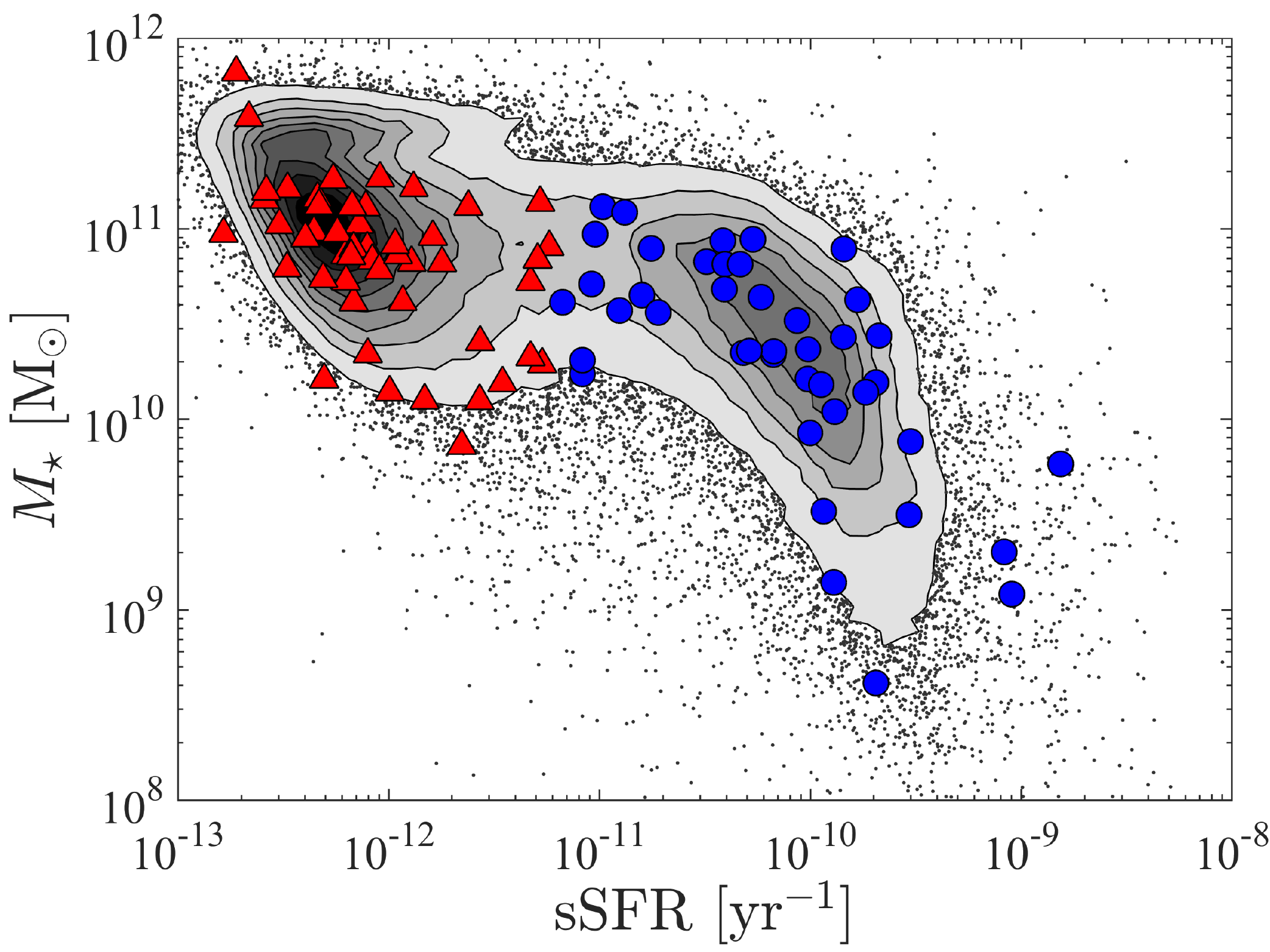} \\
 \end{tabular}
 \caption{SN~Ia rates per unit mass, as a function of galaxy SFR and sSFR. The first row shows the mass-normalized SN~Ia rates as a function of SFR (left) and sSFR (right), while the lower row of plots shows the distributions of galaxy stellar mass as a function of SFR (left) and sSFR (right). In the upper plots, we show the rates for all galaxies as black squares, while the rates for passive and star-forming galaxies are shown as red triangles and blue circles, respectively. The \citet{2006ApJ...648..868S} rates are shown as white diamonds and their re-scaling of the \citet{2005A&A...433..807M} measurements are shown as right-facing triangles. Down-facing triangles are used for the \citet{2012ApJ...755...61S} rates. The thick vertical error bars denote the statistical uncertainty stemming from the Poisson uncertainty on the number of SNe~Ia in each bin; thin vertical error bars show additional systematic uncertainties; and the horizontal error bars denote the 16th and 84th percentiles of the galaxy number distribution in each bin. The solid and dashed curves are projections of the best-fitting result of the SN~Ia rate simulation shown in Fig.~\ref{fig:SNuM_Ia_mass} and of the L11 power-law fit to their SN~Ia rates in S0 galaxies, respectively. The 68.3 per cent uncertainty on the fit to the scaling of the DTD in our rate simulation is shown as the grey band around the solid curve. The contours in the lower plots show the density of galaxies in our sample in decrements of 10 per cent. We show five per cent of the galaxies in the lowest density bin as dots. The passive and star-forming SN~Ia hosts are shown as red triangles and blue circles, respectively.}
 \label{fig:SNuM_Ia_SFR}
\end{figure*}

In the upper panels of Fig.~\ref{fig:SNuM_Ia_SFR}, we show the SN~Ia rates per unit mass as a function of galaxy SFR and sSFR, where the latter were measured by the MPA-JHU Galspec pipeline from SDSS spectra, as described in Section~\ref{subsub:vistime_galprop}, below. The measured rates seem to be constant with SFR, and although the average rate in passive galaxies is lower by a factor of $\sim1.4$ than in star-forming galaxies, there is no discernible trend in either subsample. When considered as a function of sSFR, the rates look to be constant in passive galaxies, but rise with rising sSFR in star-forming galaxies (however, due to their large uncertainties, we cannot rule out a flat trend in the rates in star-forming galaxies). 

Similar trends between SN Ia rates per unit mass and sSFR were reported by \citet{2005A&A...433..807M}, \citet{2006ApJ...648..868S}, and \citet{2012ApJ...755...61S}. \citet{2005A&A...433..807M} used 136 SNe Ia from the \citet*{cappellaro1999} sample of local SNe and measured a rising trend in mass-normalized SN Ia rates in galaxies with morphologies in the range E/S0--S0a/b--Sbc/d--Irr, which can be viewed as a proxy for a rising range of sSFR values. \citet{2006ApJ...648..868S} used 125 SNe Ia from the SuperNova Legacy Survey \citep{guy2010} in the redshift range $0.2<z<0.75$ and measured a similar rise in SN Ia rates per unit mass as a function of sSFR for galaxies with ${\rm sSFR}>10^{-11}~{\rm yr^{-1}}$. \citet{2012ApJ...755...61S} used 342 SNe Ia from the SDSS-II SN survey in the redshift $0.05<z<0.25$ to show the same trend as the previous works. These rates, which are reproduced in Fig.~\ref{fig:SNuM_Ia_SFR}, are consistent with our measurements. We note that the \citet{2005A&A...433..807M} rates reproduced in Fig.~\ref{fig:SNuM_Ia_SFR} have been scaled by \citet{2006ApJ...648..868S} so that the \citet{2005A&A...433..807M} SN Ia mass-normalized rate in E/S0 galaxies matched the \citet{2006ApJ...648..868S} rate in passive galaxies (M. Sullivan, private communication). As the galaxies in our sample are in the redshift range $0.04<z<0.2$ (where the lower and upper redshift limits are the 10th and 90th percentiles of our galaxies' redshift distribution, respectively), our measurements reinforce the \citet{2012ApJ...755...61S} observations and show that this rate-sSFR correlation exists continuously for field galaxies out to $z\sim0.75$.

\citet{Kistler2011} fit the SN~Ia rate-mass correlation with a combination of a power-law DTD with an index of $-1$ and the correlation between galaxies' ages and stellar masses, whereby older galaxies tend to be more massive than younger ones \citep{2005MNRAS.362...41G}. In GM13, we repeated the \citet{Kistler2011} SN~Ia rate simulation and found it to be consistent with our measurements. Here, we repeat it once more, with a few improvements. We express the SN~Ia rate per unit mass of galaxy $i$ at cosmic time $t$ as the convolution of the SFH, $S(t)$, and a DTD, $\Psi(t)$, divided by the total stellar mass of the galaxy, $M_{\star,i}$, after mass loss due to stellar evolution:
\begin{equation}\label{eq:rate_simulation}
 R_{{\rm Ia,M},i}(t) =\frac{1}{M_{\star,i}} \int\limits_{0}^{\Delta t} S(t')\Psi(t-t')dt',
\end{equation}
where $\Delta t = t_g - t$ is the elapsed time between the galaxy's formation time, $t_g$, and the lookback time, $t$, to the galaxy's redshift, for which we use the value measured by the SDSS DR9 pipeline. For each of the galaxies in our sample, we use the stellar mass measured by Galspec to draw a galaxy age, $t_g$, from a Gaussian distribution centred on the median values in table 2 from \citet{2005MNRAS.362...41G}, with the 16/84 per cent values acting as the distribution's lower and upper standard deviations. Following \citet{2005MNRAS.362...41G}, we use an exponential SFH of the form $e^{-\alpha t}$, with indices $\alpha$ drawn from a uniform distribution between 0 and 1. The SFH is scaled to produce the galaxy's formed mass, $M_f$, over the period of time $\Delta t$. We assume a power-law DTD with index $-1$, with the amplitude $\Psi_{1\rm Gyr}$ at $t=1$~Gyr left as a free parameter. Equation~\ref{eq:rate_simulation} thus becomes:
\begin{equation}\label{eq:rate_simulation_all}
 R_{{\rm Ia,M},i}(t) = \frac{M_f}{M_\star} \Psi_{1\rm Gyr} \int\limits_{0}^{\Delta t} e^{-\alpha t'} (t-t')^{-1} dt'.
\end{equation}
As our measurements cover mainly old galaxies in the mass range $\sim 10^9$--$10^{12}~{\rm M_\odot}$, we set $M_f/M_\star=2.3$ \citep{2003MNRAS.344.1000B}. The star that ends up exploding as a SN Ia is thought to be a carbon-oxygen white dwarf \citep{Nugent2011,Bloom2012}. As these white dwarfs evolve from 3--8~${\rm M_\odot}$ main-sequence stars, the SN Ia DTD begins at a delay time of 40 Myr, the time it takes an 8~$M_\odot$ star to evolve into a carbon-oxygen white dwarf. Finally, we bin the resultant rates according to stellar mass, SFR, and sSFR to produce simulations of the rate-mass, rate-SFR, and rate-sSFR correlations. The scaling of the DTD is derived by fitting the simulated SN Ia rates to the measured rates as a function of stellar mass. The best-fitting scaling of the SN~Ia DTD, with $\chi^2_r=0.7$ for three degrees of freedom (DOF), is $(0.064^{+0.012}_{-0.012})\times10^{-12}~{\rm M_\odot^{-1}~yr^{-1}}$, where the uncertainty is the 68.3 per cent confidence region, defined as the range of scaling values that result in a $\chi^2$ value that is $\pm1$ from the minimal $\chi^2$ value \citep{1992nrca.book.....P}. This value for the DTD scaling is consistent with the value of $(0.070^{+0.016}_{-0.016})\times10^{-12}~{\rm M_\odot^{-1}~yr^{-1}}$ found by GM13. The results of our rate simulation are shown as solid curves in Fig.~\ref{fig:SNuM_Ia_mass} and in the upper panels of Fig.~\ref{fig:SNuM_Ia_SFR}.

At first glance, the form of the simulated rates vs. SFR is surprising. While the simulated rates are consistent with the measured rates, they present a marked dependency on SFR that is not readily apparent from the measurements. Specifically, the simulated mass-normalized SN Ia rates decline up to ${\rm SFR}\sim 0.1~{\rm M_\odot~yr^{-1}}$, rise in the range $\sim 0.1$--$1~{\rm M_\odot~yr^{-1}}$, and decline again for ${\rm SFR}\gtsim1~{\rm M_\odot~yr^{-1}}$. Similarly, the simulated rates vs. sSFR are consistent with the measured rates and seem to show a nearly constant rate until ${\rm sSFR}\sim2\times10^{-11}~{\rm yr^{-1}}$, when the simulated rate begins to rise. 

The behavior of these rate-SFR and rate-sSFR correlations can be explained once again by invoking a combination of a  $t^{-1}$-shaped DTD, where $t$ is the delay time, and the known correlation between galaxy mass and age, along with the additional correlation between galaxy stellar mass and either SFR or sSFR. In the bottom panels of Fig.~\ref{fig:SNuM_Ia_SFR}, we show the density distributions of the galaxies in our sample in the stellar mass, $M_\star$, vs. SFR and vs. sSFR phase spaces. It is instructive to examine these plots in relation to the rates plots shown above them. At SFR values $\ltsim0.1~{\rm M_\odot~yr^{-1}}$, we sample mostly passive galaxies, and as their SFRs increase, so do their stellar masses, rising from $\sim10^{10}$ to $\sim10^{12}~{\rm M_\odot}$. In this range, as we move up in SFR, we sample more massive, and thus older, galaxies, which also means that we sample the SN Ia DTD at longer delay times, where the number of SNe Ia produced is smaller, thus resulting in a decline in the mass-normalized rate. In the SFR range $0.1$--$1~{\rm M_\odot~yr^{-1}}$, we transition from massive, passive galaxies, to low-mass, star-forming galaxies. As we move from high-mass to low-mass galaxies, the galaxies become younger and we sample the SN Ia DTD at shorter delay times, closer to the peak of the DTD, thus producing more SNe Ia and increasing the rate. Beyond $\sim1~{\rm M_\odot~yr^{-1}}$, the mass of the star-forming galaxies steadily increases, as does their age, and once again we expect the rates to decline. 

A similar explanation holds for the rate-sSFR correlation. At sSFR values of $\ltsim2\times10^{-11}~{\rm yr^{-1}}$, both the passive and star-forming galaxies are clustered evenly between stellar masses of $\sim10^{10}$--$10^{12}~{\rm M_\odot}$, leading to a flat rate. At ${\rm sSFR}\gtsim2\times10^{-11}~{\rm yr^{-1}}$, the mass of the star-forming galaxies steadily decreases, as does their age, so that we sample the SN~Ia DTD closer to its peak at short delay times, resulting in a rising rate. Thus, it is the ages of the galaxies, not their stellar masses, SFRs, or sSFRs that drive the correlations with the SN Ia rates. 

\citet*{2014MNRAS.445.1898C} coupled empirical models of galaxy mass assembly to a $t^{-1}$-like SN Ia DTD and showed that, at a given redshift, galaxies with higher stellar masses host SNe Ia from increasingly older progenitors (see their fig. 7). According to the authors, this comes about because passive galaxies have had their star formation quenched sometime in the past, so that any observed SNe Ia can only come from old progenitors. The SFHs of star-forming galaxies, on the other hand, evolve more slowly than the SN Ia DTD, so that most of their SNe Ia originate from young progenitors near the peak of the DTD. 

While our interpretation is in broad agreement with that of \citet{2014MNRAS.445.1898C}, we stress two points. First, it is the ages of the galaxies, not their correlated stellar masses, which drive the SN rates and progenitor age distribution. Second, \citet{2014MNRAS.445.1898C} see two main modes in their continuous distribution of progenitor ages as a function of stellar mass. Based on these modes, they relegate young SNe Ia to star-forming galaxies and old SNe Ia to passive galaxies, which they say are generally more massive than star-forming galaxies. We caution that using the monikers `passive' and `star-forming' may be misleading in this instance. Star-forming galaxies are observed to have a broad range of stellar masses, as can be seen in the bottom-right panel of Fig.~\ref{fig:SNuM_Ia_SFR}. At a given stellar mass, e.g., $10^{11}~{\rm M_\odot}$, within the relatively narrow redshift range of our galaxy sample, we sample both passive and star-forming galaxies. Since these galaxies have more or less the same ages, relative to their lookback times, they have the same rates per unit mass in the upper-right panel. This is in line with all SNe Ia in these galaxies coming from a progenitor population of the same age, as expected from the continuous distribution shown by \citet{2014MNRAS.445.1898C}. To select a homogeneous population of young SNe Ia, then, it will not be enough to simply select all those SNe in star-forming galaxies. Rather, the galaxy sample would need to be restricted to low-mass, young, star-forming galaxies.

The high simulated SN Ia rate at low SFR values in the left-hand panel of Fig.~\ref{fig:SNuM_Ia_SFR} is consistent with our measurements. The expected rate at SFR values $\le10^{-2}~{\rm M_\odot~yr^{-1}}$ is $0.25^{+0.06}_{-0.05}\times10^{-12}~{\rm M_\odot^{-1}~yr^{-1}}$ at a median SFR of $6.5\times10^{-3}~{\rm M_\odot~yr^{-1}}$. Given the visibility times of the 5\,708 galaxies in that SFR range, we should have detected 2--3 SNe Ia, consistent with the single SN Ia found in that range, given Poisson uncertainties of $1.0^{+2.3}_{-0.8}$ SNe.

As a control for our rate simulation, we consider how the L11 power-law fit to their mass-normalized SN Ia rates as a function of stellar mass in S0 galaxies (see their table 4) would appear in the $R_{\rm Ia,M}$ vs. SFR and $R_{\rm Ia,M}$ vs. sSFR phase spaces. We assign a rate to each galaxy based on its stellar mass and then bin the rates according to the SFR and sSFR values of the galaxies. The resultant rate-SFR and rate-sSFR correlations are shown as dashed curves in the top panels of Fig.~\ref{fig:SNuM_Ia_SFR}. The projected L11 correlations are consistent with our rate simulation. This is not surprising, as our rate-mass simulation and the L11 power-law fits have similar shapes in the mass range $10^9$--$10^{11}~{\rm M_\odot}$ covered by our SN~Ia rates. It is only at the extreme edges of the stellar mass axis ($<10^9$ and $>10^{11}~{\rm M_\odot}$) that our rate simulation begins to differ appreciably from the L11 power-law fit. The result of this difference can be seen in the the top right panel of Fig.~\ref{fig:SNuM_Ia_SFR}, where the L11 correlation for Sbc galaxies at sSFR values of $<10^{-12}~{\rm yr^{-1}}$ rises with increasing sSFR in a steeper fashion than our rate simulation, which is constant.

\subsection{The Type II supernova rate-mass correlation}
\label{subsec:rate_mass}

\begin{figure*}
 \centering
 \begin{tabular}{cc}
  \includegraphics[width=0.475\textwidth]{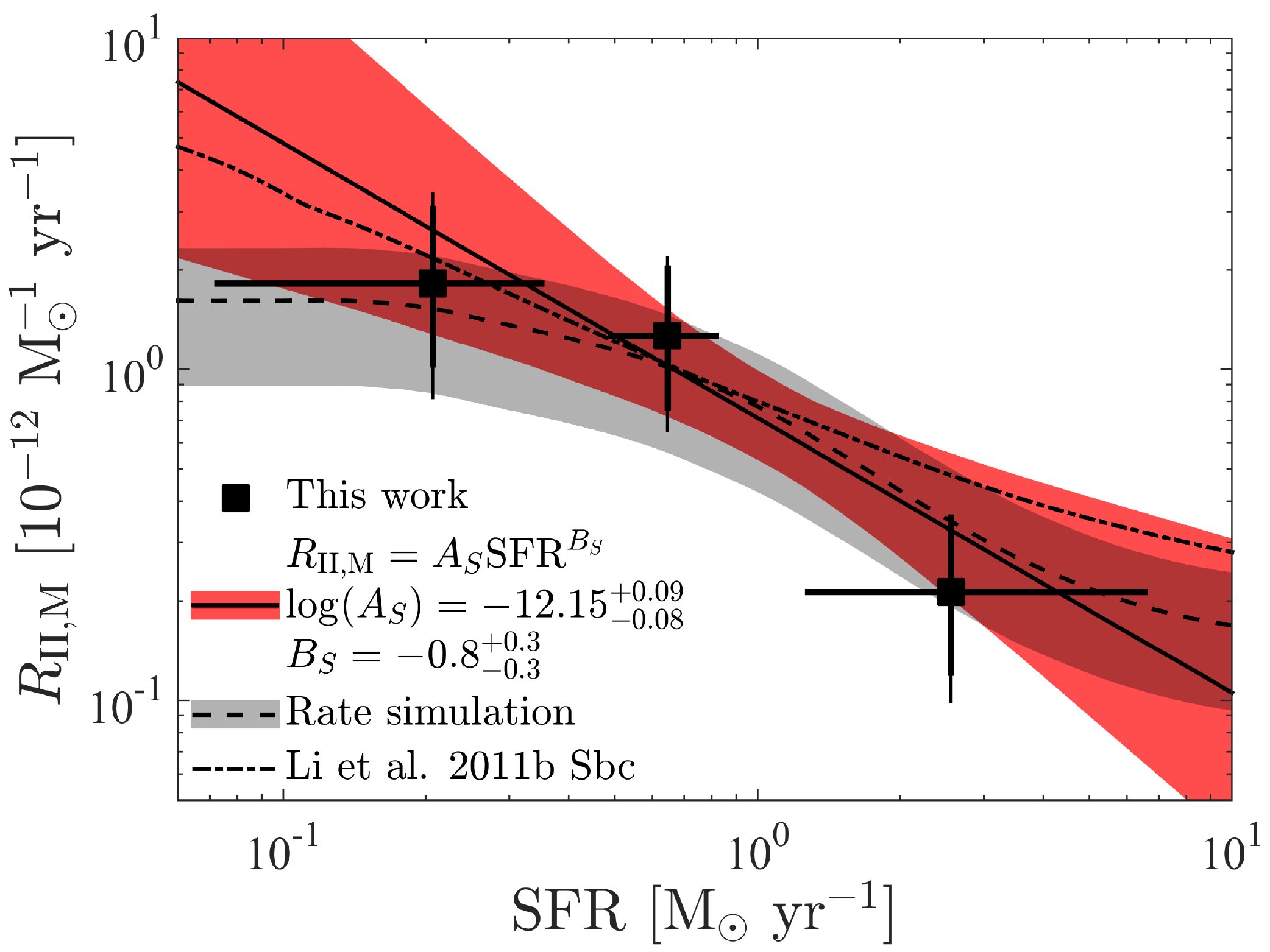} & \includegraphics[width=0.475\textwidth]{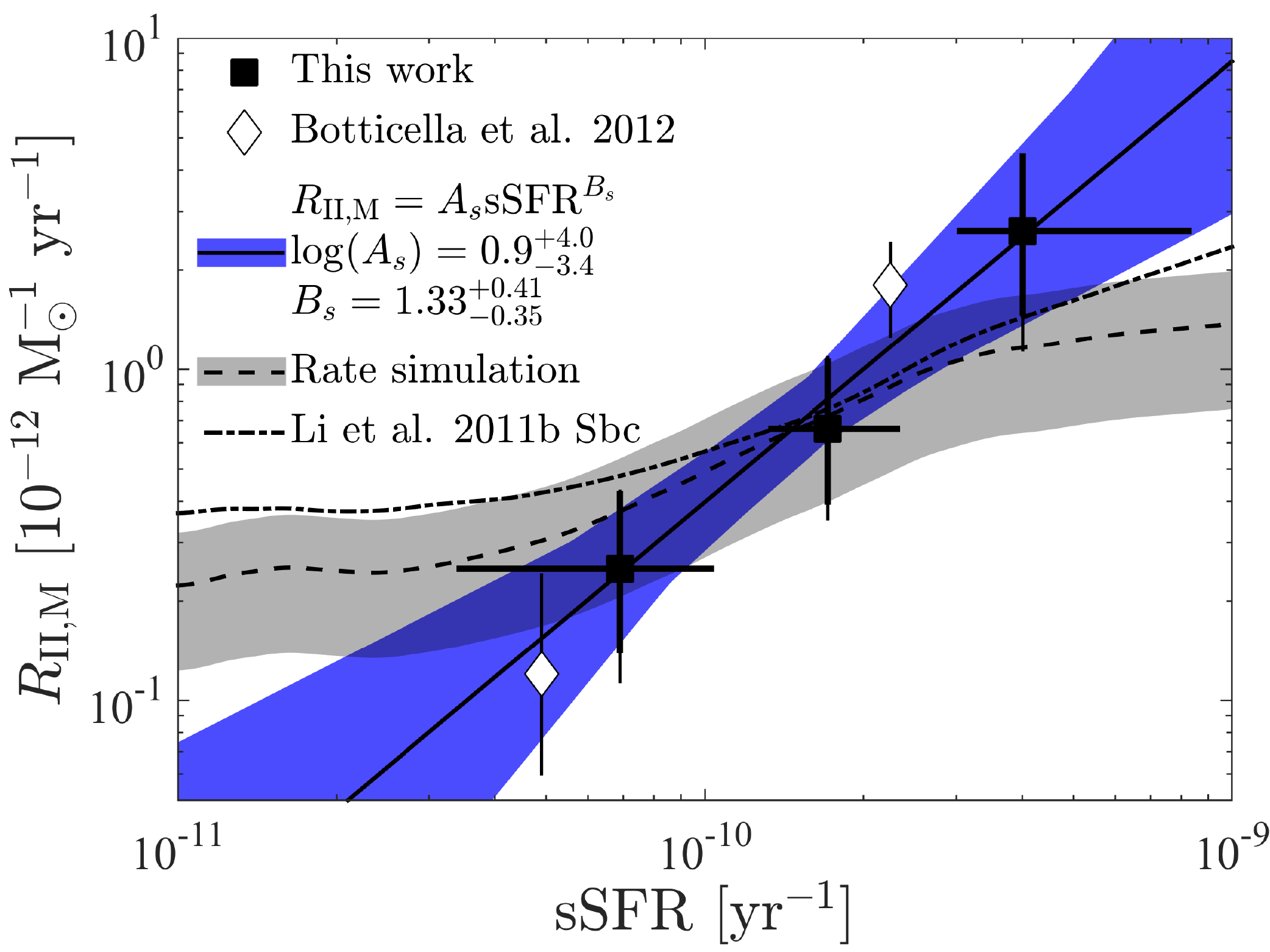} \\
  \includegraphics[width=0.475\textwidth]{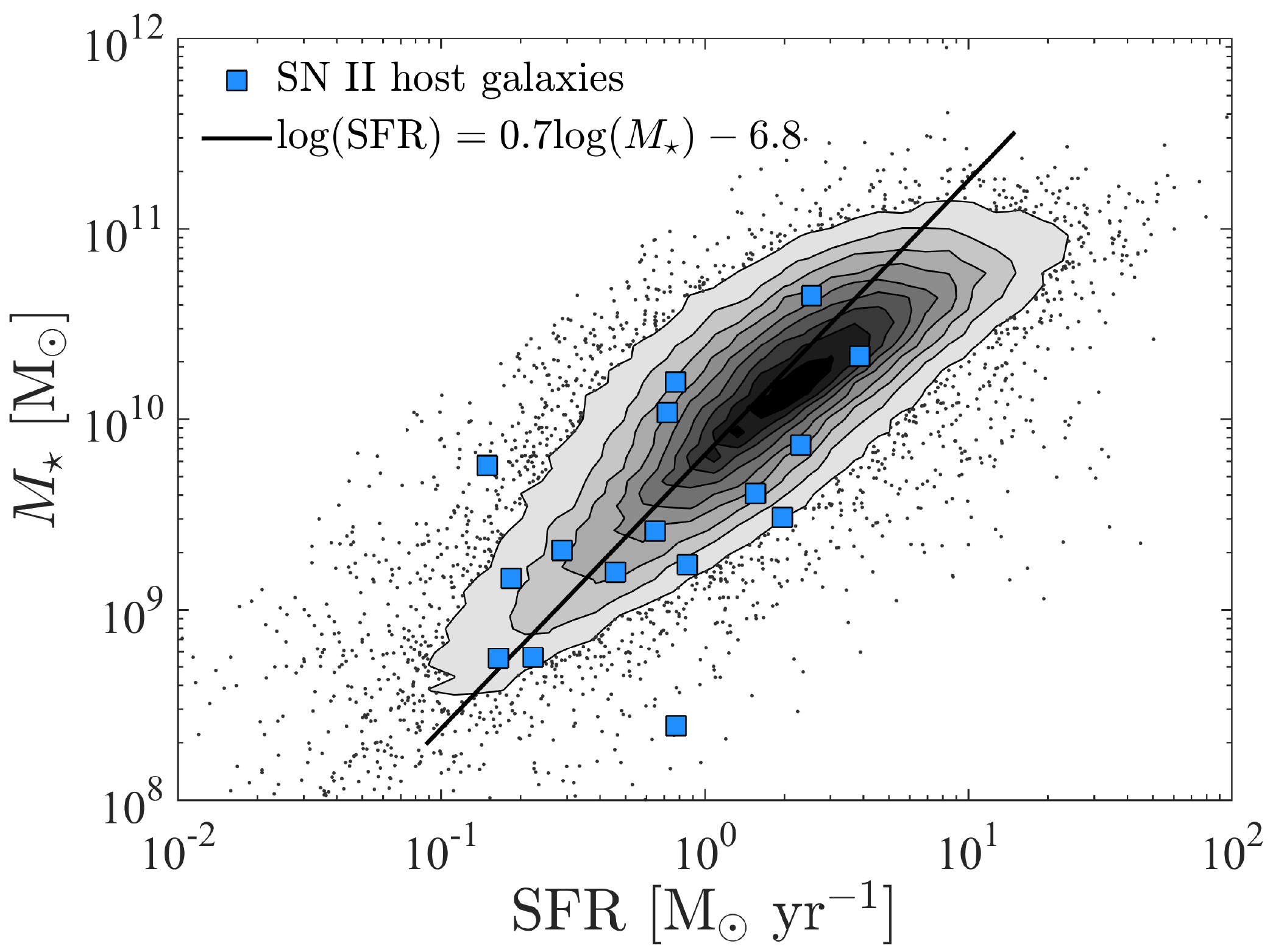} & \includegraphics[width=0.475\textwidth]{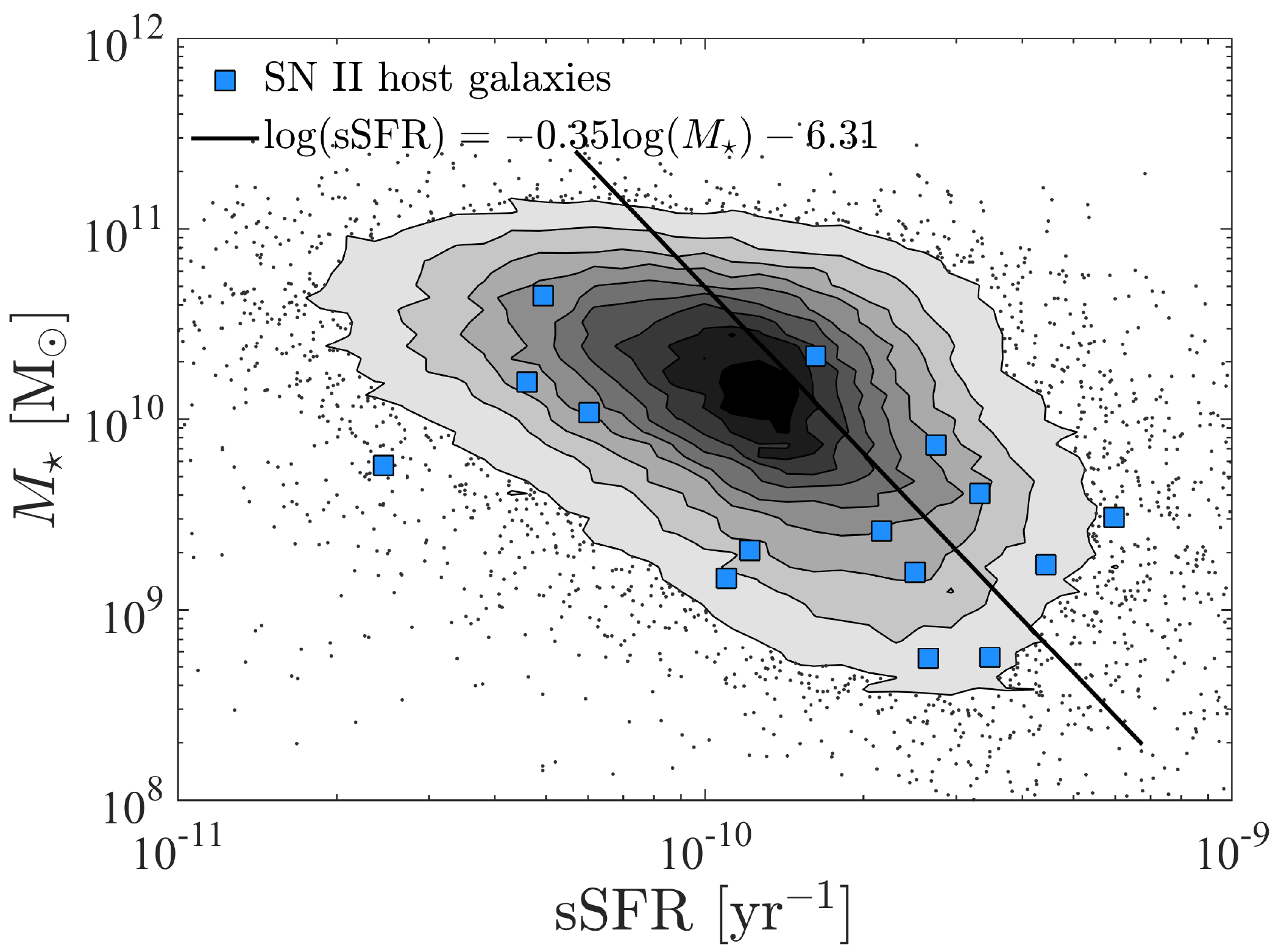} \\
  \includegraphics[width=0.475\textwidth]{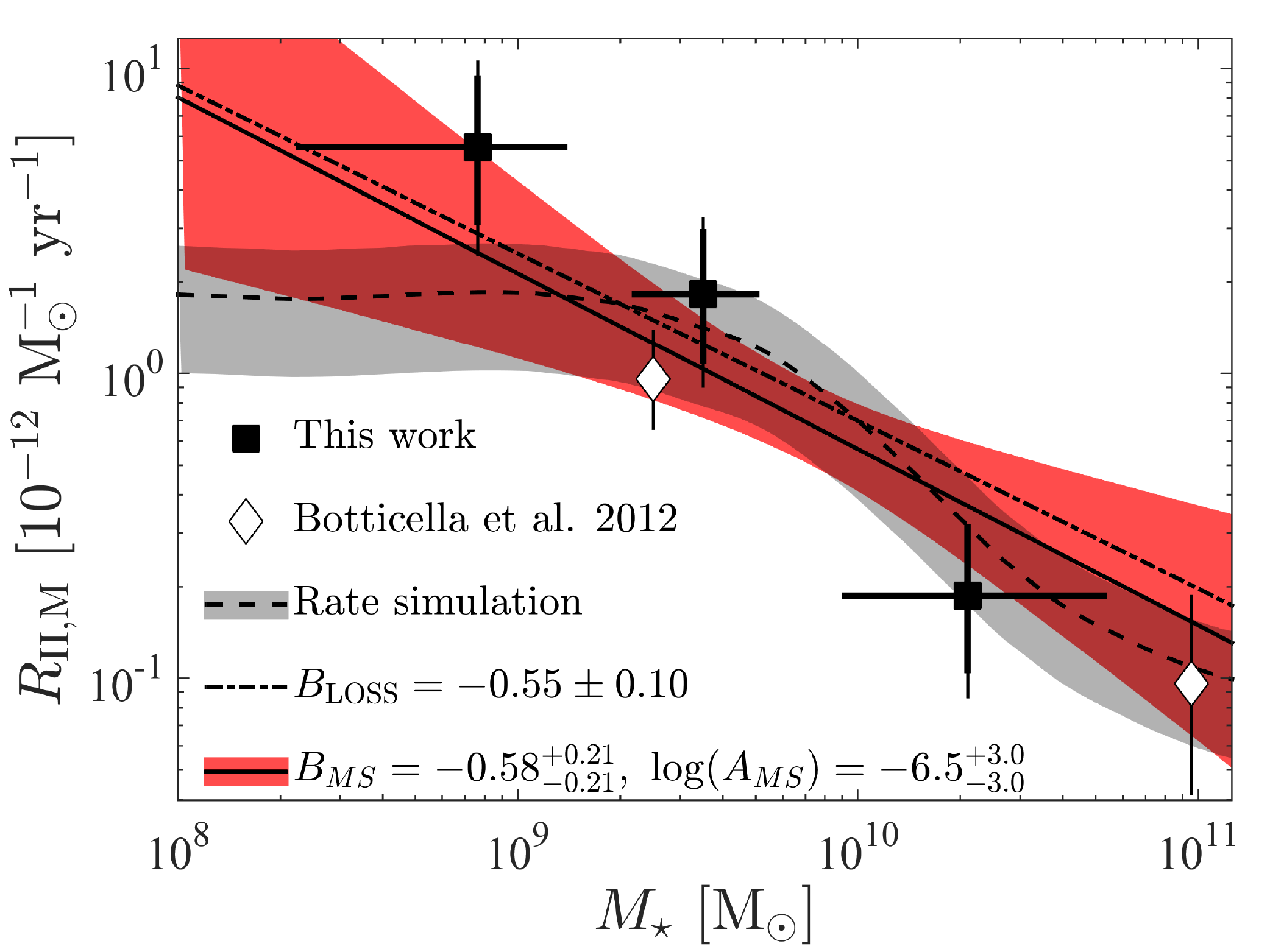} & \includegraphics[width=0.475\textwidth]{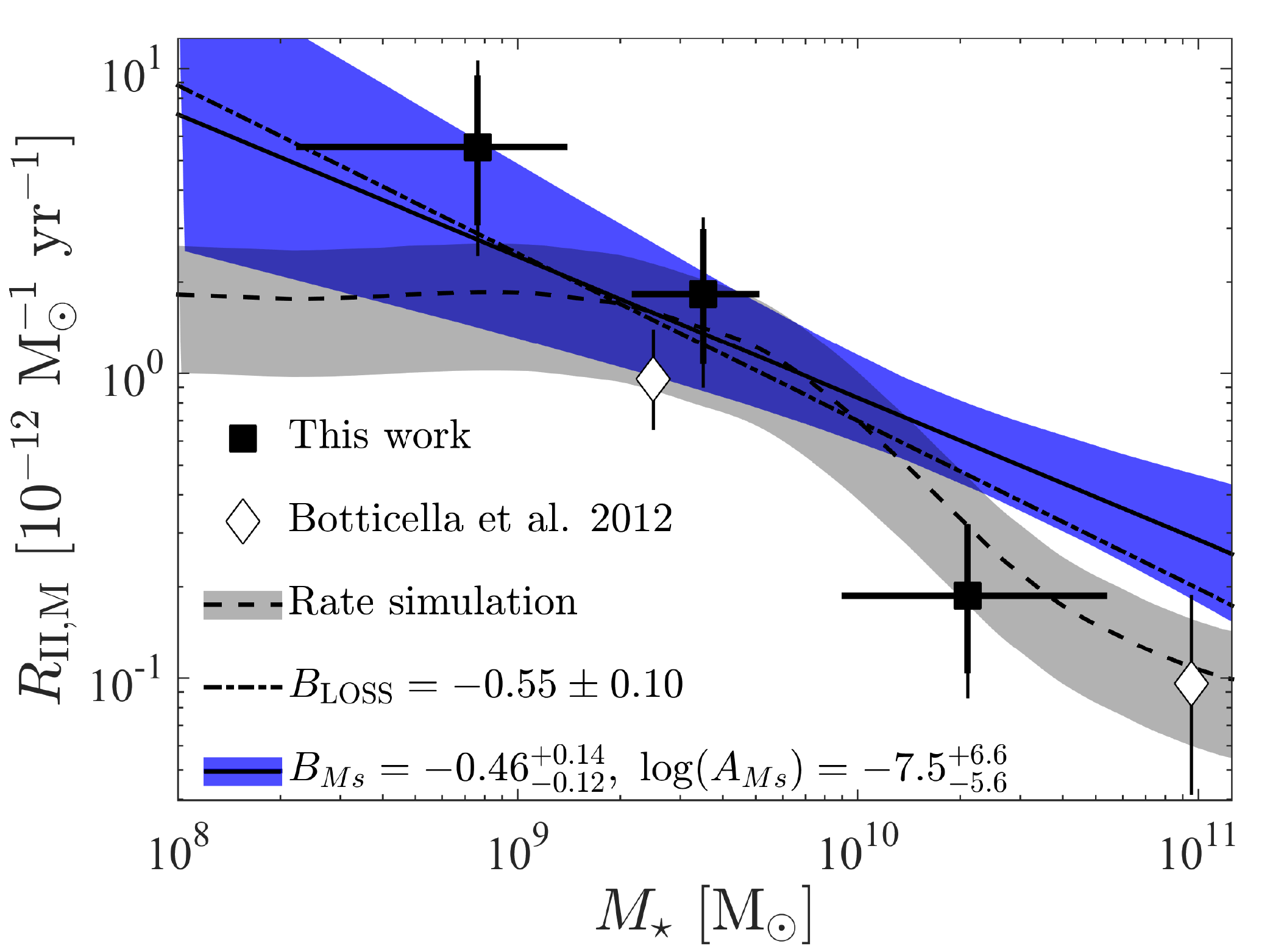} \\
 \end{tabular}
 \caption{SN~II rates per unit mass, as a function of galaxy properties. \textit{Top}: The left and right panels show SN~II rates per unit mass as a function of SFR and sSFR, respectively, as black squares. Vertical error bars are a combination of statistical (thick) and systematic (thin) uncertainties, and the horizontal error bars show the range that includes 68.3 per cent of the galaxies within the SFR or sSFR bin. CC SN rates from \citet{botticella2012}, shown as white diamonds, have been scaled down by 40 per cent to represent the contribution of SNe IIP/L alone. Best-fitting power laws are denoted by solid curves. The projection of the best-fitting rate-mass simulation from the bottom panels is shown as a dashed curve. The dot-dashed curve is the projection of the L11 power-law fit to their mass-normalized rates in Sbc galaxies, scaled down by 20 per cent to account for the abscence of SNe~IIn and IIb in our sample, and projected onto SFR and sSFR. Shaded areas show the 68.3 per cent confidence regions of the various fits, as marked. \textit{Center}: The left and right panels show correlations between galaxy stellar mass, $M_\star$, and either SFR or sSFR, respectively. Density contours of the galaxies in our sample are arranged in decrements of 10 per cent and the SN~II host galaxies are shown as blue squares. Solid curves are power-law fits to the data. \textit{Bottom}: Both panels show measurements of the SN II mass-normalized rates as a function of $M_\star$ from this work and from \citet{botticella2012}, as marked. The solid curves are the power laws predicted by Equations~\ref{eq:rate_II_SFR1}--\ref{eq:rate_II_SFR3}. The dashed curve is the best-fitting rate simulation to the rate-mass measurements. The L11 power-law fit to their measurements in Sbc galaxies is shown as the dot-dashed curve.} 
 \label{fig:SNuM_II_SFR}
\end{figure*}

We calculate mass-normalized SN II rates using the final sample of 16 SNe II described in Section~\ref{sec:sample}, the visbility times calculated in Section~\ref{subsec:vistime} for the subsample of \NIIgal\ star-forming galaxies that do not host AGNs, and Equation~\ref{eq:rates_Ia}. Because of the smaller size of the SN II sample, we derive the rates in three bins. The SN II rates per unit mass as a function of galaxy stellar mass, SFR, and sSFR are presented in Table~\ref{table:rates} and Fig.~\ref{fig:SNuM_II_SFR}.

We confirm the SN II rates per unit mass decrease with increasing galaxy stellar mass, at a median redshift of $\sim 0.075$, as originally reported by L11. Furthermore, we find similar rate-SFR and rate-sSFR correlations: the SN II rates per unit mass decrease with increasing SFR, but increase with increasing sSFR. \citet{botticella2012} also measured CC SN rates per unit mass as a function of galaxy stellar mass and sSFR. Binned into two coarse bins, their measurements, which are consistent with our own, show a possible decline of the CC SN rate per unit mass with increasing stellar mass and rise with rising sSFR. As in Section~\ref{subsec:rates_Ia}, the connection between the rate-mass correlation and the rate-SFR and rate-sSFR correlations is the known correlation between galaxy stellar mass and either SFR or sSFR. In the middle panels of Fig.~\ref{fig:SNuM_II_SFR}, we show the density distributions of the non-AGN, star-forming galaxies in our sample in the $M_\star$ vs. SFR and $M_\star$ vs. sSFR phase spaces. Because we limit ourselves to star-forming galaxies, the correlations are simpler than those in Fig.~\ref{fig:SNuM_Ia_SFR}: the SFRs of the galaxies in our sample rise as a function of stellar mass, while the sSFRs decrease. 

Because SN II progenitors are massive stars ($>8~{\rm M_\odot}$), the delay times between the formation of the progenitors and their explosions are short ($<40$ Myr), so the SN II DTD should not play as important a role in the SN II rate-mass correlation as it does for SNe Ia. Thus, we note that as $R_{\rm II,M} \propto M_\star$ and $M_\star \propto {\rm SFR}$ (or sSFR), it is not surprising that we should observe that $R_{\rm II,M} \propto {\rm SFR}$ (or sSFR). Furthermore, if the observed rate-SFR and rate-sSFR correlations are connected to the rate-mass correlation through the correlations between stellar mass and SFR or sSFR, we should expect that if we fit the rate-SFR correlation with a power law of the form $R_{\rm II,M} = A_S{\rm SFR}^{B_S}$, where $A_S$ and $B_S$ are constants, and the correlation between SFR and stellar mass as a power-law of the form ${\rm SFR} = C_S M_\star^{D_S}$, then the rate-mass correlation would be described by a power-law of the form 
\begin{equation}\label{eq:rate_II_SFR1}
 R_{\rm II,M} = A_{MS} M_\star^{B_{MS}},
\end{equation}
where
\begin{equation}\label{eq:rate_II_SFR2}
 A_{MS} = A_S C_S^{B_S};~B_{MS} = D_S B_S.
\end{equation}
In the same manner, we can fit the rate-sSFR measurements with a power law of the form $R_{\rm II,M} = A_s{\rm sSFR}^{B_s}$ and the sSFR vs. $M_\star$ correlation with a power law of the form ${\rm sSFR} = C_s M_\star^{D_s}$, which should result in a rate-mass correlation described by a power-law with parameters $A_{Ms}$ and $B_{Ms}$, where
\begin{equation}\label{eq:rate_II_sSFR}
 A_{Ms} = A_s C_s^{B_s};~B_{Ms} = D_s B_s.
\end{equation}
It is more convenient to present these equations in a linear format, as then the various free parameters are unitless. Equation~\ref{eq:rate_II_SFR1}, for example, becomes 
\begin{equation}\label{eq:rate_II_SFR3}
 {\rm log}(R_{\rm II,M}) = {\rm log}(A_{MS})+B_{MS}{\rm log}(M_\star/{\rm M_\odot}),
\end{equation}
where $R_{\rm II,M}$ is measured in units of $10^{-12}~{\rm M_\odot^{-1}~yr^{-1}}$.

The best-fitting parameters for the power-law fit to the mass-normalized SN II rates as a function of SFR, with $\chi^2_r = 1.0$ for one DOF, are ${\rm log}(A_S) = -12.15^{+0.09}_{-0.08}$ and $B_S = -0.8^{+0.3}_{-0.3}$. Likewise, the best-fitting parameters for the SN II rates per unit mass a function of sSFR, with $\chi^2_r = 0.15$ for one DOF, are ${\rm log}(A_s) = 0.9^{+4.0}_{-3.4}$ and $B_s = 1.33^{+0.41}_{-0.35}$. 

Similarly, we fit power laws to the SFR vs. $M_\star$ and sSFR vs. $M_\star$ values of all the galaxies in our subsample. Although we do not take into account their individual uncertainties, by fitting all \NIIgal\ value pairs, we take into account their scatter. We find that the correlations between SFR (or sSFR) and stellar mass can be described by ${\rm log(SFR/{\rm M_\odot~yr^{-1}})} = 0.7{\rm log}(M_\star/{\rm M_\odot})-6.8$ and ${\rm log(sSFR/{\rm yr^{-1}})} = -0.35{\rm log}(M_\star/{\rm M_\odot})-6.31$. The latter is comparable to similar fits by \citet{2007ApJS..173..267S}, who find ${\rm log(sSFR/{\rm yr^{-1}})} = -0.35{\rm log}(M_\star/{\rm M_\odot})-6.33$ and \citet{2007ApJS..173..315S}, who find ${\rm log(sSFR/{\rm yr^{-1}})} = -0.36{\rm log}(M_\star/{\rm M_\odot})-6.4$. We note that our fits, as presented in Fig.~\ref{fig:SNuM_II_SFR}, do not pass through the densest areas of the contour maps describing the distribution of galaxies in the $M_\star$ vs. SFR and sSFR phase spaces. Similar fits to mean and median values of $M_\star$ and either SFR or sSFR, where the galaxies were first divided into ten equally-spaced bins, provided similar power laws that did not pass through the densest regions of the contour maps, either. We attribute this to the effect of correlations within the data on the way the contour maps are calculated, i.e., values in separate bins affect the way contours are calculated, but they affect neither the median or mean values, nor a fit to all of the data. 

Using Equations~\ref{eq:rate_II_SFR1}--\ref{eq:rate_II_SFR3}, and propagating the uncertainties of the power-law parameters, the predicted values of the parameters of the power-law rate-mass correlations are ${\rm log}(A_{MS}) = -6.5^{+3.0}_{-3.0}$ and $B_{MS} = -0.58^{+0.21}_{-0.21}$ from the rate-SFR correlation and ${\rm log}(A_{Ms}) = -7.5^{+6.6}_{-5.6}$ and $B_{Ms} = -0.46^{+0.14}_{-0.12}$ from the rate-sSFR correlation. The predicted power-law rate-mass correlations, presented in the bottom panels of Fig.~\ref{fig:SNuM_II_SFR}, are consistent with the measured SN~II rates per unit mass vs. stellar mass. Moreover, both of the slopes $B_{MS}$ and $B_{Ms}$ are consistent with the value of the slope from the L11 power-law fit, $B_{\rm LOSS} = -0.55\pm0.10$. 

We can also fit a power law of the form $R_{\rm II,M}=A_M M_\star^{B_M}$ directly to the SN II mass-normalized rates as a function of stellar mass. With a reduced $\chi_r^2=0.52$ for one DOF, the best-fitting values of the free parameters are ${\rm log}(A_M)=-2.0^{+2.6,+4.8}_{-2.7,-5.0}$ and $B_M=-1.03^{+0.28,+0.53}_{-0.27,-0.50}$, where the uncertainties are the 68.3 and 95.4 per cent confidence regions defined by the range of values, for each parameter, that satisfy $\chi^2=\chi^2_{\rm min}+\Delta(\chi^2)$, where $\Delta(\chi^2)=1$ or 4, respectively \citep{1992nrca.book.....P}. The slope of this fit is steeper than that found by both L11 and by our predicted values above, but consistent at the 95.4 per cent uncertainty limit. We ascribe the difference between the values to the SN II rate measurement in the most massive galaxies, which biases the slope to a steeper value. This could be caused by Poisson fluctuations due to small-number statistics. Alternatively, we note that L11 separated their rates according to the morphology of the SN host galaxies and the resulting power-law fits had the same slope but a range of scalings. As we do not separate our galaxies in this manner, it is possible that a different composition of galaxy morphologies in the most massive bins, relative to the other bins, is pulling the SN II rate down.

Next, we test whether the rate simulation we conducted in Section~\ref{subsec:rates_Ia} to explain the SN Ia rate-mass correlation can also explain the correlations between the SN II mass-normalized rates and $M_\star$, SFR, and sSFR. We repeat this simulation using a simple model of a SN II DTD, which we assume to be uniform in the range 9--40 Myr, and zero at any other time, so that Equation~\ref{eq:rate_simulation_all} becomes
\begin{equation}\label{eq:rate_simulation_II}
 R_{\rm II,M} = \frac{M_f}{M_\star} \frac{\Psi_{\rm II}}{\alpha} \left(1-e^{-\alpha \Delta t}\right),
\end{equation}
where the scaling of the DTD, $\Psi_{\rm II}$, is a free parameter. The range of delay times was chosen as these are the zero-age main sequence lifetimes of $8$--$20~{\rm M_\odot}$ stars \citep*{2002RvMP...74.1015W}, which have been observed to be progenitors of SNe IIP/L (see review by \citealt{Smartt2009review}). Although the true SN II DTD will most likely not be a uniform probability density function, given the large uncertainties of our measurements, our simplified DTD model should be good enough to test whether the rate-simulation can broadly explain our measurements. 

We fit the mass-normalized SN II rates as a function of stellar mass and find that the best-fitting value of the DTD scaling is $\Psi_{\rm II} = (0.81^{+0.36}_{-0.36})\times10^{-21}~{\rm M_\odot^{-1}~yr^{-2}}$, with $\chi_r^2 = 1.9$ for two DOF. The high $\chi^2_r$ value is due to a bad fit between the measured rate in the lowest-mass bin and the simulated rates, which plateau below $\sim3\times10^9~{\rm M_\odot}$ due to the form of the \citet{2005MNRAS.362...41G} age-mass relation and our assumption that all SFHs are declining exponential functions. Thus, although the discrepancy between the measured rate and the rate-simulation fit could be the result of Poisson noise due to small-number statistics, it could also mean that a more realistic simulation, using the actual SFHs of the galaxies, is necessary. This could be achieved with the \vespa\ SFHs, but, as mentioned in Section~\ref{sec:galaxies}, although such SFHs are available for $\sim70$ per cent of the dwarf galaxies in our sample, the remaining 30 per cent host half of the SN II sample (and are thus responsible for most of the SNe in the two low-mass bins). 

The simulated rate-SFR and rate-sSFR correlations, obtained by re-binning the simulated rates, are consistent with the measured rates. Taking a higher value for the progenitor mass, e.g., 40 M$_\odot$, and thus a lower limit on the delay times of $\sim6$ Myr \citep{2002RvMP...74.1015W}, makes no appreciable difference to the fit. The best-fitting form of the simulated SN II rate-mass correlation and its projections onto the $R_{\rm II,M}$ vs. SFR and $R_{\rm II,M}$ vs. sSFR phase spaces are shown in the bottom and top panels of Fig.~\ref{fig:SNuM_II_SFR}, respectively. 

Finally, we also test the L11 power-law fit to their mass-normalized SN II rates vs. stellar mass in Sbc galaxies. As before, the L11 power-law fit and our simulated rates depart from one another at low and high stellar masses ($\ltsim10^9~{\rm M_\odot}$ and $\gtsim10^{10}~{\rm M_\odot}$), as well as at low SFR values of $\ltsim0.1~{\rm M_\odot~yr^{-1}}$. 

\subsection{The core-collapse supernova volumetric rate}
\label{subsec:rate_vol}

Here, we follow the calculation outlined in section 5.3 of GM13 to convert the SN II rates per unit mass derived in the previous Section into a volumetric rate. As the SN Ia rate per unit mass measured here, averaged over all stellar masses and redshifts, is identical to the GM13 value, we do not repeat its conversion into a volumetric rate.

As in GM13, we use the \citet{2012MNRAS.421..621B} galaxy stellar mass function, which requires us to limit our galaxy sample to the range $M\ge10^8$ M$_\odot$. This cut limits the star-forming galaxy sample used to measure the SN II mass-normalized rates to 212\,713 galaxies (98.9 per cent of the full sample), and thus has a negligible effect on the resulting rate. None of our SN-hosting galaxies are excluded by this cut. For $R(M)$ in equation 8 of GM13, we use our own power-law fit to the rate-mass correlation, as well as the one from L11, which has a more precise slope.  

\begin{figure}
 \begin{center}
  \includegraphics[width=0.475\textwidth]{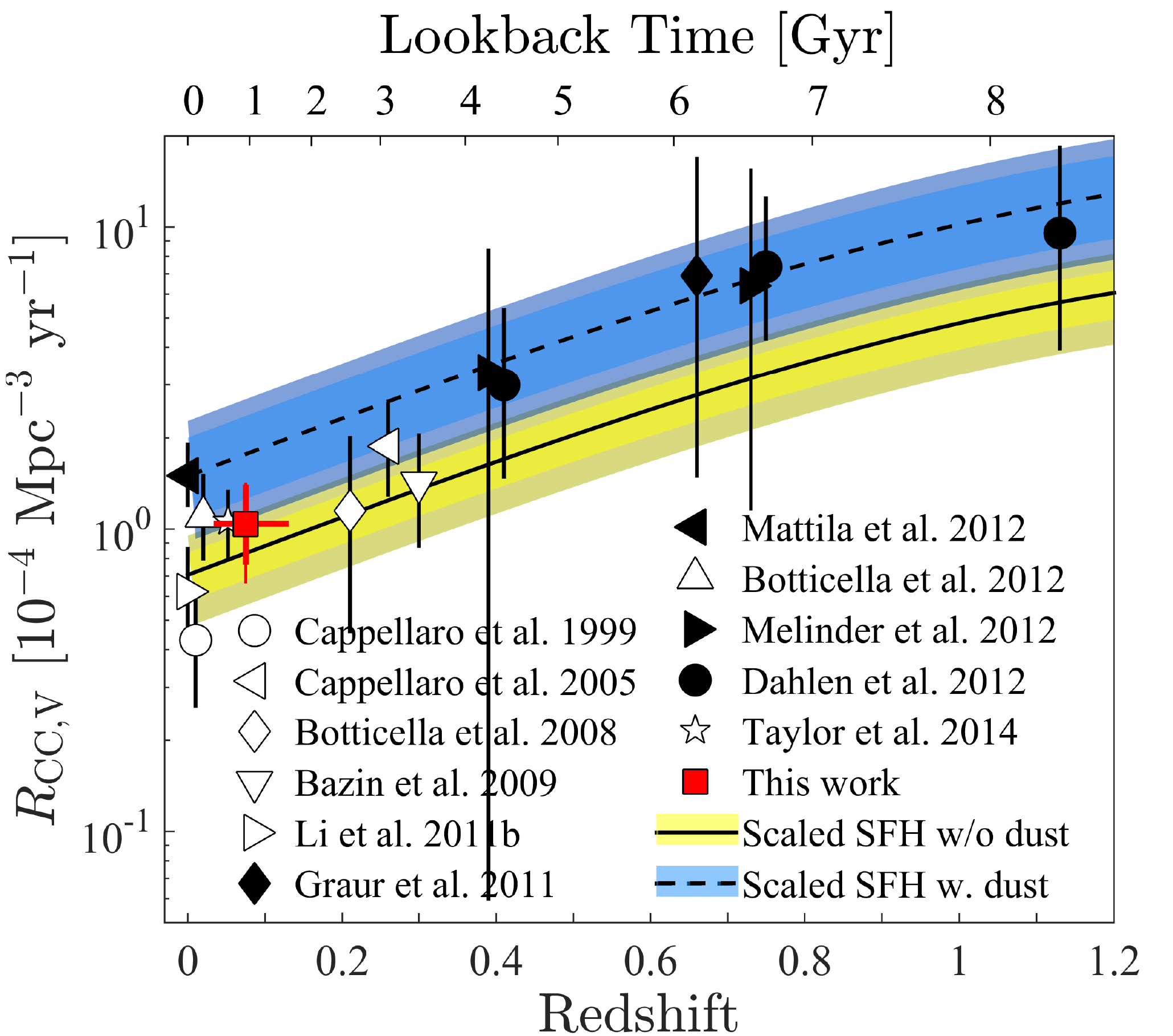}
  \caption{CC~SN volumetric rates as a function of redshift. The rate measured in this work is shown as a red square. The thick vertical error bars denote the statistical uncertainty, while the thin vertical error bar shows the added systematic uncertainty. The horizontal error bar delineates the 68.3 per cent redshift range of the star-forming, non-AGN, galaxies in our sample. Rates from the literature that did not include a correction for the fraction of CC~SNe missed in highly star-forming galaxies are shown as open symbols, as marked, while rates that have been corrected for this missing fraction are shown as black symbols, as marked. All vertical error bars include statistical and systematic uncertainties. The solid and dashed curves are the \citet{Behroozi2012} cosmic SFH scaled to fit the uncorrected and corrected rates, respectively. The dark and light shaded areas around the curves are the 68.3 per cent confidence regions around the best-fitting scalings and account for both the statistical and systematic uncertainties, respectively, of the rates used in each fit. The redshifts of the \citet{botticella2012} and \citet{2012ApJ...757...70D} measurements have been shifted by 0.01, and the \citet{2014ApJ...792..135T} measurement by $-0.01$, to improve their visibility in the plot.}
  \label{fig:rate_vol_CC}
 \end{center}
\end{figure}

The volumetric SN II rate, at a median redshift of $z=0.075^{+0.050}_{-0.040}$ (where the uncertainties are the 16th and 84th percentiles of the redshift distribution of the galaxies in the sample), is $R_{\rm II,V} =$~\RvolII\ when using the L11 fit to the rate-mass correlation, where `stat' stands for the statistical uncertainty and `sys' for systematic uncertainties. Assuming that SNe IIP/L make up $\sim60$ per cent of all CC SNe \citep{li2011LF,2012IAUS..279...34A}, we can derive a volumetric CC SN rate of $R_{\rm CC,V}=$~\RvolCC. This rate, together with other CC SN rates from the literature, is listed in Table~\ref{table:rates_CC} and shown in Fig.~\ref{fig:rate_vol_CC}. 

When using our power-law fit to the SN II mass-normalized rates as a function of stellar mass, the resulting SN II volumetric rate is $0.50^{+0.16}_{-0.13}~{\rm (stat)}~^{+0.02}_{-0.05}~{\rm (sys)}~\times 10^{-4}~{\rm yr^{-1}~ Mpc^{-3}}$. Although this value is systematically lower by 19 per cent than the volumetric rate derived with the L11 fit, the two are consistent at the 68.3 per cent confidence level. 

Using a sample of 89 CC SNe discovered during the imaging-based SDSS-II Supernova Survey, \citet{2014ApJ...792..135T} measured a CC SN volumetric rate in the same redshift range probed here of $R_{\rm CC,V}(z=0.072)=1.06\pm0.11~{\rm (stat)}~\pm0.15~{\rm (sys)}~\times10^{-4}~{\rm yr^{-1}~Mpc^{-3}}$, consistent with our measurement.

\begin{table}
\caption{Volumetric CC SN rate measurements}\label{table:rates_CC}
\begin{tabular}{cccl}
\hline
\hline
Redshift & $N_{\textrm{CC}}$ & Rate  & Reference \\
\hline

0 & 440 & $0.62^{+0.07,+0.17}_{-0.07,-0.15}$ & \citet{li2011rates} \\

$<0.0026^a$ & 14 & $1.1^{+0.4}_{-0.3}$ & \citet{botticella2012} \\

$<0.0035^a$ & 35 & $>1.5^{+0.4}_{-0.3}$ & \citet{2012ApJ...756..111M} \\

$<0.0066^a$ & 92 & $>0.96$ & \citet{smartt2009mnras} \\

0.01 & 67 & $0.43\pm 0.17$ & \citet{cappellaro1999}$^b$ \\

0.072 & 89 & $1.06^{+0.11,+0.15}_{-0.11,-0.15}$ & \citet{2014ApJ...792..135T} \\

\textbf{0.075} & \textbf{16} & $\mathbf{1.04^{+0.33,+0.04}_{-0.26,-0.11}}$ & \textbf{This work}$^c$ \\ 

0.21 & 44.95$^d$ & $1.15^{+0.43,+0.42}_{-0.33,-0.36}$ & \citet{botticella2008} \\

0.26 & 31.2$^d$ & $1.88^{+0.71}_{-0.58}$ & \citet{cappellaro2005}$^b$ \\

0.3 & 17 & $2.51^{+0.88,+0.75}_{-0.75,-1.86}$ & \citet{dahlen2004}$^e$ \\

0.3 & 117 & $1.63^{+0.34,+0.37}_{-0.34,-0.28}$ & \citet{bazin2009} \\

0.39 & 3 & $3.29^{+3.08,+1.98}_{-1.78,-1.45}$ & \citet{melinder2012} \\

0.39 & 9 & $3.00^{+1.28,+1.04}_{-0.94,-0.57}$ & \citet{2012ApJ...757...70D} \\

0.66 & 8.7 & $6.9^{+9.9}_{-5.4}$ & \citet{Graur2011} \\

0.7 & 17 & $3.96^{+1.03,+1.92}_{-1.06,-2.60}$ & \citet{dahlen2004}$^e$ \\

0.73 & 5 & $6.40^{+5.30,+3.65}_{-3.12,-2.11}$ & \citet{melinder2012} \\

0.73 & 25 & $7.39^{+1.86,+3.20}_{-1.52,-1.60}$ & \citet{2012ApJ...757...70D} \\

1.11 & 11 & $9.57^{+3.76,+4.96}_{-2.80,-2.80}$ & \citet{2012ApJ...757...70D} \\

\hline

\multicolumn{4}{l}{\textit{Note.} Rates are measured in units of $10^{-4}$ yr$^{-1}$ Mpc$^{-3}$. Where} \\
\multicolumn{4}{l}{reported, the statistical errors are followed by systematic errors,} \\
\multicolumn{4}{l}{and separated by commas.} \\
\multicolumn{4}{l}{$^a$\citet{botticella2012}, \citet{2012ApJ...756..111M}, and} \\
\multicolumn{4}{l}{\citet{smartt2009mnras} consider CC~SNe within 11, 15, and 28 Mpc,} \\
\multicolumn{4}{l}{respectively.} \\
\multicolumn{4}{l}{$^b$Rates have been converted to volumetric rates using equation~\ref{eq:volconv}.} \\
\multicolumn{4}{l}{$^c$Converted from the SN~II rate of $0.621^{+0.197,+0.024}_{-0.154,-0.063}$ by assuming} \\
\multicolumn{4}{l}{that SNe IIP/L account for 60 per cent of all CC SNe.} \\
\multicolumn{4}{l}{$^d$\citet{botticella2008} and \citet{cappellaro2005} found a total} \\
\multicolumn{4}{l}{of 86 and 31.2 SN candidates of all types, respectively.} \\
\multicolumn{4}{l}{$^e$Superseded by \citet{2012ApJ...757...70D}.}

\end{tabular}
\end{table}

\citet{2011ApJ...738..154H} compared the volumetric CC SN rates and the cosmic SFH and found that the scaling factor required to match the two was a factor of $\sim2$ smaller than what they expected from the initial mass function (IMF). This meant that too few CC SNe were detected to account for the explosive death of all stars with masses $>8~{\rm M_\odot}$. This discrepancy may be due to SN surveys systematically missing, and not accounting for, some fraction of CC SNe that explode in the dusty environments of highly star-forming galaxies. Several surveys have attempted to compensate for this missing fraction (\citealt*{2007MNRAS.377.1229M}; \citealt{Graur2011,2012ApJ...756..111M,melinder2012,2012ApJ...757...70D}). Here, we demonstrate that not only can these new rates account for the observed discrepancy, but also that the rate we measure here, which is not corrected for this missing fraction, falls exactly where expected.

First, we repeat the calculation of the expected scaling factor, but use a different cosmic SFH fit \citep*{Behroozi2012} and IMF in order to remain consistent with previous rates-measurement works such as GM13, \citet*{Maoz2010clusters}, \citet{Maoz2010loss}, and \citet*{Maoz2012sdss}. We use the `diet' Salpeter IMF from \citet{2001ApJ...550..212B}; see section 5 of \citealt{Graur2014} for more details. These choices result in a scaling factor of $A = 0.0093~{\rm SNe~M_\odot^{-1}}$, similar to the \citet{2011ApJ...738..154H} value of $0.0088~{\rm SNe~M_\odot^{-1}}$.

Next, we fit all CC SN rate measurements as they appear in Table~\ref{table:rates_CC}, excluding the rates measured by \citet{dahlen2004}, which have been superseded by the rates in \citet{2012ApJ...757...70D}, and the rate from \citet{smartt2009mnras}, which is an estimate of the lower limit of the CC SN rate. Where necessary, we have corrected rates to reflect the value of $h=0.7$ used in this work. Rates that were originally reported in units of SNuB ($10^{-12}~{\rm yr^{-1}~L_{B,\odot}}$) were converted to volumetric rates by means of the \citet{botticella2008} redshift-dependent luminosity density function,
\begin{equation}\label{eq:volconv}
 j_B(z)=(1.03+1.76\, z)\times 10^8~\rm{L}_{B,\odot}~\textrm{Mpc}^{-3}.
\end{equation}
When fitting all CC SN rate measurements in Fig.~\ref{fig:rate_vol_CC}, the best-fitting scaling factor is $A_{all} = 0.0058^{+0.0012}_{-0.0012}~{\rm (stat)}~^{+0.0014}_{-0.0011}~{\rm (sys)}$, with a reduced $\chi^2$ value $\chi_r^2 = 1.2$ for 14 DOF. This is indeed lower by a factor of $\sim2$ than the value we would expect from the IMF. However, when we divide the CC SN rates into those with (`dust') and without (`no dust') the correction for the missing CC SNe in highly star-forming galaxies, the resultant scaling factors are $A_{dust} = 0.0104^{+0.0035}_{-0.0030}~{\rm (stat)}~^{+0.0019}_{-0.0011}~{\rm (sys)}$ and $A_{no~dust} = 0.0050^{+0.0008}_{-0.0009}~{\rm (stat)}~^{+0.0009}_{-0.0007}~{\rm (sys)}$, with $\chi^2_r = 0.03$ and $1.0$ for six and seven DOF, respectively. The low $\chi_r^2$ value for the fit to the dust-corrected CC SN rates is due to their large statistical, as well as systematic, uncertainties, which, in turn, are due to the small samples of CC SNe observed at the relatively high redshifts of their respective surveys. The scaling factor $A_{dust}$ is consistent with the value we expect from the IMF.

\subsection{Sources of systematic uncertainty}
\label{subsec:systematics}
There are several sources of systematic uncertainty in the calculation of the visibility time that can propagate into the SN rates. Here, we examine several such sources: the choice of LF, the criterion chosen to divide between passive and star-forming galaxies, and the fractions of SNe~IIP and IIL in our sample. We also discuss potential biases in the galaxy properties measured by the MPA-JHU Galspec pipeline. In the case of SNe Ia, as we discussed in Section~\ref{sec:sample}, four SNe may have exploded outside the area covered by the fibre aperture, and five may have been misclassified SNe Ic. In Table~\ref{table:sysbins}, we detail into which SN Ia rate bins we allocate these SNe. Tables~\ref{table:errors_mass}--\ref{table:errors_sSFR} summarize the uncertainty budgets of the various measurements. Though we test several sources of systematic uncertainty, our measurements are limited by the statistical uncertainties due to the size of the SN sample.

\subsubsection{Luminosity functions}
\label{subsub:vistime_LF}
We use various LFs to test what systematic effect they might have on the derived rates. For the SN~Ia rates, we use three LFs. The first, from \citet{yasuda2010}, assumes that any colour variation in SNe~Ia is due to host-galaxy extinction with $R_V=3.1$, as is the average value in our Galaxy (although it appears likely that colour variation among SNe Ia may be due to a combination of intrinsic colour scatter and reddening by dust; e.g., \citealt{Chotard2011,2014ApJ...780...37S}; \citealt*{2014ApJ...797...75M}). We also test a second \citet{yasuda2010} LF, which assumes $R_V=1.92$, an average value closer to those of SN Ia host galaxies \citep{2008A&A...487...19N,2009ApJS..185...32K}, and the two \citet{li2011LF} LFs for E--Sa and Sb--Irr galaxies, both of which were not corrected for host-galaxy extinction. Likewise, for the SN~II visibility time, we use the extinction-corrected LFs used by \citet{Graur2014} and \citet{2014AJ....148...13R}, as well as the \citet{li2011LF} LFs for SNe IIP and IIL. Because the \citet{li2011LF} LFs have not been corrected for host-galaxy extinction, when we use them in the visibility-time calculation, we cannot redden the spectra before they are redshifted, as we cannot break the degeneracy between the intrinsic luminosity of the SN and any extinction it might have suffered. The choice of LF has, at most, a 15 per cent effect on the measured rates, $\sim2$--4 times smaller than the statistical uncertainty due to the size of the SN sample.

\subsubsection{Galaxy-type criterion}
\label{subsub:ssfr_criterion}
In Sections~\ref{sec:method} and \ref{subsec:vistime}, we and measured the detection efficiency of SNe II and the visibility times of the galaxies in our sample, respectively, by assuming (as in \citealt{2006ApJ...648..868S}) that galaxies could be classified as passive if their sSFR obeyed ${\rm log(sSFR/{\rm yr^{-1}})}<-12$ and star-forming if ${\rm log(sSFR/{\rm yr^{-1}})}\ge-12$. However, in Section~\ref{subsec:rates_Ia}, we saw that a better classification criterion for the galaxies in our sample is ${\rm log(sSFR/{\rm yr^{-1}})}=-11.2$. This new criterion has no effect on the SN detection efficiencies used here, as those measure the probability of detecting a SN given the S/N ratio of the data and the contrast between the SN and the galaxy. The SN Ia visibility times, however, are affected, as we have used different stretch distributions for passive and star-forming galaxies. To enable direct comparison with the GM13 rates, we keep the original visbility times, but calculate new values for the $\sim150\,000$ galaxies ($\sim20.3$ per cent of the galaxy sample) in the range $-12<{\rm log(sSFR/{\rm yr^{-1}})}<-11.2$ and derive systematic uncertainties from the resulting rates. These uncertainties are presented in Tables~\ref{table:errors_mass}--\ref{table:errors_sSFR} under `Galaxy-type criterion.'

\begin{table}
 \center
 \caption{Allocation of systematic uncertainties of the mass-normalized SN~Ia rates}
 \begin{tabular}{lccc}
  \hline
  \hline
  Plate-MJD-fibre & Stellar mass              & SFR                       & sSFR \\
                  & $(10^{10}~{\rm M_\odot})$ & $({\rm M_\odot~yr^{-1}})$ & $(10^{-12}~{\rm yr^{-1}})$ \\
  \hline
  \multicolumn{4}{c}{SNe~Ia that exploded outside the fibre aperture} \\
  0646-52523-183 & 3.7 (2,2,0)  & 0.8 (3,1,0)  & 19 (3,1,0) \\
  0767-52252-123 & 0.8 (1,1,0)  & 2.6 (4,2,0)  & 300 (4,3,0) \\
  1452-53112-120 & 0.04 (1,1,0) & 0.1 (2,1,0)  & 210 (4,3,0) \\
  1665-52976-155 & 0.7 (1,0,1)  & 0.02 (1,0,1) & 2.2 (2,0,3) \\
  \multicolumn{4}{c}{SNe~Ic possibly misclassified as SNe~Ia} \\
  0498-51984-102 & 13 (4,0,3)  & 0.1 (2,0,3)  & 0.8 (1,0,2) \\
  1059-52618-552 & 0.6 (1,1,0) & 10 (4,3,0)   & 1500 (4,3,0) \\
  1574-53476-461 & 14 (4,0,3)  & 0.85 (3,0,3) & 5.2 (2,0,3) \\
  1645-53172-349 & 18 (4,0,3)  & 0.1 (2,0,3)  & 0.5 (1,0,1) \\
  1946-53432-030 & 2.2 (2,2,0) & 1.2 (3,2,0)  & 50 (3,2,0) \\
  \hline
  \multicolumn{4}{l}{\textit{Note.} the numbers in parentheses represent the mass-normalized} \\
  \multicolumn{4}{l}{SN Ia rate bin in which each SN is included. The bins for rates} \\
  \multicolumn{4}{l}{measured in all galaxies, as well as star-forming and passive} \\
  \multicolumn{4}{l}{galaxies, are separated by commas. A zero value means that the} \\
  \multicolumn{4}{l}{specific SN is not included in either the star-forming or passive} \\
  \multicolumn{4}{l}{galaxies subsamples.}
 \end{tabular}
 \label{table:sysbins}
\end{table}

\subsubsection{SN IIP and IIL fractions}
\label{subsub:vistime_frac}
When calculating the visibility time of SNe II, we need to simulate what fraction of our SN~II sample would be composed of either SNe IIP or IIL. As we cannot break the degeneracy between SN IIP and IIL spectra without light curves (see section~4 of GM13), we rely on the population fractions measured by L11. For a magnitude-limited survey, such as ours, L11 calculated that SNe~IIP and IIL would make up 39.4 and 27.5 per cent of the SN~II population (in which SNe IIn and IIb were also included, at 10.1 and 23.0 per cent, respectively). However, these fractions are based on a local sample of SNe that were observed in bright, massive galaxies. Our SN~II sample is at a median redshift of 0.075 and half of it is observed in dwarf galaxies, where the CC~SN subtype fractions may be quite different (\citealt{2010ApJ...721..777A}, but see \citealt{2014AN....335..841T} for a different conclusion). Based on the L11 results, we calculate the SN~II visibility time by assigning SN IIP light curves to half of our galaxy sample and SN IIL light curves to the other half. However, as we do not know the real distribution of SNe IIP and IIL in the SDSS galaxy sample, we also test the extreme cases in which we use only either SN IIP or SN IIL light curves. Although these are unrealistic assumptions, they still result in a negligible effect on the measured SN II rates of $\sim2$--15 per cent, more than four times smaller than the statistical uncertainties.

\begin{table*}
 \center
 \caption{Uncertainty percentages for SN rates per unit mass vs. stellar mass}
 \begin{tabular}{lcccc}
  \hline
  \hline
  \multicolumn{5}{c}{SN~Ia rates per unit mass in all galaxies} \\
              & \multicolumn{4}{c}{Mass range $(10^{10}~{\rm M_\odot})$} \\
  Uncertainty           & $M_*<2.1$ & $2.1\le M_*<5.3$ & $5.3\le M_*<9.1$ & $M_*\ge9.1$ \\
  Poisson               & $+26,-21$ & $+26,-21$ & $+26,-21$ & $+26,-21$ \\
  Luminosity function   & $+9.5,-0$ & $+7.8,-0$ & $+8.7,-0$ & $+14,-3.7$ \\
  Extra-aperturial SNe  & $+0,-13$  & $+0,-4.5$ & $+0,-0$   & $+0,-0$ \\
  Misclassification     & $+0,-4.3$ & $+0,-4.5$ & $+0,-0$   & $+0,-13$ \\
  Galaxy-type criterion & $+3,-0$   & $+5.6,-0$ & $+6.9,-0$ & $+7.8,0$ \\
  \multicolumn{5}{c}{SN~Ia rates per unit mass in star-forming galaxies} \\
  Uncertainty           & $M_*<1.7$ & $1.7\le M_*<4.3$ & $M_*\ge4.3$ & \\
  Poisson               & $+35,-26$ & $+35,-26$ & $+35,-26$ & \\
  Luminosity function   & $+10,-0$  & $+13,-0$  & $+15,-0$  & \\
  Extra-aperturial SNe  & $+0,-14$  & $+0,-7.1$ & $+0,-0$   & \\
  Misclassification     & $+0,-7.1$ & $+0,-7.1$ & $+0,-0$   & \\
  Galaxy-type criterion & $+0,-0$   & $+0,-0$   & $+0,-0$   & \\
  \multicolumn{5}{c}{SN~Ia rates per unit mass in passive galaxies} \\
  Uncertainty           & $M_*<6.2$ & $6.2\le M_*<9.6$ & $M_*\ge9.6$ & \\
  Poisson               & $+32,-25$ & $+32,-25$   & $+31,-24$  & \\
  Luminosity function   & $+6.5,-0$ & $+8.8,-0.5$ & $+15,-7.7$ & \\
  Extra-aperturial SNe  & $+0,-6.3$ & $+0,-0$     & $+0,-0$    & \\
  Misclassification     & $+0,-0$   & $+0,-0$     & $+0,-18$   & \\
  Galaxy-type criterion & $+11,-0$  & $+11,-0$    & $+10,-0$   & \\
  \multicolumn{5}{c}{SN~II rates per unit mass in non-AGN star-forming galaxies} \\
  Uncertainty         & $M_*<0.17$ & $0.17\le M_*<0.58$ & $M_*\ge0.58$ & \\
  Poisson             & $+68,-43$ & $+60,-40$ & $+68,-43$   & \\
  Luminosity function & $+5.6,-0$ & $+3.1,-0$ & $+0,-4.9$   & \\
  SN~IIP/IIL fraction & $+17,-13$ & $+13,-11$ & $+4.9,-4.9$ & \\
  \hline
  \multicolumn{5}{l}{\textit{Note.} All uncertainties are reported as percentages of the rates.}
 \end{tabular}
 \label{table:errors_mass}
\end{table*}

\begin{table*}
 \center
 \caption{Uncertainty percentages for SN rates per unit mass vs. SFR}
 \begin{tabular}{lcccc}
  \hline
  \hline
  \multicolumn{5}{c}{SN~Ia rates per unit mass in all galaxies} \\
              & \multicolumn{4}{c}{SFR range $({\rm M_\odot~yr^{-1}})$} \\
  Uncertainty & ${\rm SFR}<0.063$ & $0.063\le{\rm SFR}<0.19$ & $0.19\le{\rm SFR}<1.6$ & ${\rm SFR}\ge1.6$ \\
  Poisson               & $+26,-21$   & $+26,-21$  & $+26,-21$ & $+26,-21$ \\
  Luminosity function   & $+8.6,-6.8$ & $+12,-5.1$ & $+11,-0$  & $+15,-0$  \\
  Extra-aperturial SNe  & $+0,-4.3$   & $+0,-4.5$  & $+0,-4.3$ & $+0,-4.3$ \\
  Misclassification     & $+0,-0$     & $+0,-9.1$  & $+0,-8.7$ & $+0,-4.3$ \\
  Galaxy-type criterion & $+2.6,-0$   & $+9.2,-0$  & $+13,-0$  & $+0.4,-0$ \\
  \multicolumn{5}{c}{SN~Ia rates per unit mass in star-forming galaxies} \\
  Uncertainty & ${\rm SFR}<1.2$ & $1.2\le{\rm SFR}<2.6$ & ${\rm SFR}\ge2.6$ & \\
  Poisson               & $+35,-26$ & $+35,-26$ & $+35,-26$ & \\
  Luminosity function   & $+12,-0$  & $+14,-0$  & $+15,-0$  & \\
  Extra-aperturial SNe  & $+0,-14$  & $+0,-7.1$ & $+0,-0$   & \\
  Misclassification     & $+0,-0$   & $+0,-7.1$ & $+0,-7.1$ & \\
  Galaxy-type criterion & $+0,-0$   & $+0,-0$   & $+0,-0$   & \\
  \multicolumn{5}{c}{SN~Ia rates per unit mass in passive galaxies} \\
  Uncertainty & ${\rm SFR}<0.05$ & $0.05\le{\rm SFR}<0.1$ & ${\rm SFR}\ge0.1$ & \\
  Poisson               & $+32,-25$   & $+32,-25$  & $+31,-24$ & \\
  Luminosity function   & $+8.2,-6.6$ & $+11,-7.6$ & $+13,-0$  & \\
  Extra-aperturial SNe  & $+0,-6.3$   & $+0,-0$    & $+0,-0$   & \\
  Misclassification     & $+0,-0$     & $+0,-0$    & $+0,-18$  & \\
  Galaxy-type criterion & $+2.3,-0$   & $+5.0,-0$  & $+21,-0$  & \\
  \multicolumn{5}{c}{SN~II rates per unit mass in non-AGN star-forming galaxies} \\
  Uncertainty         & ${\rm SFR}<0.4$ & $0.4\le{\rm SFR}<0.9$ & ${\rm SFR}\ge0.9$ & \\
  Poisson             & $+68,-43$ & $+60,-40$  & $+68,-43$   & \\
  Luminosity function & $+3.8,-0$ & $+0.5,-0$  & $+0,-5.6$   & \\
  SN~IIP/IIL fraction & $+15,-12$ & $+12,-8.8$ & $+4.0,-3.2$ & \\
  \hline
  \multicolumn{5}{l}{\textit{Note.} All uncertainties are reported as percentages of the rates.}
 \end{tabular}
 \label{table:errors_SFR}
\end{table*}

\subsubsection{Galaxy properties}
\label{subsub:vistime_galprop}
In this work, we rely on the stellar masses, SFRs, and sSFRs measured for each galaxy by the MPA-JHU Galspec pipeline. Since these measurements are done on the SDSS galaxy spectra, the parameters of the SN host galaxies could be systematically affected by contamination by the SN light. 

\begin{table*}
 \center
 \caption{Uncertainty percentages for SN rates per unit mass vs. sSFR}
 \begin{tabular}{lcccc}
  \hline
  \hline
  \multicolumn{5}{c}{SN~Ia rates per unit mass in all galaxies} \\
              & \multicolumn{4}{c}{sSFR range $(10^{-12}~{\rm yr^{-1}})$} \\
  Uncertainty & ${\rm sSFR}<0.74$ & $0.74\le{\rm sSFR}<5.1$ & $5.1\le{\rm sSFR}<60$ & ${\rm sSFR}\ge60$ \\
  Poisson               & $+26,-21$ & $+26,-21$ & $+26,-21$ & $+26,-21$ \\
  Luminosity function   & $+12,-13$ & $+10,-0$  & $+14,-0$  & $+13,-0$  \\
  Extra-aperturial SNe  & $+0,-0$   & $+0,-4.5$ & $+0,-4.3$ & $+0,-8.7$ \\
  Misclassification     & $+0,-8.7$ & $+0,-4.5$ & $+0,-4.3$ & $+0,-4.3$ \\
  Galaxy-type criterion & $+0,-0$   & $+22,-0$  & $+1.8,-0$ & $+0,-0$   \\
  \multicolumn{5}{c}{SN~Ia rates per unit mass in star-forming galaxies} \\
  Uncertainty & ${\rm sSFR}<40$ & $40\le{\rm sSFR}<120$ & ${\rm sSFR}\ge120$ & \\
  Poisson               & $+35,-26$ & $+35,-26$ & $+35,-26$ & \\
  Luminosity function   & $+14,-0$  & $+14,-0$  & $+13,-0$  & \\
  Extra-aperturial SNe  & $+0,-7.1$ & $+0,-0$   & $+0,-14$  & \\
  Misclassification     & $+0,-0$   & $+0,-7.1$ & $+0,-7.1$ & \\
  Galaxy-type criterion & $+0,-0$   & $+0,-0$   & $+0,-0$   & \\
  \multicolumn{5}{c}{SN~Ia rates per unit mass in passive galaxies} \\
  Uncertainty & ${\rm sSFR}<0.6$ & $0.6\le{\rm sSFR}<1.2$ & ${\rm sSFR}\ge1.2$ & \\
  Poisson               & $+32,-25$ & $+32,-25$  & $+31,-24$ & \\
  Luminosity function   & $+12,-13$ & $+11,-7.1$ & $+14,-0$  & \\
  Extra-aperturial SNe  & $+0,-0$   & $+0,-0$    & $+0,-5.9$ & \\
  Misclassification     & $+0,-6.3$ & $+0,-6.3$  & $+0,-5.9$ & \\
  Galaxy-type criterion & $+0,-0$   & $+5.6,-0$  & $+29,-0$  & \\
  \multicolumn{5}{c}{SN~II rates per unit mass in non-AGN star-forming galaxies} \\
  Uncertainty         & ${\rm sSFR}<120$ & $120\le{\rm sSFR}<280$ & ${\rm sSFR}\ge280$ & \\
  Poisson             & $+68,-43$   & $+60,-40$   & $+68,-43$   & \\
  Luminosity function & $+0,-6.2$   & $+0,-4.6$   & $+0,-8.3$   & \\
  SN~IIP/IIL fraction & $+3.8,-5.0$ & $+4.4,-2.2$ & $+1.6,-4.9$ & \\ 
  \hline
  \multicolumn{5}{l}{\textit{Note.} All uncertainties are reported as percentages of the rates.}
 \end{tabular}
 \label{table:errors_sSFR}
\end{table*}

In Galspec, stellar masses are computed using the Bayesian methodology and model grids of \citet{2003MNRAS.341...33K} and the photometry of the galaxy: fibre magnitudes for the stellar mass within the fibre aperture and model magnitudes for the stellar mass of the entire galaxy. The spectra are used only to correct the photometry for the small contribution due to nebular emission lines. SFRs are computed both within the SDSS fibre aperture, using spectral emission lines \citep{2004MNRAS.351.1151B}, and outside the aperture using galaxy photometry \citep{2007ApJS..173..267S}. The sSFRs are calculated by combining the likelihood distributions of the stellar masses and SFRs \citep{2004MNRAS.351.1151B}. 

According to this methodology, the stellar masses measured within the fibre, as well as any galaxy properties dependent on measurements of the spectral emission lines, could be affected by contaminating SN light. The spectral emission lines are measured by first fitting the continuum with a stellar population model composed of the \citet{2003MNRAS.344.1000B} stellar population libraries. Any remaining residuals after subtracting this model are removed using a sliding 200-pixel median filter. Finally, the various emission lines are fit with separate Gaussians simultaneously, with the requirement that all lines of a specific element have the same line width and velocity offset \citep{2004ApJ...613..898T}. The removal of the continuum should, in principle, mitigate any contamination by the SN light. 

The stellar masses within the fibre aperture are only used in Equation~\ref{eq:rates_Ia}, above, where they are multiplied by the visibility times of the galaxies within mass bin $i$ (which is deilneated according to the total stellar mass of the galaxies), and summed to produce the denominator of the mass-normalized SN rate in bin $i$. As each bin contains of the order of $10^4$--$10^5$ galaxies, any systematic error in the fibre masses of the 91 and 16 SNe Ia and SN II host galaxies will have a negligible effect on the final result.

\begin{figure}
 \includegraphics[width=0.475\textwidth]{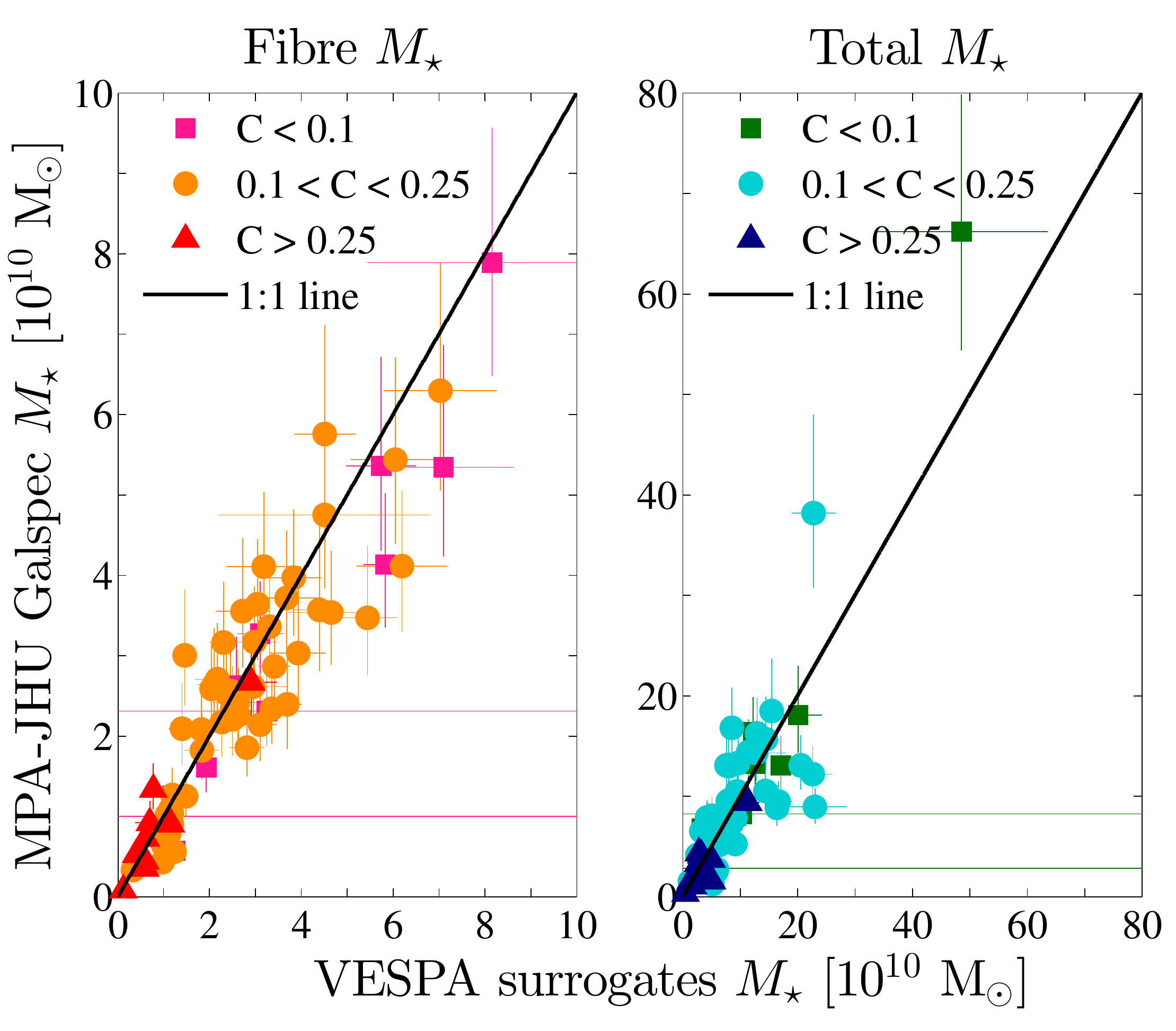}
 \caption{Comparison between the fibre (left) and total (right) stellar masses derived by the MPA-JHU Galspec pipeline used here and the \vespa-derived masses of the surrogate galaxies chosen by GM13 of the subsample of SN~Ia host galaxies shared by both works. Squares denote measurements for original host galaxies where the SN light contributed less than ten per cent of the total light (i.e., ${\mathrm C}<0.1$). Circles and triangles denote measurements for galaxies where $0.1<{\mathrm C}<0.25$ and ${\mathrm C}>0.25$, respectively. The horizontal and vertical error bars denote the \vespa\ and Galspec uncertainties, defined according to equation 26 in \citet{Tojeiro2009} for \vespa\ and the 16th and 84th percentiles of the Galspec mass probability density function of each galaxy. Most measurements fall along the solid 1:1 curve, showing that the Galspec masses are insensitive to SN light contamination.}
 \label{fig:mass_comp}
\end{figure}

In GM13, we avoided any systematic effects posed by the way \vespa\ measured the stellar masses and SFHs from the galaxy spectra by identifying `surrogate' galaxies that had nearly identical spectra to those of the SN host galaxies once the SN signal was removed. As 69 of the SNe Ia in our current sample were also detected in GM13, we can compare the stellar masses of these host galaxies, as measured by \vespa\ and Galspec. Fig.~\ref{fig:mass_comp} shows this comparison for stellar masses measured both inside the fibre aperture and for the entire galaxy. We separate the SNe according to the contrast, C, between the SN and galaxy light. In GM13, we did not choose surrogate galaxies for host galaxies where the SN light contributed less than 10 per cent of the total flux in the spectrum. The scatter around the 1:1 line is consistent with the uncertainties of the measurements, which reassures us that the stellar masses mesured by Galspec, and thus, the SFRs and sSFRs as well, are not systematically affected by the contaminating SN light.


\section{Discussion and Testable Predictions}
\label{sec:discuss}

\subsection{Testing the supernova rate correlations}
\label{subsec:disucss_correlations}

The connections shown in Figures~\ref{fig:SNuM_Ia_mass} and \ref{fig:SNuM_Ia_SFR} between SN rates and various galaxy properties lead us to conclude that the SN Ia rate correlations can all be explained as a combination of the SN Ia DTD (a power law with an index of $-1$), the redshifts at which the galaxies in the survey are observed, and galaxy downsizing, whereby older galaxies tend to be more massive than younger ones. The ages of the galaxies, not their stellar masses, SFRs, or sSFRs, seem to be the dominant galaxy property that affects the SN Ia rates. To check whether this is indeed the case, one could isolate a sample of SNe Ia that exploded in galaxies with the same stellar masses and redshifts but with different ages. The age-mass relation measured by \citealt{2005MNRAS.362...41G} shows a wide dispersion of ages in any given mass bin (where the dispersion is larger than what one would expect from the statistical uncertainties of the measurements alone). Thus, in a given mass bin, we predict that the SN rates per unit mass will be higher in the younger, rather than the older, galaxies.

To effectively test this prediction, one would require a large sample of SNe in a large sample of galaxies with independent measurements of their stellar masses and ages. This could be done, for example, by: a) extending the work of \citet{2005MNRAS.362...41G}, originally done with a sample of $~44\,000$ galaxies from SDSS DR2, to all of the galaxies in SDSS DR8, and using the SN Ia sample from this work; b) discovering SNe among the $\sim1.5$ million spectra from SDSS-III BOSS and using the galaxy properties derived by the Portsmouth Group pipeline \citep{2013MNRAS.435.2764M,2013MNRAS.431.1383T}; or c) measuring ages and stellar masses for the galaxies monitored by LOSS and using their SN sample.

We have shown, in Fig.~\ref{fig:SNuM_II_SFR}, that the correlations between SN II rates and galaxy properties, combined with the correlations between galaxy stellar mass and either SFR or sSFR can predict rate-mass correlations that are consistent with the measured rates. This is due to the short delay times between the formation and explosion of the progenitor stars of SNe II, as we show by extending the SN Ia rate simulation to SNe II, assuming that the SFHs of the galaxies can be described by declining exponential functions, convolved with a simplified model for the SN II DTD: a uniform probability to explode 9--40 Myr after the formation of the probed stellar population. 

In the case of SNe II, then, it is not only the age of the galaxy that drives the rate correlations, but also the shape of the galaxy's SFH. If we assume that all SFHs decline over time, then the SN II DTD would always come into effect at the lowest point of the SFH. Thus, older galaxies will, on average, have lower SN rates than younger galaxies. Galaxies at the same global age, but with higher star-formation rates in the previous tens of millions of years, would be expected to exhibit higher SN II rates. To test this, one could either measure detailed SFHs for the galaxies in a SN survey, or preferably connect the SNe directly to the stellar populations in which they originated. One way to accomplish this would be to use integral-field unit spectroscopy to collect the spectra of isolated star-forming regions in hundreds of thousands of galaxies; in other words, a spectroscopic galaxy survey similar to SDSS but with higher spatial resolution. The SDSS-IV project Mapping Nearby Galaxies at APO \citep{2015ApJ...798....7B} is an example of such a survey, though it is not expected to discover many SNe, as it only intends to survey $\sim10\,000$ galaxies. 

The mass-normalized rates measured here are also consistent with the L11 power-law fits. Due to the shape of the \citet{2005MNRAS.362...41G} galaxy mass-age relation, the SN Ia rates per unit mass plateau at low ($\ltsim10^9~{\rm M_\odot}$) and high ($\gtsim10^{12}~{\rm M_\odot}$) galaxy stellar masses, providing a testable deviation from the power-law fits. The SN II simulated rates deviate in a similar manner from the L11 power-law fits at $\ltsim2\times10^9~{\rm M_\odot}$. To measure rates in the low-mass range, one could either search for SNe at high redshifts, where the galaxies are on average younger than in the local Universe (e.g., \citealt{Graur2014,2014AJ....148...13R}), or in dwarf galaxies. The high-mass end, on the other hand, could be targeted by applying our method to discover SNe in galaxy spectra to the $\sim1.5$ million BOSS galaxy spectra, where the full-width-at-half-maximum of the galaxy mass distribution is 1--4$\times10^{11}~{\rm M_\odot}$, as opposed to $0.1$--$1.7\times10^{11}~{\rm M_\odot}$ in SDSS.

\subsection{Metallicity effects}
\label{subsec:disucss_metals}

\citet{2005MNRAS.362...41G} measured a correlation between stellar metallicity (from galaxy absorption lines) and stellar mass: the metallicity of the galaxy increases with increasing stellar mass. This correlation has the same form as that between galaxy age and stellar mass, which means that younger galaxies have lower metallicities than older galaxies. \citet{2004ApJ...613..898T} found a similar correlation between galaxy stellar mass and gas-phase oxygen abundance for SDSS star-forming galaxies. \citet{2010MNRAS.408.2115M} showed that this correlation is part of a more general `fundamental metallicity relation' between stellar mass, metallicity, and SFR, and hypothesized that this relation could be explained by the effect of infalling, pristine gas on SFR and the expulsion of enriched gas by the latter (see also \citealt{2014arXiv1412.2139C,2015ApJ...800...91H}). \citet{2014MNRAS.438.1391P} found that PTF SN Ia host galaxies and general field galaxies follow the same correlation between stellar mass and metallicity as well as the fundamental metallicity relation, suggesting any metallicity effect on the SN Ia rate would be small.

\citet{Kistler2011} attempted to test whether the correlation between metallicity and stellar mass, combined with a power-law DTD, could explain the L11 rate-mass correlation. In this case, galaxies with lower metallicity are expected to exhibit a higher SN Ia rate as the lower metallicity allows the formation of relatively more massive white dwarfs, given the same initial stellar mass (e.g., \citealt*{2008A&A...487..625M,2011PASJ...63L..31M}). Such white dwarfs would then either require less time to reach the critical mass at which they explode, or provide more binary white-dwarf systems that, upon merger due to loss of energy to gravitational waves, would once again be massive enough to trigger carbon burning and the subsequent explosion. Alternatively, \citet{1999ApJ...513..861U} claim that the main-sequence stellar mass required to produce carbon-oxygen white dwarfs decreases with decreasing metallicity, which would potentially lead to more SN Ia progenitors in lower-metallicity environments, and thus to higher SN Ia rates. \citet{Kistler2011} concluded that due to the similar shapes of the metallicity and galaxy age correlations with stellar mass, the two were degenerate and provided consistent fits to the LOSS SN Ia rates.  If both galaxy age and metallicity affect SN Ia rates, we might expect our simulated SN Ia rates, which do not take into account the correlation between metallicity and stellar mass, to deviate from the measured rates. Yet, no such deviation is apparent in Figures~\ref{fig:SNuM_Ia_mass} and \ref{fig:SNuM_Ia_SFR}.

By measuring the metallicities of the host galaxies of a sample of SNe II discovered in PTF and contrasting them with the distribution of galaxy metallicities in the SDSS (from the subsample of galaxies that have Galspec metallicities), \citet{2013ApJ...773...12S} found that SNe II were not biased by the metallicity of their host galaxies. Thus, we would not expect any discrepancies between the measured and simulated SN II rates. However, due to small-number statistics, the uncertainties of our rates are too large to either detect or exclude such deviations. Furthermore, in this work we cannot test possible correlations between SN rates and galaxy metallicity. Although Galspec measures oxygen abundances, it does so only for a small subset of the SDSS galaxies. Of the galaxies in our sample, only 193\,591 ($\sim26$ per cent of the sample) galaxies of all types and 28\,742 star-forming galaxies ($\sim13$ per cent of the star-forming galaxy sample) have such measurements. Only nine SN Ia and four SN II host galaxies are included in these subsamples.   

Whereas the proposed metallicity effects on the SN Ia DTD, and subsequently the SN Ia rates, are too small to test with the SN samples presented here and in L11, corellations between metallicity and other galaxy properties should have a more profound effect on the rates of stripped-envelope CC SNe (stripped SNe), i.e., SNe Ib, IIb, Ic, and broad-lined Ic (see reviews by \citealt{1997ARA&A..35..309F} and \citealt{2011AN....332..434M} for SN types). There are indications that the oxygen abundances measured directly at the explosion sites of SNe Ib and SNe Ic are statistically different (\citealt{Modjaz2011,2011A&A...530A..95L,2012ApJ...759..107K,2012ApJ...758..132S}; for a meta-analysis see \citealt{2012IAUS..279..207M}), with both SN Ib and SN Ic metallicities higher than those of SNe II (\citealt*{2008ApJ...673..999P}; \citealt{2010MNRAS.407.2660A}). These observations can be explained by metallicity-dependent winds, wherein higher-metallicity progenitors have more of their envelopes stripped by stronger and faster winds, though other factors such as binary interaction need to be included for stripped SNe (\citealt*{1992ApJ...391..246P}; \citealt{2012Sci...337..444S,2014ApJ...782....7D}). Furthermore, the rarer broad-lined SNe Ic appear to prefer sites and host galaxies with lower oxygen abundances than `normal' SNe Ic \citep{2010ApJ...721..777A,Modjaz2011,2012ApJ...759..107K,2012ApJ...758..132S}, with broad-lined SNe Ic connected with gamma-ray bursts occupying the lowest metallicity sites systematically, but not exclusively. Thus, any DTD models fit to stripped SN rates would need to include the metallicity dependence for each stripped SN subtype.

One could repeat the work done here for stripped SNe by using the stripped SN rates measured by L11, together with measurements of the SFHs of the LOSS galaxies and independent measurements of their metallicities. However, as we note above, in order to probe the short delay times of CC SNe, one would require SFHs with high temporal resolution (of the order of millions of years). While some of the LOSS galaxies already have SFHs measured with \vespa\ \citep{Maoz2010loss}, their resolution is not high enough (see figure 1 of \citealt{Tojeiro2009}), so that new SFHs would need to be measured for all the LOSS galaxies.


\section{SUMMARY AND CONCLUSIONS}
\label{sec:summary}

Using a method developed by GM13 to detect and classify SNe in galaxy spectra, we have detected 91 SNe Ia and 16 SNe II in \Ngal\ galaxies of all types and \NIIgal\ star-forming galaxies without AGNs, respectively, in SDSS DR9. Of these SNe, 15 SNe Ia and 8 SNe II are new discoveries reported here for the first time. We have used these SN samples to measure mass-normalized SN rates as a function of galaxy stellar mass, SFR, and sSFR, using the values measured by the MPA-JHU Galspec pipeline. 

The total SN Ia and SN II rates per unit mass decrease with increasing stellar mass, as first reported by \citet{2006ApJ...648..868S} for SNe Ia and L11 for all SN types. However, our SN Ia rates per unit mass in passive galaxies are consistent with being independent of galaxy stellar mass, similarly to the SNLS SN Ia rates observed by \citet{2006ApJ...648..868S}. 

We also confirm the correlation between SN Ia rates per unit mass and sSFR. This correlation was first noted by \citet{2005A&A...433..807M}, based on a sample of local SNe, and confirmed by \citet{2006ApJ...648..868S}, based on the SNLS SN sample in the redshift range $0.2<z<0.75$, and \citet{2012ApJ...755...61S}, using the $0.05<z<0.25$ SDSS-II SN sample. Our rates, measured in the intermediate redshift range $0.04<z<0.2$, reinforce the \citet{2012ApJ...755...61S} rates, which bridge the gap between the earlier surveys. A similar correlation exists for SNe II. Finally, we show for the first time a third correlation between SN rates per unit mass and galaxy SFR.

The mass-normalized SN Ia and SN II rates, averaged over all masses and redshifts in their respective galaxy samples, are $R_{\rm Ia,M}(z=0.1) = $~\SNuMIaall\ and $R_{\rm II,M}(z=0.075) = $~\SNuMII. Taking into account the mass distribution of our galaxy sample, we have converted the mass-normalized SN II rate into a volumetric rate at $z=0.075$ of $R_{\rm II,V} = $~\RvolII\ (the volumetric SN Ia rate was previously measured by us in GM13). Assuming that SNe IIP and IIL account for 60 per cent of all CC SNe, the volumetric rate of all CC SNe is $R_{\rm CC,V} = $~\RvolCC. 

We perform a detailed analysis of possible sources of systematic uncertainty that might affect our SN rates and find that the dominant source of uncertainty is the statistical uncertainty associated with the size of our SN sample.

We argue that the correlations shown here between SN rates and galaxy stellar mass, SFR, and sSFR can all be explained by a combination of the redshifts at which the galaxies are observed, their ages and SFHs, and the SN DTD. This explanation was first proposed for SNe Ia by \citet{Kistler2011}, but we extend it to SNe II and show that it can explain not only the correlation between SN rates and galaxy stellar mass, but also the correlations with SFR and sSFR. We provide several ways to test whether the age of the SN host galaxy is the main galaxy property behind the SN Ia correlations and whether SFH is the main galaxy property driving the SN II correlations. However, we note that metallicity, which has a similar correlation with stellar mass as the age of the galaxy, is also expected to affect the rates, at least for SNe Ia and stripped SNe. 


\section*{Acknowledgments}
We thank the following for useful discussions and comments: Michael Blanton, Jenny Greene, Saurabh Jha, Dan Maoz, Amy Reines, Michael M. Shara, and David Zurek, as well as the referee, Mark Sullivan.

FBB is supported by a James Arthur fellowship at the Center for Cosmology and Particle Physics at NYU. MM is supported in parts by the NSF CAREER award AST-1352405 and by NSF award AST-1413260.

Funding for SDSS-III has been provided by the Alfred P. Sloan Foundation, the Participating Institutions, the National Science Foundation, and the U.S. Department of Energy Office of Science. The SDSS-III web site is http://www.sdss3.org/.

SDSS-III is managed by the Astrophysical Research Consortium for the Participating Institutions of the SDSS-III Collaboration including the University of Arizona, the Brazilian Participation Group, Brookhaven National Laboratory, Carnegie Mellon University, University of Florida, the French Participation Group, the German Participation Group, Harvard University, the Instituto de Astrofisica de Canarias, the Michigan State/Notre Dame/JINA Participation Group, Johns Hopkins University, Lawrence Berkeley National Laboratory, Max Planck Institute for Astrophysics, Max Planck Institute for Extraterrestrial Physics, New Mexico State University, New York University, Ohio State University, Pennsylvania State University, University of Portsmouth, Princeton University, the Spanish Participation Group, University of Tokyo, University of Utah, Vanderbilt University, University of Virginia, University of Washington, and Yale University.

This research has made use of NASA's Astrophysics Data System Bibliographic Services.


\section*{Appendix A: Notes on individual supernovae}
\label{sec:appA}

Here, we describe the differences between our final SN sample, presented in Section~\ref{sec:sample}, and the GM13 sample. We also detail which SNe were first discovered in other works.

Of the SNe in our final sample, seven SNe II (44 per cent) and 69 SNe Ia (76 per cent) were previously reported in GM13. SN 0487-51869-328, reported in GM13, and SN 0487-51876-322, reported here, are the same SN II, separated by seven days. Here, we report the second detection of this SN. While \snid\ classified both spectra as belonging to a SN~II at maximum light, the SVD portion of our detection code correctly classifed SN 0487-51876-322 as belonging to an older SN than SN 0487-51869-328 (24 and 19 days after maximum light, respectively). Yet, both of these ages are consistent given our method's uncertainty at detecting the age of SNe~II of $\pm33$~d (the large uncertainty is due to the plateau phase that lasts $\sim100$ d, during which SN II spectra show little variance; see section 3.4 of GM13). The two remaining SNe~II from GM13, namely 0864-52320-082 and 1755-53386-516, appear in host galaxies that were excluded from our sample by the selection criteria in Section~\ref{sec:galaxies}. 

Five of the 90 SNe Ia from GM13 were hosted by galaxies included in our sample but were not detected in this work: SNe 0604-52079-209, 0606-52365-412, 1788-54468-126, 2499-54176-550, and 2594-54177-348. We attribute these non-detections to the different reductions applied to the original SDSS spectral data between DR7 and DR9. For example, SN 2499-54176-550 was not detected here because our detection code identified only eight spectral features in the data, while we require a minimum of ten features for detection. SN 2594-54177-348, on the other hand, was detected and classified as a SN~Ia by the SVD phase, but as a SN~IIP by \snid, leading to its rejection from the SN~Ia sample. Conversely, seven of the SNe~Ia reported here, but not in GM13, are hosted by galaxies included in the \vespa\ galaxy sample we probed in GM13 (SNe 0304-51609-436, 0646-52523-183, 0767-52252-123, 1700-53502-359, 1747-53075-177, 2118-53820-468, and 1574-53476-461). We attribute these non-detections to similar reasons. SN 0304-51609-436, for example, had an erroneous redshift of 0.71 in DR7, which was corrected to 0.13 by the DR9 reduction pipeline. 

Five of the SNe Ia detected here but not in GM13 were previously discovered in other works that searched for SNe in SDSS galaxy spectra. SNe 0305-51613-575 and 0472-51955-247 were discovered by \citet{Madgwick2003} (the latter was also reported in \citealt{Krughoff2011}), and SNe 1059-52618-553, 1266-52709-024, and 1304-52993-552 were discovered by \citet{Tu2010}. Two more SNe Ia were detected by the SDSS collaboration: SNe 1452-53112-120 and 1700-53502-359 were reported as SN2004cn and SN2005ca, respectively, by \citet{2004IAUC.8359....1C} and \citet{2005IAUC.8530....1S}. Some of the SNe II reported here have also been reported in \citet{2013ApJ...775..116R}, where they were classified by us. These include SNe 0475-51965-626, 1207-52672-512, 1459-53117-022, 1684-53239-484, and 2593-54175-334. Of these, all but the first were previously reported in GM13. 

\end{document}



\newpage
\begin{table*}
 \caption{SNe detected in SDSS DR9 galaxies.}\label{table:SNe_Ia}
 \begin{tabular}{l c c c c c c c c c c c c c}
  \hline
  \hline
  \multicolumn{1}{c}{SDSS Name} & Plate-MJD-fibre & Date & $z$ & Age$_1$ & Age$_2$ & $s$ & $\R_{\rm SN}$ & $\R_{\rm H}$ & $\chi^2_{\rm gal}$ & $\chi^2_{\rm SN}$ & Type \\
  \multicolumn{1}{c}{(1)} & (2) & (3) & (4) & (5) & (6) & (7) & (8) & (9) & (10) & (11) & (12) \\
  \hline

  J153856.21+474546.4$^a$ & 1167-52738-214 & 09/04/03 & 0.070 & -9 & -11 & 0.91 & 20.57 & 19.11 & 1.4 & 1.0 & Ia \\

  J122728.10+422028.5     & 1452-53112-120 & 17/04/04 & 0.024 & 43 & 38 & 1.00 & 20.47 & 19.54 & 1.4 & 1.1 & Ia \\

  J112148.00+125250.6$^a$ & 1605-53062-528 & 27/02/04 & 0.101 & 0 & -2 & 1.01 & 20.22 & 18.63 & 1.7 & 1.1 & Ia \\

  J095842.45+200817.2$^a$ & 2371-53762-404 & 27/01/06 & 0.039 & 123 & 123 & $\cdots$ & 20.48 & 19.66 & 1.1 & 1.0 & Ia \\

  J101800.48--000158.0     & 0271-51883-171 & 05/12/00 & 0.065 & 27 & 43 & 0.91 & 19.55 & 18.11 & 2.4 & 1.2 & Ia \\

  J124733.40+000557.1     & 0291-51928-076 & 19/01/01 & 0.086 & 28 & 24 & 1.07 & 20.04 & 17.87 & 1.7 & 1.0 & Ia \\

  J141852.38+005318.6$^a$ & 0304-51609-436 & 06/03/00 & 0.129 & -1 & -7 & 0.90 & 19.47 & 17.99 & 1.9 & 1.2 & Ia \\

  J143014.07+003035.3     & 0305-51613-575 & 10/03/00 & 0.096 & 3 & -2 & 1.10 & 19.63 & 18.34 & 2.0 & 1.3 & Ia \\

  J152734.99--000334.1     & 0313-51673-154 & 09/05/00 & 0.044 & 33 & 38 & 1.11 & 19.93 & 17.42 & 1.6 & 1.3 & Ia \\

  J114341.24--012837.0     & 0328-52282-570 & 08/01/02 & 0.125 & 27 & 31 & 0.85 & 19.91 & 17.99 & 1.5 & 1.1 & Ia \\

  J173228.53+560425.4     & 0358-51818-181 & 01/10/00 & 0.122 & 21 & 19 & 1.05 & 20.17 & 18.22 & 1.6 & 1.1 & Ia \\

  J005505.63+001104.5     & 0394-51812-554 & 25/09/00 & 0.146 & -5 & -6 & 1.14 & 20.07 & 18.08 & 1.3 & 1.1 & Ia \\

  J080421.29+464713.2     & 0438-51884-166 & 06/12/00 & 0.187 & 16 & 12 & 0.88 & 21.23 & 18.17 & 1.1 & 1.0 & Ia \\

  J080312.61+473649.6     & 0438-51884-462 & 06/12/00 & 0.117 & 2 & -5 & 0.91 & 19.51 & 18.07 & 2.1 & 1.1 & Ia \\

  J092229.14+575429.2     & 0452-51911-319 & 02/01/01 & 0.063 & 9 & 11 & 0.84 & 19.43 & 17.94 & 1.8 & 1.1 & Ia \\

  J091138.38--004253.9    & 0472-51955-247 & 15/02/01 & 0.070 & 6 & -2 & 1.03 & 18.71 & 18.31 & 4.7 & 1.3 & Ia \\

  J095153.06+010605.8     & 0480-51989-024 & 21/03/01 & 0.063 & 2 & -4 & 1.01 & 18.77 & 17.77 & 4.5 & 1.6 & Ia \\

  J103849.48+040056.2     & 0578-52339-314 & 06/03/02 & 0.129 & 6 & 3 & 1.05 & 19.58 & 17.87 & 1.8 & 1.3 & Ia \\

  J161713.38+482827.8     & 0622-52054-011 & 25/05/01 & 0.104 & 16 & 88 & 1.05 & 19.89 & 18.21 & 1.7 & 1.3 & Ia \\

  J232650.82--095632.8$^a$ & 0646-52523-183 & 06/09/02 & 0.052 & -5 & -5 & 1.28 & 19.99 & 17.28 & 1.4 & 1.3 & Ia \\

  J010939.83+000346.9$^a$ & 0694-52209-152 & 27/10/01 & 0.078 & 21 & 59 & 0.89 & 20.59 & 17.42 & 1.5 & 1.3 & Ia \\

  J222710.20+133958.0     & 0738-52521-360 & 04/09/02 & 0.150 & 27 & 22 & 1.05 & 20.20 & 18.14 & 1.1 & 1.0 & Ia \\

  J232306.16+134000.6     & 0745-52258-092 & 15/12/01 & 0.041 & 54 & 45 & 0.85 & 20.18 & 17.39 & 1.1 & 0.9 & Ia \\

  J083906.35+434244.2     & 0762-52232-067 & 19/11/01 & 0.125 & 31 & 32 & 1.06 & 20.39 & 18.04 & 1.0 & 0.9 & Ia \\

  J092620.08+502157.2$^a$ & 0767-52252-123 & 09/12/01 & 0.060 & 3 & -1 & $\cdots$ & 20.38 & 17.83 & 1.2 & 1.2 & Ia \\

  J114447.93+041652.3$^a$ & 0838-52378-021 & 14/04/02 & 0.104 & 2 & -1 & 0.98 & 19.63 & 18.30 & 1.5 & 1.1 & Ia \\

  J121739.58+051924.3     & 0844-52378-462 & 14/04/02 & 0.103 & 31 & 30 & 1.06 & 20.28 & 17.85 & 1.1 & 1.0 & Ia \\

  J102958.55+533006.5     & 0905-52643-213 & 04/01/03 & 0.137 & 42 & 38 & 1.04 & 20.62 & 18.25 & 1.0 & 0.9 & Ia \\

  J140516.18--014240.7     & 0915-52443-543 & 18/06/02 & 0.054 & 27 & 32 & 1.05 & 19.37 & 17.44 & 1.5 & 1.0 & Ia \\

  J112900.54+484359.3     & 0966-52642-221 & 03/01/03 & 0.074 & 3 & -0 & 0.91 & 19.45 & 18.27 & 1.8 & 1.1 & Ia \\

  J124724.64+534350.9     & 1038-52673-135 & 03/02/03 & 0.153 & 12 & 11 & 0.87 & 20.59 & 19.09 & 1.2 & 1.0 & Ia \\

  J074734.48+272647.3     & 1059-52618-144 & 10/12/02 & 0.063 & 33 & 32 & 1.04 & 19.44 & 17.51 & 1.8 & 0.9 & Ia \\

  J161921.65+410523.6     & 1171-52753-185 & 24/04/03 & 0.038 & 10 & 9 & 0.89 & 18.04 & 16.15 & 3.8 & 1.1 & Ia \\

  J084903.48+055015.8     & 1189-52668-239 & 29/01/03 & 0.126 & 3 & 4 & 0.76 & 20.01 & 18.07 & 1.3 & 1.0 & Ia \\

  J081118.42+260958.0     & 1205-52670-632 & 31/01/03 & 0.144 & 12 & 12 & $\cdots$ & 20.23 & 17.58 & 1.3 & 1.1 & Ia \\

  J081647.02+251731.6     & 1266-52709-024 & 11/03/03 & 0.140 & -1 & -5 & 0.91 & 19.92 & 18.20 & 1.5 & 1.2 & Ia \\

  J123625.46+503641.7     & 1278-52735-425 & 06/04/03 & 0.106 & 6 & 1 & 0.96 & 19.53 & 17.64 & 1.5 & 1.0 & Ia \\

  J144058.73+450750.8     & 1289-52734-413 & 05/04/03 & 0.074 & 8 & 12 & 0.94 & 19.53 & 16.57 & 1.0 & 0.8 & Ia \\

  J083909.65+072431.5     & 1298-52964-304 & 21/11/03 & 0.047 & 32 & 43 & 0.96 & 19.08 & 17.53 & 1.8 & 1.1 & Ia \\

  J093749.93+101138.2     & 1304-52993-552 & 20/12/03 & 0.094 & 16 & 17 & 1.04 & 19.59 & 18.80 & 3.3 & 1.2 & Ia \\

  J113412.69+581543.6     & 1310-53033-459 & 29/01/04 & 0.122 & 16 & 15 & 1.05 & 19.55 & 17.66 & 1.8 & 1.0 & Ia \\

  J140758.92+542147.1     & 1324-53088-169 & 24/03/04 & 0.067 & 52 & 44 & 1.07 & 19.88 & 17.60 & 1.4 & 1.0 & Ia \\

  J162011.08+380641.2     & 1337-52767-480 & 08/05/03 & 0.130 & 2 & 4 & 0.84 & 19.81 & 18.12 & 1.5 & 1.0 & Ia \\

  J152045.08+364842.3     & 1400-53470-234 & 10/04/05 & 0.103 & 6 & -4 & 1.02 & 18.90 & 16.58 & 1.8 & 0.9 & Ia \\

  J151654.43+370726.4     & 1400-53470-351 & 10/04/05 & 0.116 & -6 & -2 & 1.14 & 20.14 & 17.17 & 1.1 & 1.0 & Ia \\

  J154857.37+335725.3     & 1403-53227-456 & 10/08/04 & 0.128 & 31 & 33 & 0.84 & 20.49 & 18.36 & 1.2 & 1.0 & Ia \\

  J114827.45+420755.3     & 1445-53062-067 & 27/02/04 & 0.086 & -1 & 4 & 0.90 & 19.79 & 17.39 & 1.3 & 0.9 & Ia \\

  J154024.75+325157.2     & 1581-53149-470 & 24/05/04 & 0.054 & 30 & 40 & 0.91 & 18.98 & 17.69 & 3.3 & 1.0 & Ia \\

  J102250.11+114210.7     & 1598-53033-380 & 29/01/04 & 0.102 & 2 & -1 & 0.93 & 19.66 & 18.00 & 1.6 & 1.0 & Ia \\

  J032108.86+411510.9$^a$ & 1665-52976-155 & 03/12/03 & 0.016 & 21 & 15 & 1.05 & 20.54 & 16.70 & 1.7 & 1.5 & Ia \\

  J131630.13+124037.0     & 1697-53142-506 & 17/05/04 & 0.151 & 3 & 4 & 0.99 & 19.92 & 17.66 & 1.3 & 1.0 & Ia \\

  J133432.73+110756.7     & 1700-53502-302 & 12/05/05 & 0.095 & 3 & -5 & 1.19 & 19.57 & 16.97 & 1.3 & 1.1 & Ia \\

  J133238.59+114833.2     & 1700-53502-359 & 12/05/05 & 0.150 & 10 & 8 & 0.98 & 20.49 & 18.91 & 1.3 & 1.1 & Ia \\

  J143610.56+122642.3     & 1710-53504-488 & 14/05/05 & 0.085 & 30 & 24 & 0.92 & 20.35 & 17.93 & 1.2 & 1.0 & Ia \\

  J102934.76+131637.3$^a$ & 1747-53075-177 & 11/03/04 & 0.090 & 41 & 33 & 1.02 & 20.53 & 17.79 & 1.2 & 1.0 & Ia \\

  J113225.25+143712.7     & 1755-53386-309 & 16/01/05 & 0.082 & 21 & 21 & 1.09 & 19.31 & 17.13 & 1.7 & 1.0 & Ia \\

  J082933.46+085205.3     & 1758-53084-523 & 20/03/04 & 0.112 & 2 & 3 & 0.85 & 19.63 & 17.75 & 1.6 & 1.0 & Ia \\

  J125626.85+101339.8     & 1791-54266-583 & 15/06/07 & 0.107 & 12 & 10 & 1.02 & 19.30 & 17.66 & 2.0 & 1.2 & Ia \\

  J130543.99+093446.7     & 1793-53883-040 & 28/05/06 & 0.055 & 52 & 46 & 1.01 & 19.68 & 17.03 & 1.4 & 1.1 & Ia \\

  J133112.78+075726.4     & 1801-54156-371 & 25/02/07 & 0.123 & 12 & 13 & 1.00 & 19.81 & 17.79 & 1.4 & 1.1 & Ia \\

  J134137.62+055215.7     & 1803-54152-260 & 21/02/07 & 0.059 & 30 & 37 & 0.84 & 20.03 & 17.10 & 1.3 & 1.0 & Ia \\

  J140309.73+060754.3$^a$ & 1808-54176-107 & 17/03/07 & 0.117 & 3 & -4 & 1.14 & 19.26 & 18.01 & 2.2 & 1.2 & Ia \\

  \end{tabular}
\end{table*}
  
\begin{table*}
 \begin{tabular}{l c c c c c c c c c c c c c}
  \multicolumn{11}{l}{{\bf Table 1} -- {\it continued}} \\
  \hline
  \hline
  \multicolumn{1}{c}{SDSS Name} & Plate-MJD-fibre & Date & $z$ & Age$_1$ & Age$_2$ & $s$ & $\R_{\rm SN}$ & $\R_{\rm H}$ & $\chi^2_{\rm gal}$ & $\chi^2_{\rm SN}$ & Type \\
  \multicolumn{1}{c}{(1)} & (2) & (3) & (4) & (5) & (6) & (7) & (8) & (9) & (10) & (11) & (12) \\
  \hline
  
  J145257.20+310427.0     & 1843-53816-491 & 22/03/06 & 0.094 & 52 & 54 & 1.09 & 20.20 & 17.71 & 1.5 & 1.1 & Ia \\

  J093747.44+281715.3     & 1944-53385-434 & 15/01/05 & 0.153 & 2 & -2 & 0.86 & 20.02 & 18.51 & 1.6 & 1.0 & Ia \\

  J100326.49+320758.7     & 1949-53433-080 & 04/03/05 & 0.166 & 3 & -2 & 1.06 & 20.26 & 18.08 & 1.1 & 0.9 & Ia \\

  J102230.50+354034.9     & 1957-53415-232 & 14/02/05 & 0.128 & 21 & 17 & 0.95 & 20.24 & 18.00 & 1.2 & 0.9 & Ia \\

  J135439.29+280952.9$^a$ & 2118-53820-468 & 26/03/06 & 0.073 & 12 & 9 & 0.85 & 20.02 & 17.45 & 1.1 & 1.0 & Ia \\

  J152245.61+194220.3     & 2159-54328-161 & 16/08/07 & 0.109 & 12 & 12 & 0.87 & 19.71 & 17.65 & 1.2 & 0.9 & Ia \\

  J153812.10+250244.2     & 2165-53917-406 & 01/07/06 & 0.067 & 27 & 43 & 1.07 & 19.25 & 17.69 & 1.3 & 0.9 & Ia \\

  J160142.54+203436.8     & 2173-53874-154 & 19/05/06 & 0.123 & 2 & -1 & 0.87 & 20.31 & 17.80 & 1.4 & 1.2 & Ia \\

  J112600.18+260313.0     & 2218-53816-295 & 22/03/06 & 0.158 & 6 & -5 & 1.02 & 20.21 & 17.96 & 1.5 & 1.2 & Ia \\

  J114438.44+295323.6     & 2222-53799-480 & 05/03/06 & 0.076 & 12 & 15 & 0.93 & 19.31 & 18.22 & 3.3 & 1.1 & Ia \\

  J103357.18+202025.6     & 2376-53770-183 & 04/02/06 & 0.087 & 2 & -5 & 1.00 & 18.89 & 16.90 & 2.0 & 1.0 & Ia \\

  J084828.08+142523.5     & 2429-53799-033 & 05/03/06 & 0.069 & 38 & 36 & 1.03 & 19.75 & 17.71 & 1.5 & 1.1 & Ia \\

  J084943.93+121755.3     & 2430-53815-267 & 21/03/06 & 0.051 & 33 & 38 & 1.03 & 19.46 & 18.78 & 2.2 & 1.1 & Ia \\

  J130554.98+174937.8     & 2603-54479-486 & 14/01/08 & 0.078 & 28 & 26 & 1.06 & 20.90 & 18.56 & 1.2 & 1.0 & Ia \\

  J120549.05+185858.5     & 2609-54476-295 & 11/01/08 & 0.168 & 3 & -4 & 1.18 & 20.00 & 18.07 & 1.5 & 1.1 & Ia \\

  J124332.24+185745.3     & 2614-54481-257 & 16/01/08 & 0.168 & 9 & 5 & 1.00 & 20.30 & 18.39 & 1.2 & 1.0 & Ia \\

  J132301.39+243023.6     & 2664-54524-468 & 28/02/08 & 0.073 & 6 & 8 & 0.94 & 19.15 & 17.63 & 2.4 & 1.3 & Ia \\

  J140622.31+162900.4     & 2744-54272-561 & 21/06/07 & 0.014 & 40 & 36 & 0.99 & 17.38 & 15.24 & 4.6 & 0.9 & Ia \\

  J142608.24+152501.9$^a$ & 2746-54232-635 & 12/05/07 & 0.053 & 2 & -5 & 0.78 & 18.58 & 17.11 & 3.7 & 1.5 & Ia \\

  J143518.10+153206.4     & 2747-54233-613 & 13/05/07 & 0.107 & 12 & 8 & 0.88 & 19.90 & 17.89 & 1.6 & 1.1 & Ia \\

  J153651.78+120806.4     & 2754-54240-593 & 20/05/07 & 0.094 & 27 & 26 & 0.90 & 19.92 & 18.09 & 1.9 & 1.1 & Ia \\

  J141224.51+170023.4     & 2758-54523-082 & 27/02/08 & 0.174 & 6 & 1 & 0.93 & 20.26 & 18.28 & 1.4 & 1.1 & Ia \\

  J153253.97+130703.5     & 2768-54265-233 & 14/06/07 & 0.073 & 27 & 24 & $\cdots$ & 19.65 & 17.28 & 1.3 & 1.1 & Ia \\

  J140954.70+193941.5     & 2771-54527-005 & 02/03/08 & 0.077 & 18 & 18 & 1.10 & 20.04 & 16.92 & 1.2 & 1.0 & Ia \\

  J141058.32+645050.8     & 0498-51984-102 & 16/03/01 & 0.140 & 21 & 0 & 1.01 & 20.48 & 18.09 & 1.2 & 1.0 & Ia/Ic \\

  J074933.15+275729.2     & 1059-52618-553 & 10/12/02 & 0.122 & 2 & 5 & $\cdots$ & 19.49 & 18.77 & 3.1 & 2.2 & Ia/Ic \\

  J162333.74+252420.7$^a$ & 1574-53476-461 & 16/04/05 & 0.190 & -0 & 8 & $\cdots$ & 19.86 & 18.53 & 1.2 & 1.1 & Ia/Ic \\

  J142855.54+350456.4     & 1645-53172-349 & 16/06/04 & 0.121 & 3 & 38 & $\cdots$ & 19.71 & 16.98 & 1.3 & 1.1 & Ia/Ic \\

  J095313.01+305122.4$^a$ & 1946-53432-030 & 03/03/05 & 0.045 & 3 & 22 & $\cdots$ & 19.24 & 17.91 & 2.1 & 1.2 & Ia/Ic \\
  
  J075813.33+440108.1$^a$ & 0437-51876-322 & 28/11/00 & 0.047 & 24 & 0 & $\cdots$ & 19.98 & 17.46 & 1.6 & 1.3 & II \\

  J093313.94+015858.7     & 0475-51965-626 & 25/02/01 & 0.031 & 64 & 20 & $\cdots$ & 19.30 & 18.39 & 2.9 & 1.6 & II \\

  J073753.33+315331.0$^a$ & 0541-51959-057 & 19/02/01 & 0.098 & 4 & -3 & $\cdots$ & 20.65 & 18.44 & 1.6 & 1.4 & II \\

  J114447.10+535501.4$^a$ & 1015-52734-019 & 05/04/03 & 0.062 & 19 & 44 & $\cdots$ & 20.02 & 18.94 & 1.9 & 1.4 & II \\

  J082449.94+293644.1     & 1207-52672-512 & 02/02/03 & 0.040 & 87 & 36 & $\cdots$ & 20.23 & 18.06 & 1.4 & 1.0 & II \\

  J122742.37+095728.5     & 1231-52725-553 & 27/03/03 & 0.070 & 34 & 49 & $\cdots$ & 20.15 & 18.41 & 1.2 & 1.0 & II \\

  J163305.64+350600.9$^a$ & 1339-52767-327 & 08/05/03 & 0.035 & -4 & -4 & $\cdots$ & 19.40 & 17.57 & 1.6 & 1.2 & II \\

  J161252.72+305058.5     & 1406-52876-528 & 25/08/03 & 0.048 & 27 & 20 & $\cdots$ & 20.15 & 18.19 & 1.3 & 1.0 & II \\

  J131307.11+460554.3     & 1459-53117-022 & 22/04/04 & 0.030 & 35 & 50 & $\cdots$ & 19.89 & 19.33 & 2.3 & 1.6 & II \\

  J162244.78+323933.0     & 1684-53239-484 & 22/08/04 & 0.041 & 62 & 36 & $\cdots$ & 19.84 & 17.01 & 1.9 & 1.3 & II \\

  J170626.69+232409.9$^a$ & 1689-53177-325 & 21/06/04 & 0.063 & 250 & 310 & $\cdots$ & 21.95 & 17.37 & 1.5 & 1.2 & II \\

  J074820.66+471214.2$^a$ & 1737-53055-369 & 20/02/04 & 0.062 & -7 & -6 & $\cdots$ & 19.05 & 18.05 & 1.7 & 1.4 & II \\

  J133057.65+365921.2$^a$ & 2102-54115-072 & 15/01/07 & 0.058 & 64 & 96 & $\cdots$ & 20.61 & 19.41 & 1.7 & 1.0 & II \\

  J120323.91+351932.9     & 2103-53467-081 & 07/04/05 & 0.028 & 19 & 34 & $\cdots$ & 19.22 & 17.48 & 1.7 & 0.8 & II \\

  J103134.64+190407.1     & 2593-54175-334 & 16/03/07 & 0.034 & 35 & 113 & $\cdots$ & 19.97 & 19.68 & 1.8 & 1.2 & II \\

  J131503.77+223522.7$^a$ & 2651-54507-488 & 11/02/08 & 0.023 & 250 & 51 & $\cdots$ & 21.20 & 17.36 & 1.6 & 1.2 & II \\
  
  \hline
  \multicolumn{12}{l}{$^a$SNe first discovered in this work.} \\
  \multicolumn{12}{l}{(1) -- SDSS name, composed of right ascension and declination (J2000).} \\
  \multicolumn{12}{l}{(2) -- SDSS DR9 Plate, MJD, and fibre in which the SN was discovered.} \\
  \multicolumn{12}{l}{(3) -- Date on which the SN was captured, in dd/mm/yy.} \\
  \multicolumn{12}{l}{(4) -- SN host-galaxy redshift.} \\
  \multicolumn{12}{l}{(5) and (6) -- SN age, in days, as measured by SVD and \snid, respectively [with uncertainties of $\pm6$ d ($\pm33$ d) for SNe Ia (II)].} \\ 
  \multicolumn{12}{l}{(7) -- SN stretch, as measured with the \salt\ templates. All stretches have an uncertainty of $^{+0.10}_{-0.14}$.} \\
  \multicolumn{12}{l}{(8) and (9) -- SN and host-galaxy \R-band magnitudes.} \\
  \multicolumn{12}{l}{(10) and (11) -- Reduced $\chi^2$ value of galaxy and galaxy+transient fits.} \\
  \multicolumn{12}{l}{(12) -- SN type.}
 \end{tabular}
\end{table*}

\newpage
\begin{figure*}
 \begin{minipage}{\textwidth}
  \vspace{0.2cm}
  \begin{tabular}{cc} 
   \includegraphics[width=0.475\textwidth]{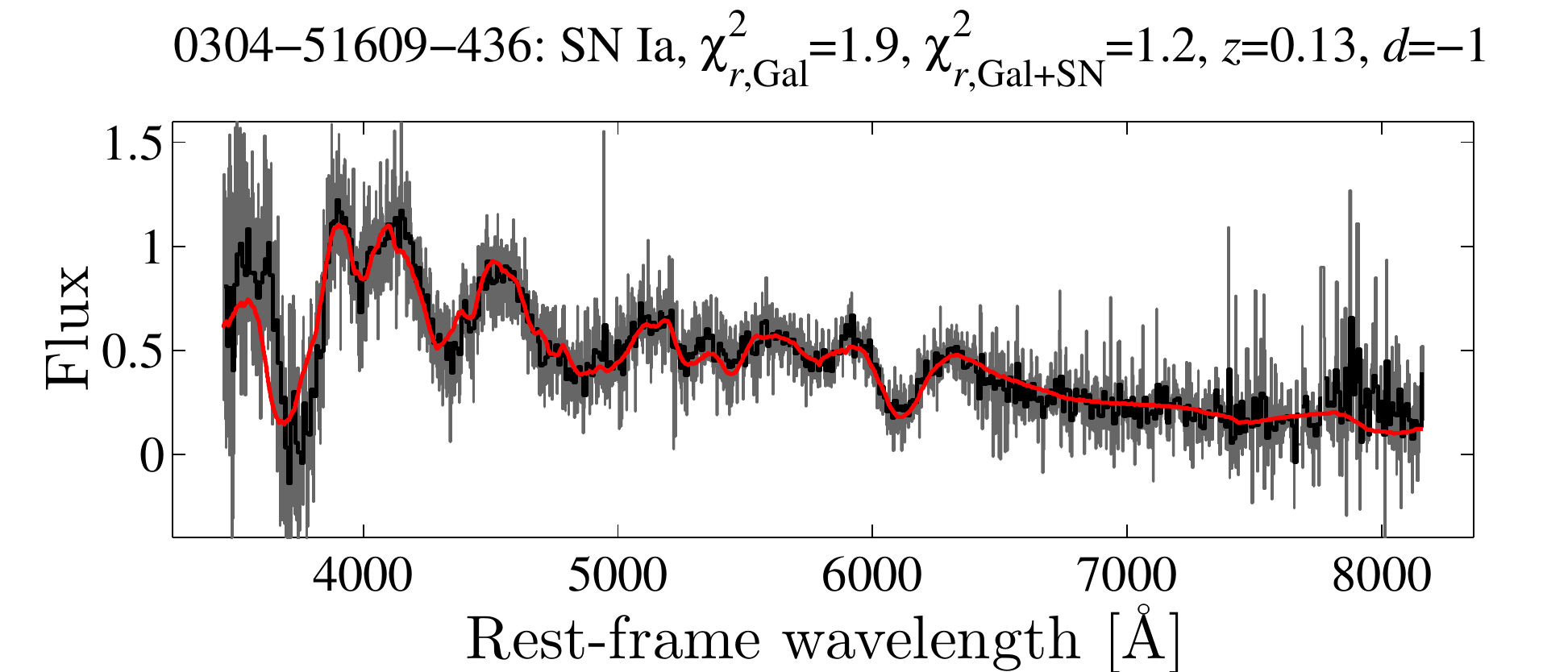} &
   \includegraphics[width=0.475\textwidth]{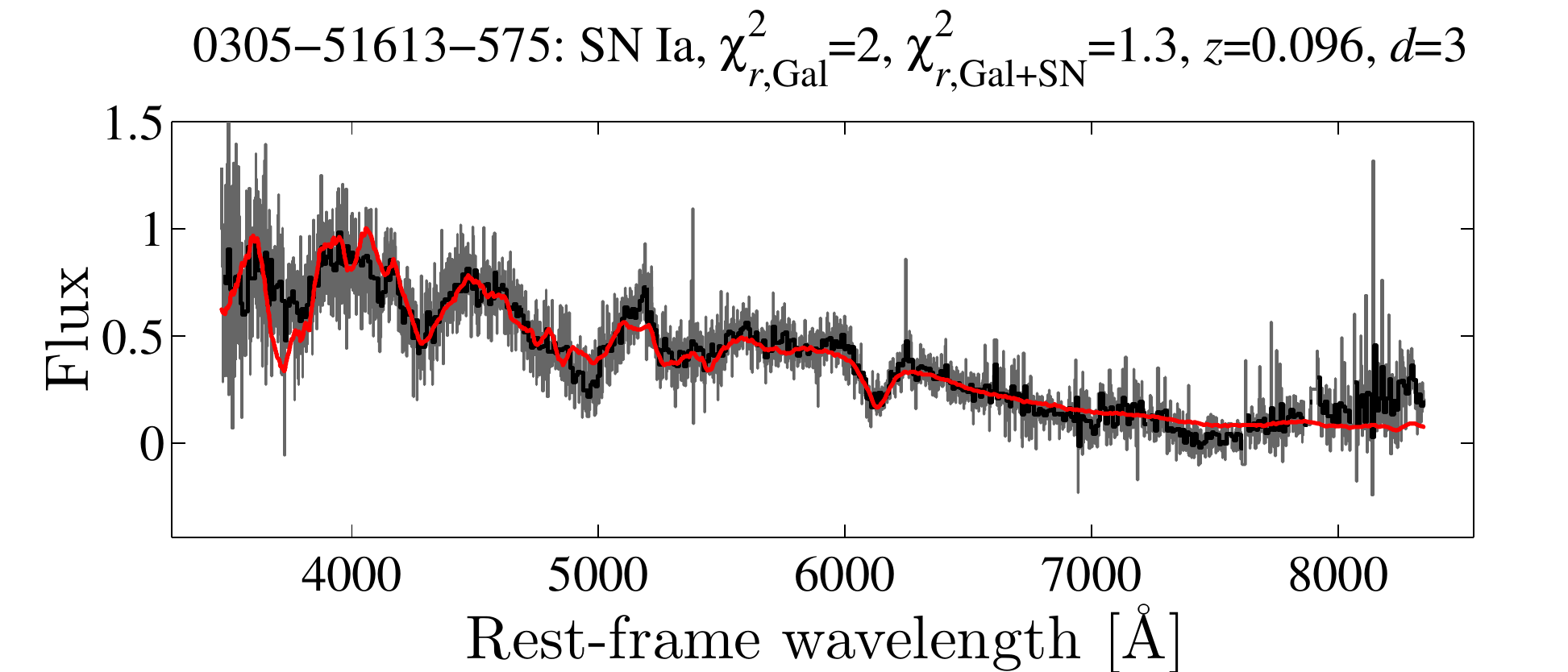} \\ 
   \includegraphics[width=0.475\textwidth]{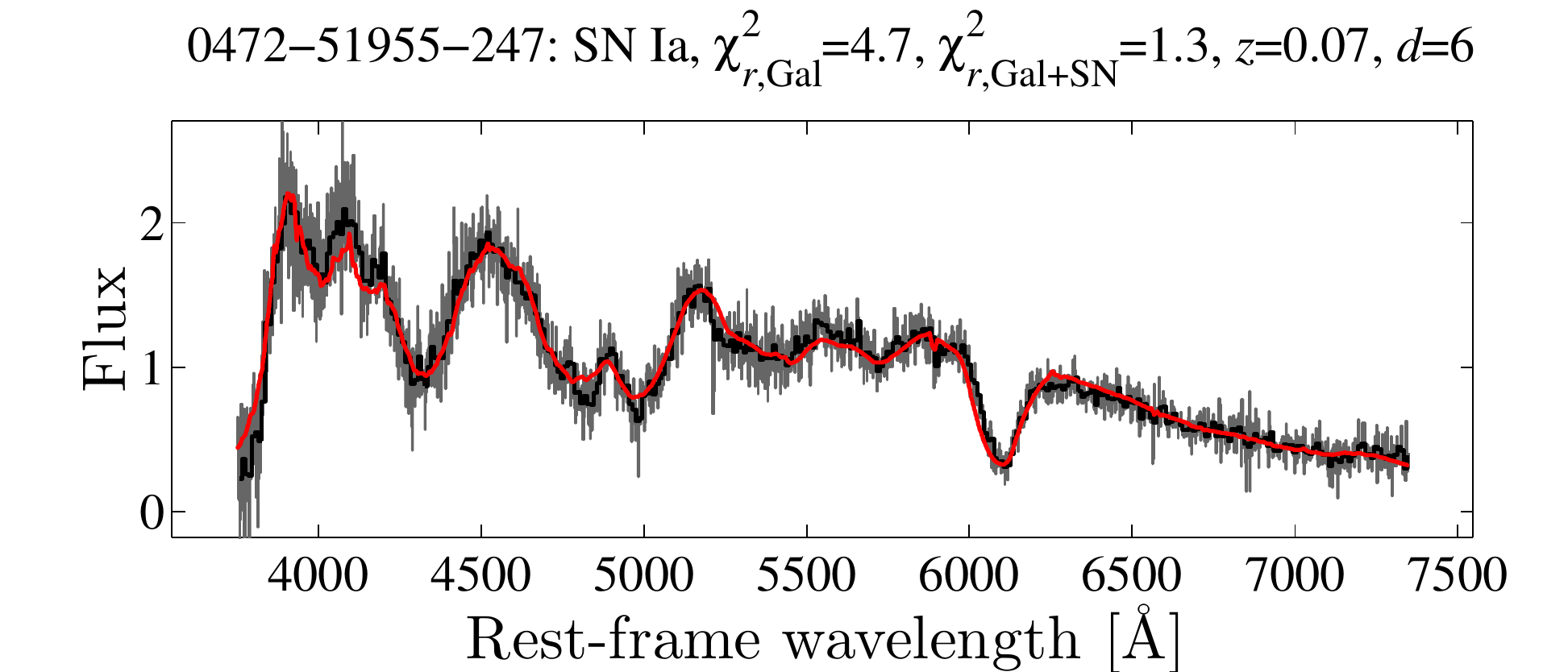} &
   \includegraphics[width=0.475\textwidth]{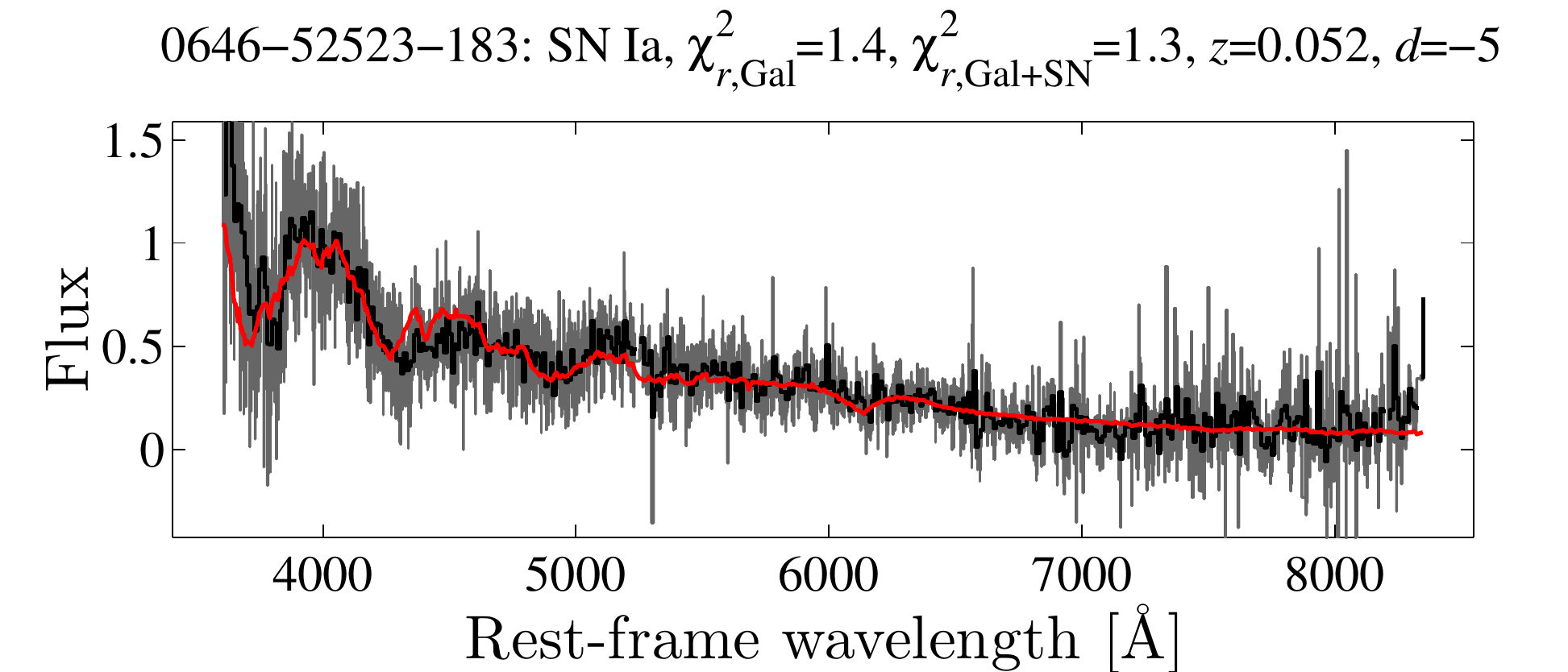} \\ 
   \includegraphics[width=0.475\textwidth]{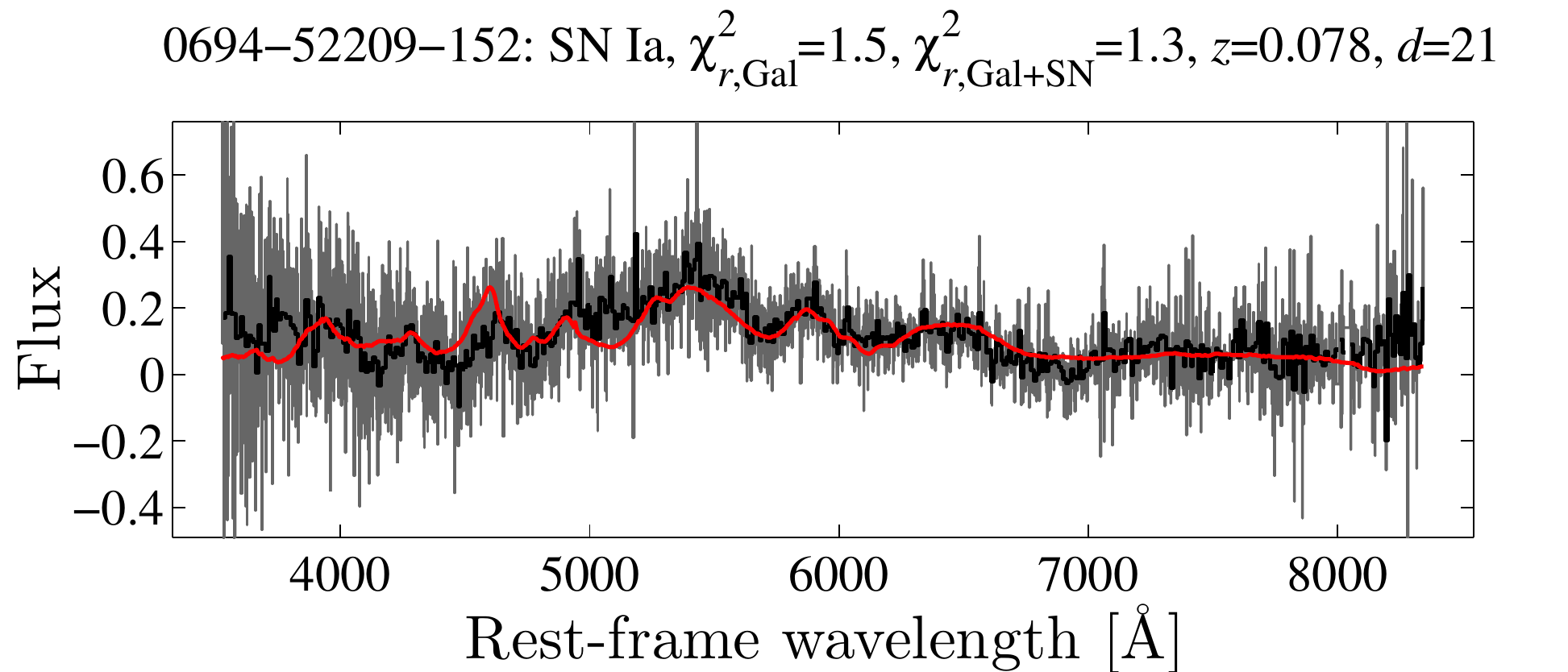} &
   \includegraphics[width=0.475\textwidth]{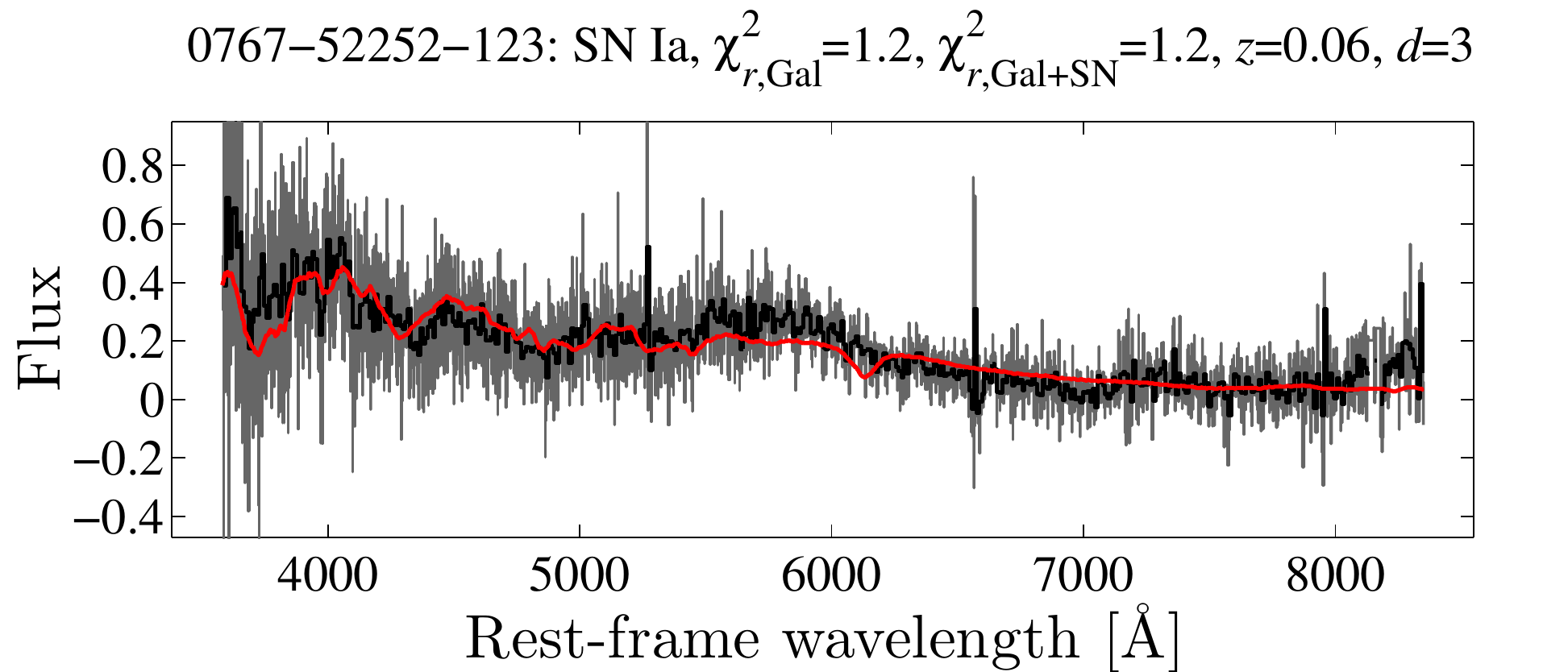} \\ 
   \includegraphics[width=0.475\textwidth]{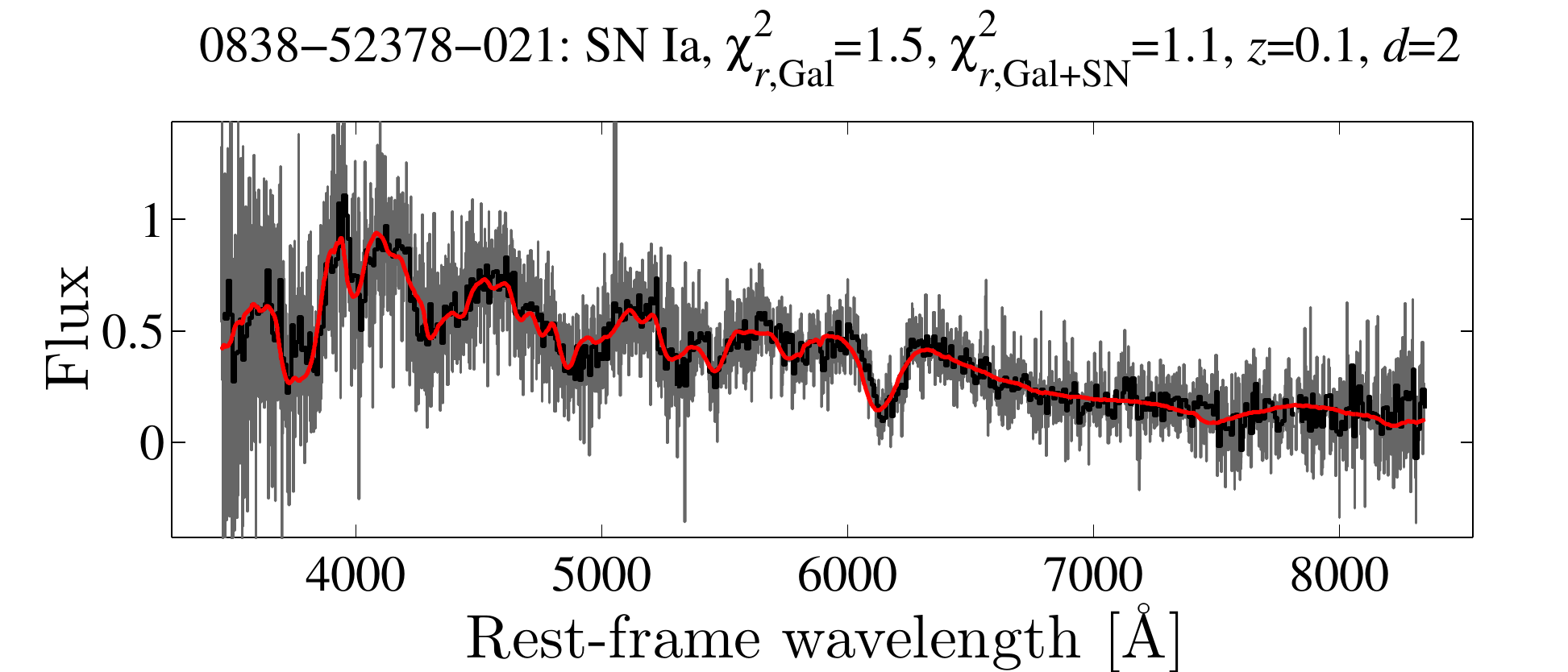} & 
   \includegraphics[width=0.475\textwidth]{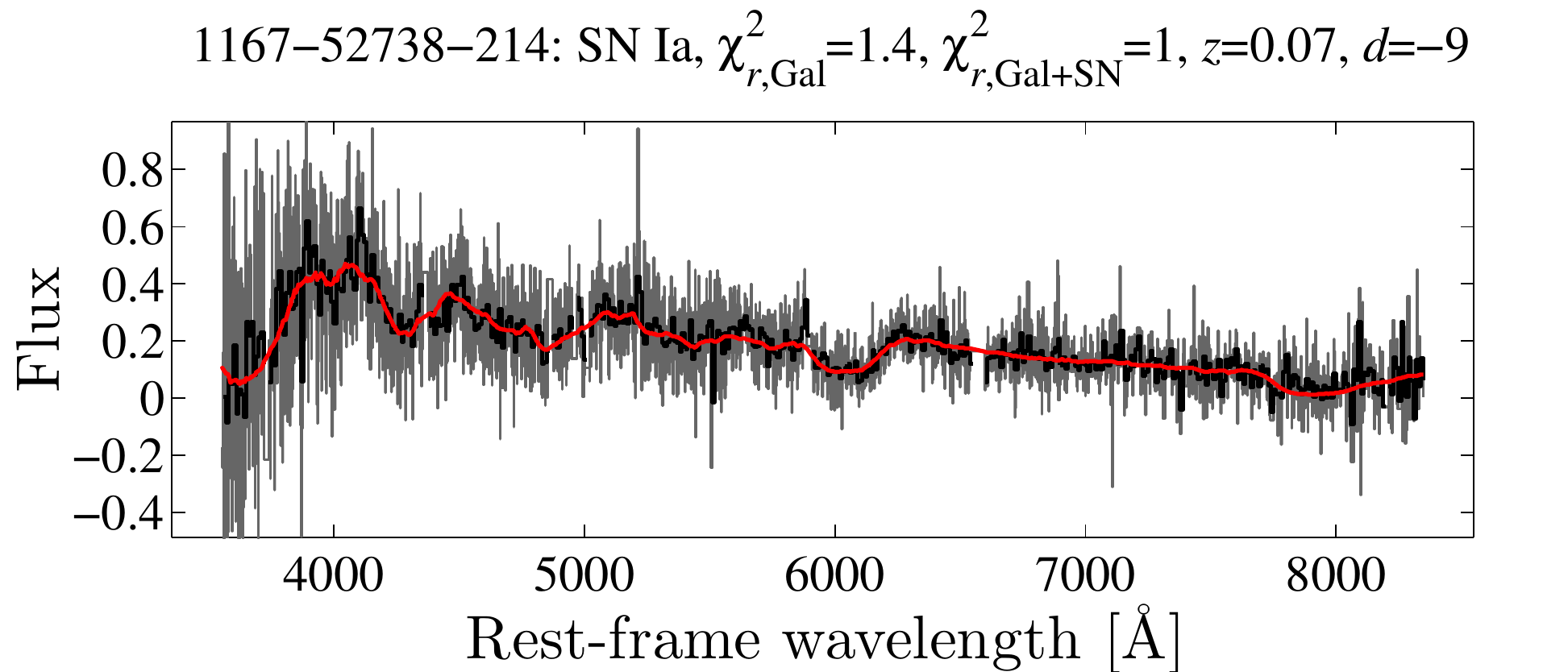} \\
   \includegraphics[width=0.475\textwidth]{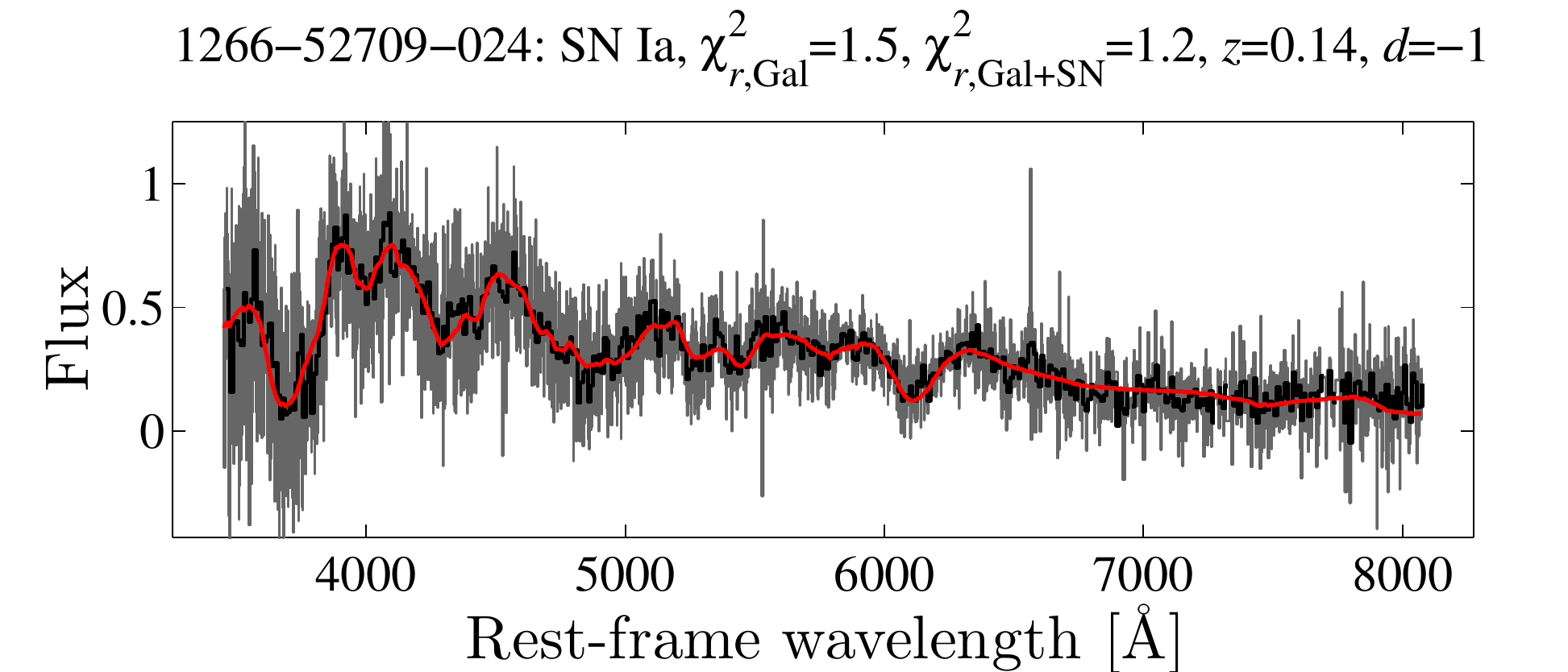} & 
   \includegraphics[width=0.475\textwidth]{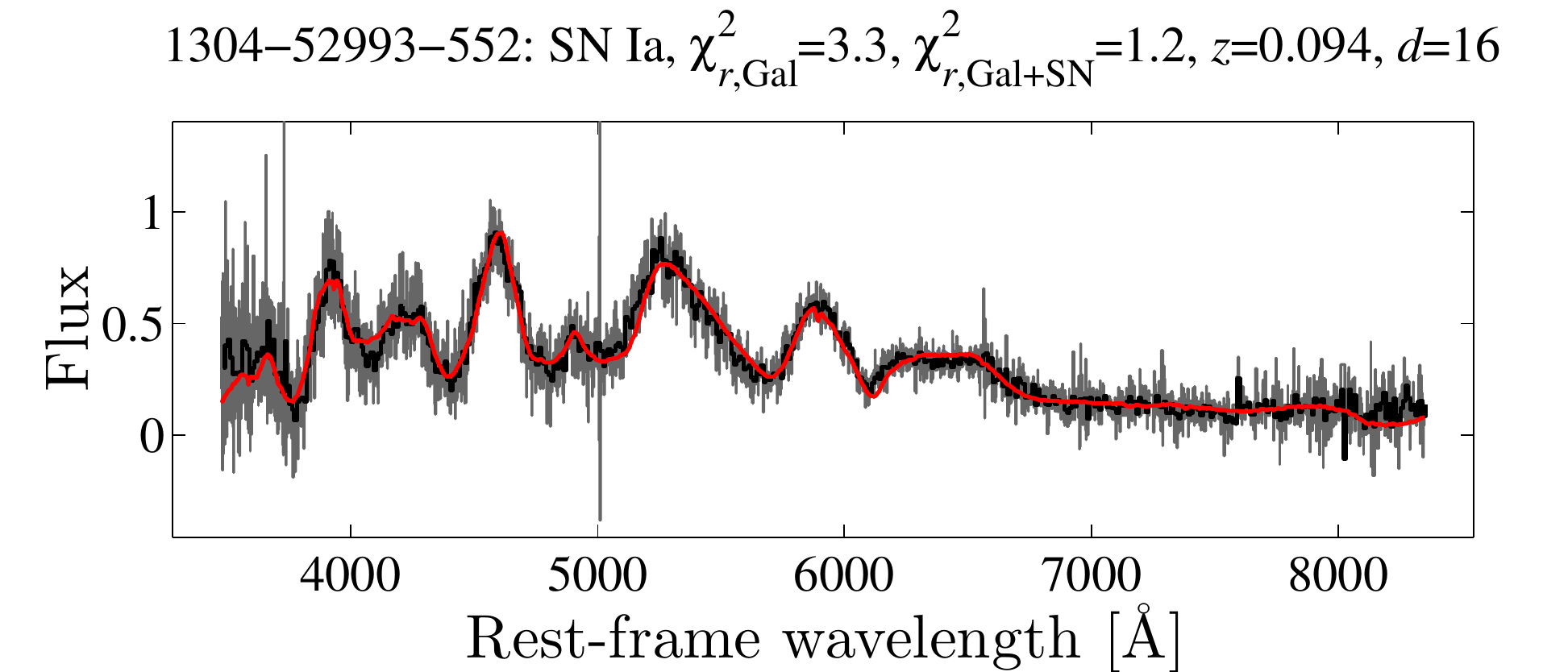} \\ 
   \end{tabular}
  \caption{SNe~Ia and SNe~II detected in this work and not previously reported by GM13. The residual spectrum, obtained by first fitting galaxy eigenspectra and transient templates to the original spectrum, and then subtracting the resulting galaxy model, is shown in grey. In black is the same residual, binned into 10~\AA{} bins. The best-fitting SN~Ia or SN II template is overlaid in red or blue, respectively. The flux is in units of $10^{-16}$~erg cm$^{-2}$~s$^{-1}$~\AA{}$^{-1}$. The title of each panel details the plate, MJD, and fibre in which it was discovered; the SN classification; the reduced $\chi^2$ value obtained when fitting only galaxy eigenspectra to the spectrum, $\chi^2_r$(Gal); the reduced $\chi^2$ obtained from the best-fitting combination of galaxy eigenspectra and transient templates, $\chi^2_r$(Gal+SN); the redshift of the SN-host galaxy, $z$; and the SVD-derived age, $d$. The spectra of the SNe shown here, as well as those from GM13, can be found on WISeREP.
  SNe 0305-51613-575 and 0472-51955-247 were previously reported by Madgwick et al. (2003). SNe 1266-52709-024 and 1304-52993-552 were reported by Tu et al. (2010).}
  \label{fig:sn_all_1} 
 \end{minipage}
\end{figure*}

\begin{figure*}
 \begin{minipage}{\textwidth}
  \vspace{0.2cm}
  \begin{tabular}{cc}
   \includegraphics[width=0.475\textwidth]{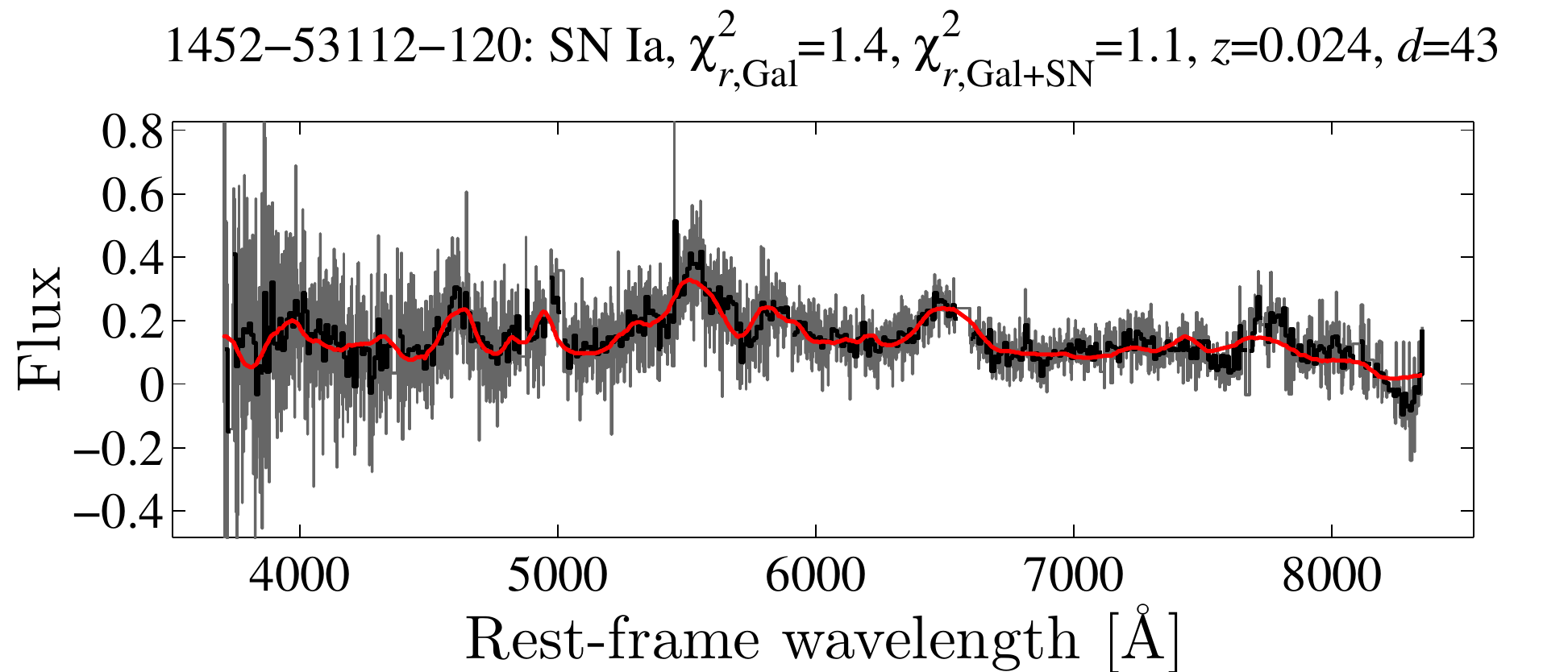} & 
   \includegraphics[width=0.475\textwidth]{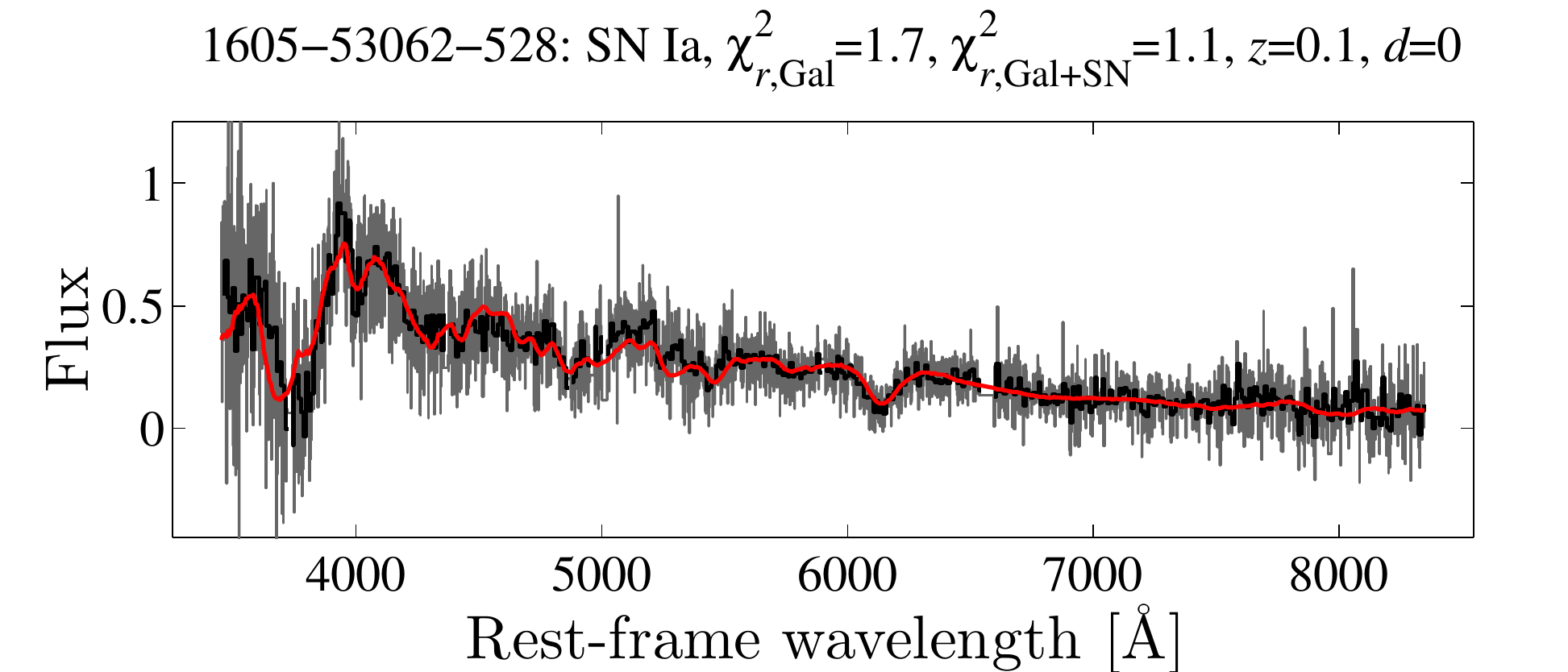} \\ 
   \includegraphics[width=0.475\textwidth]{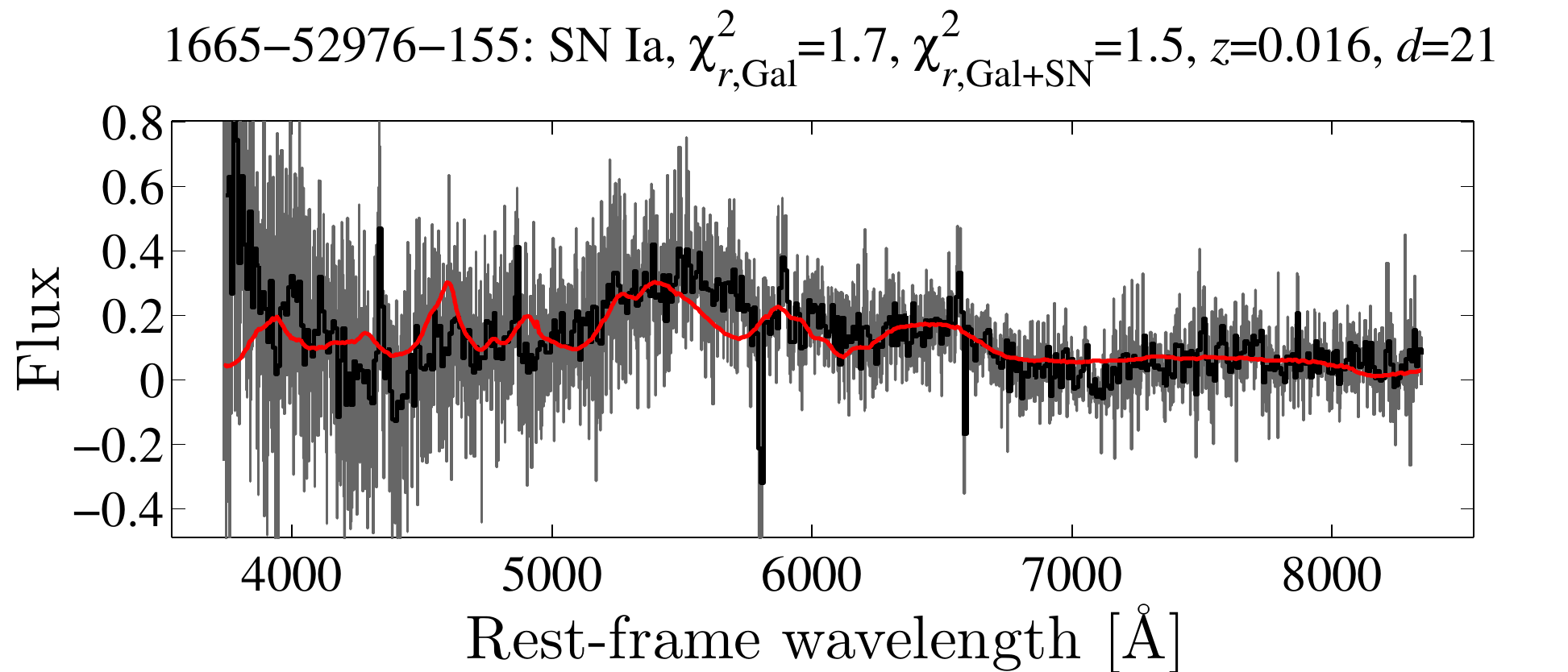} &
   \includegraphics[width=0.475\textwidth]{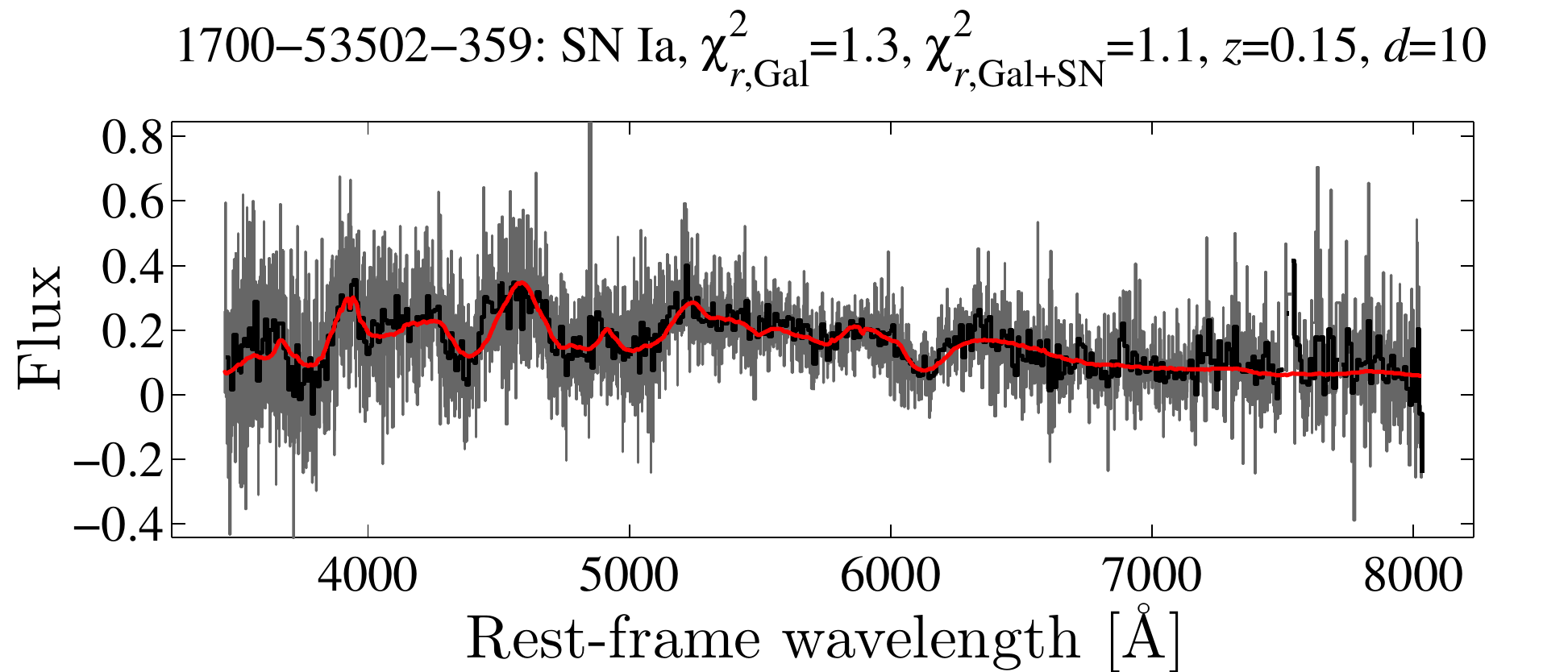} \\ 
   \includegraphics[width=0.475\textwidth]{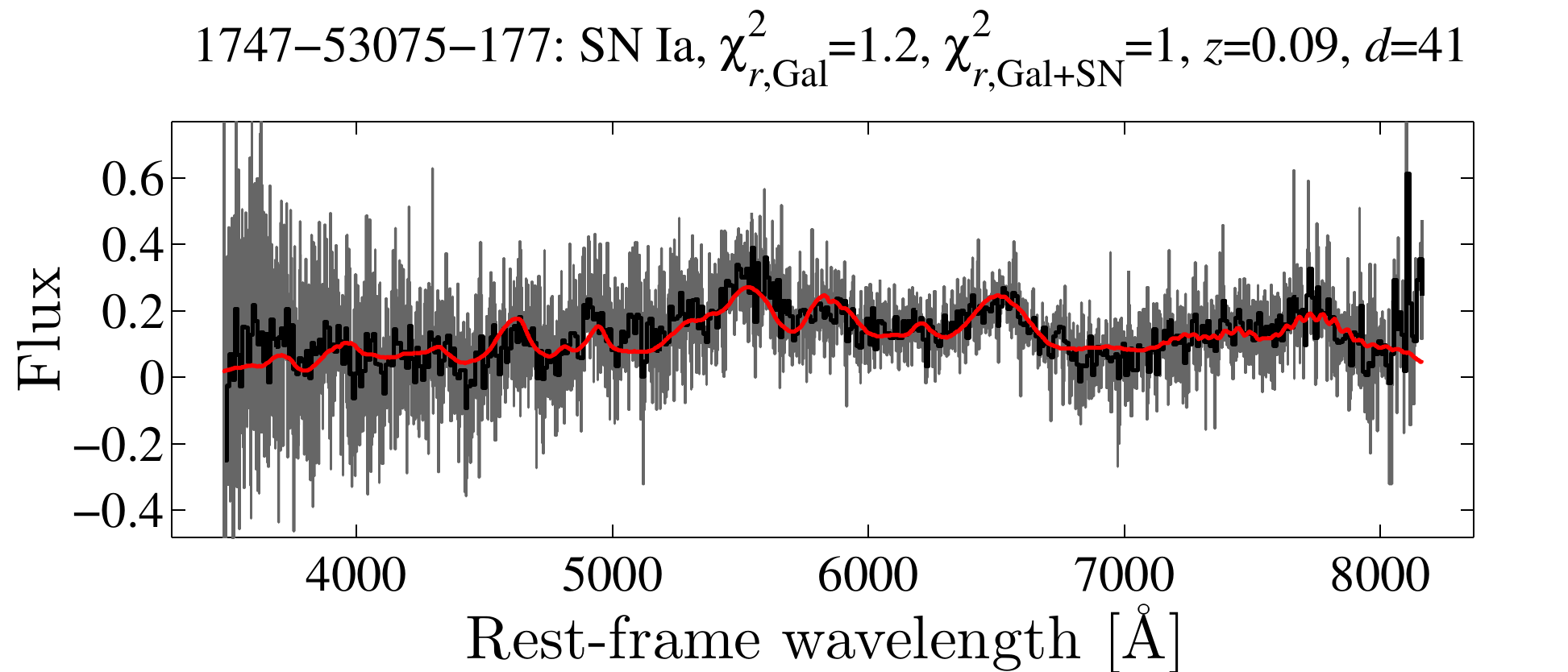} & 
   \includegraphics[width=0.475\textwidth]{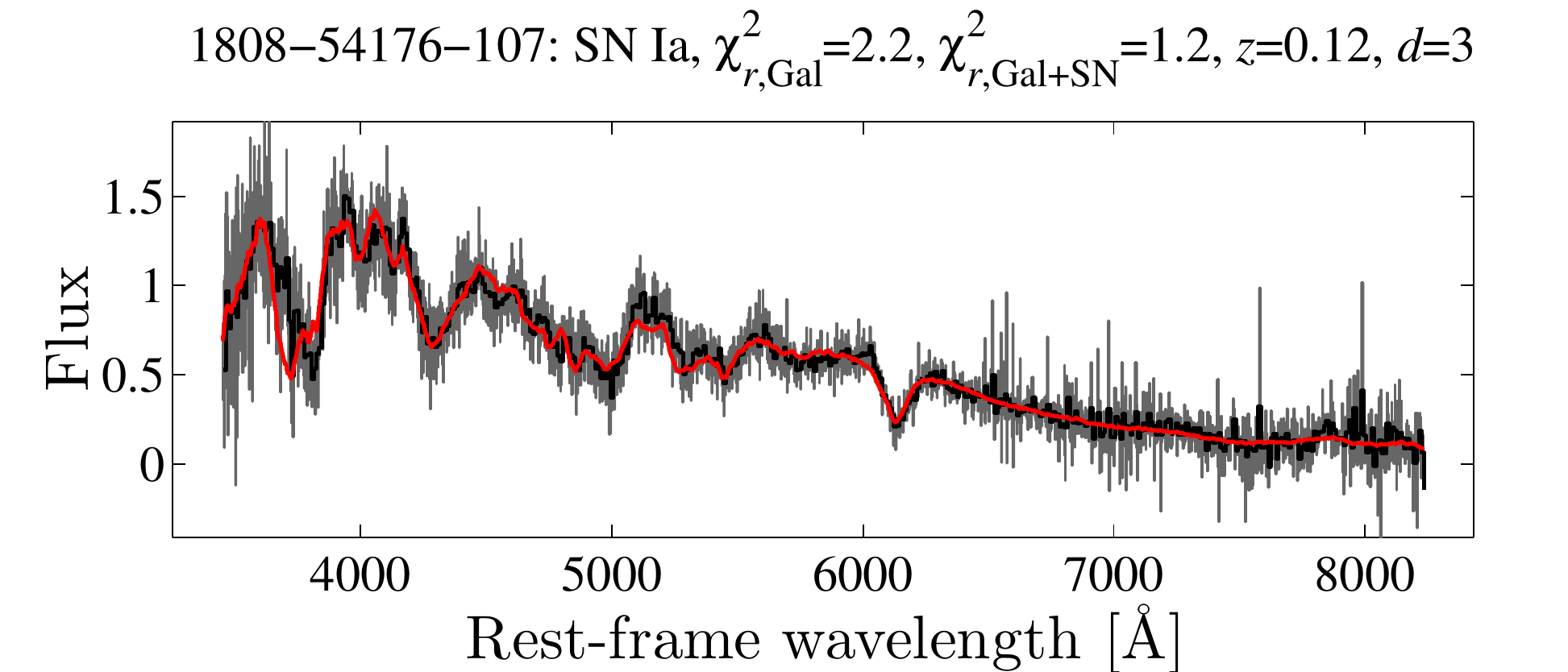} \\
   \includegraphics[width=0.475\textwidth]{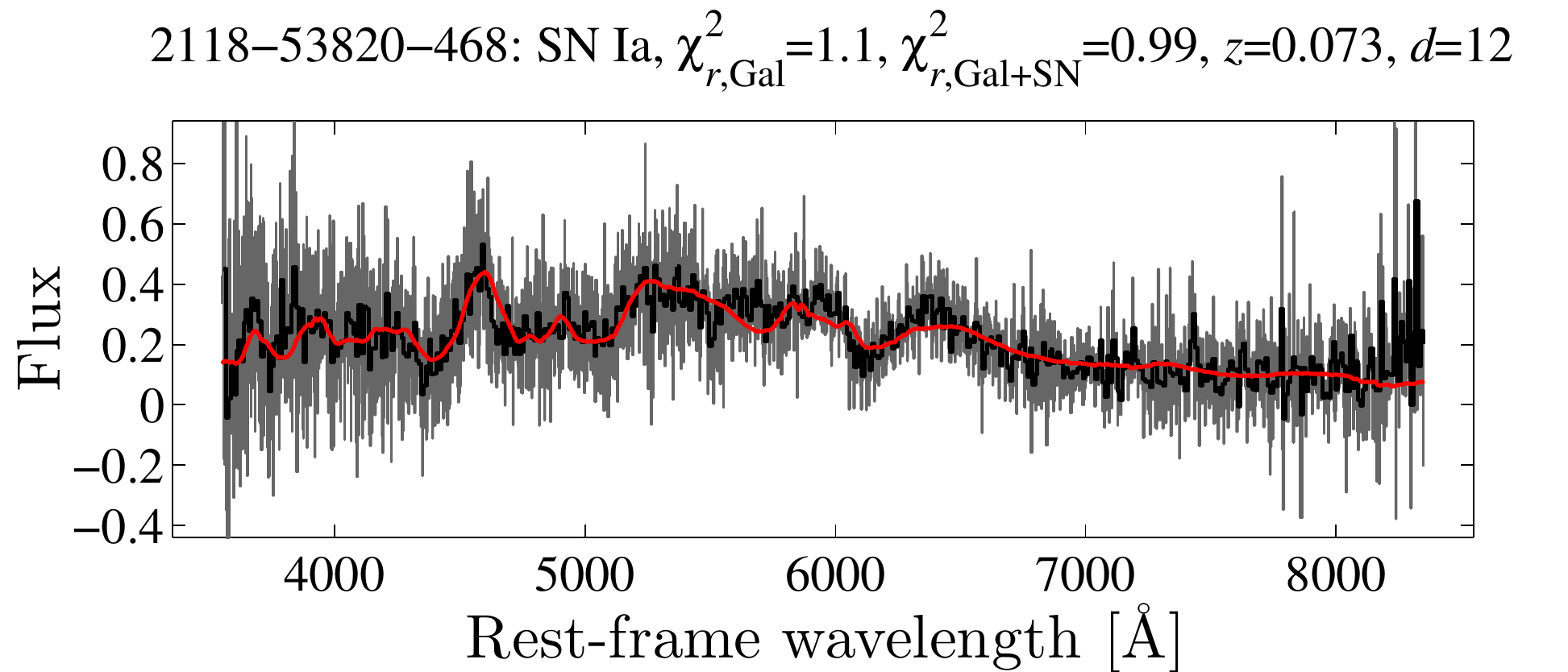} &
   \includegraphics[width=0.475\textwidth]{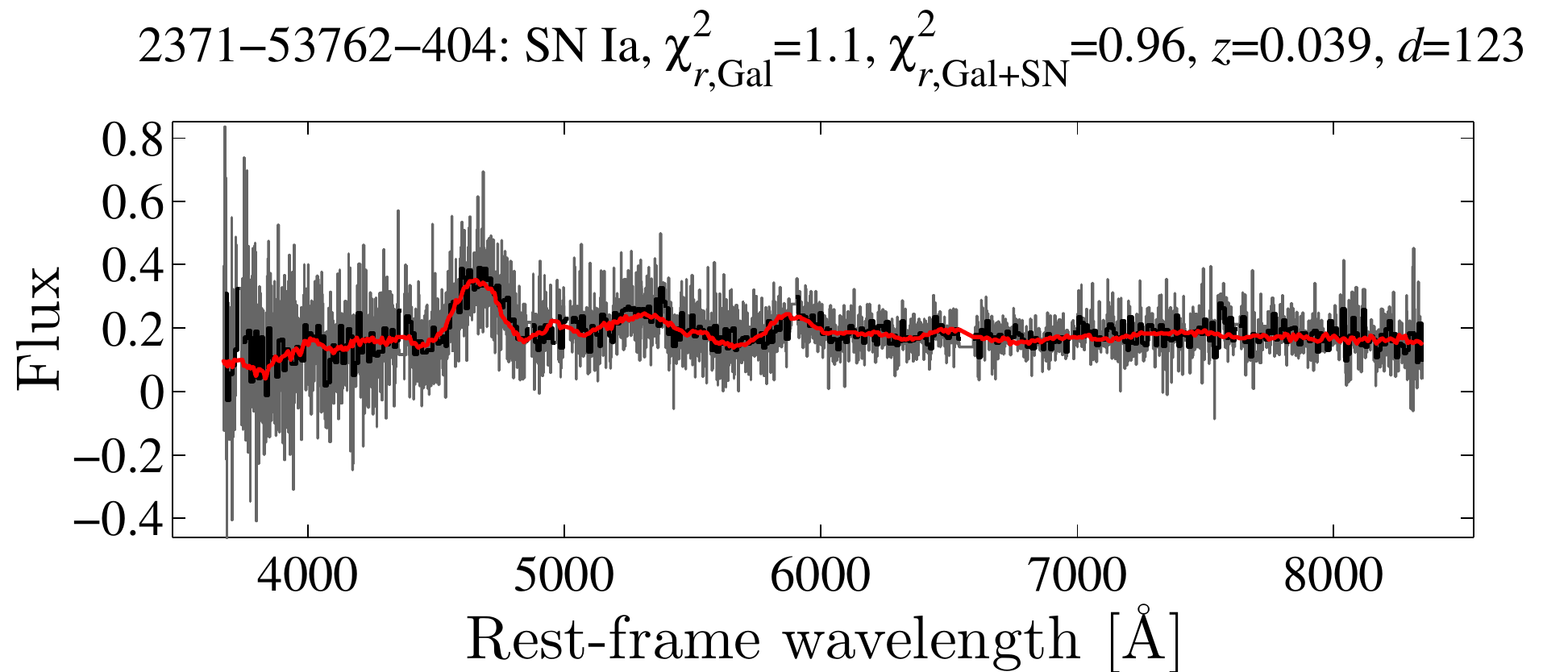} \\
   \includegraphics[width=0.475\textwidth]{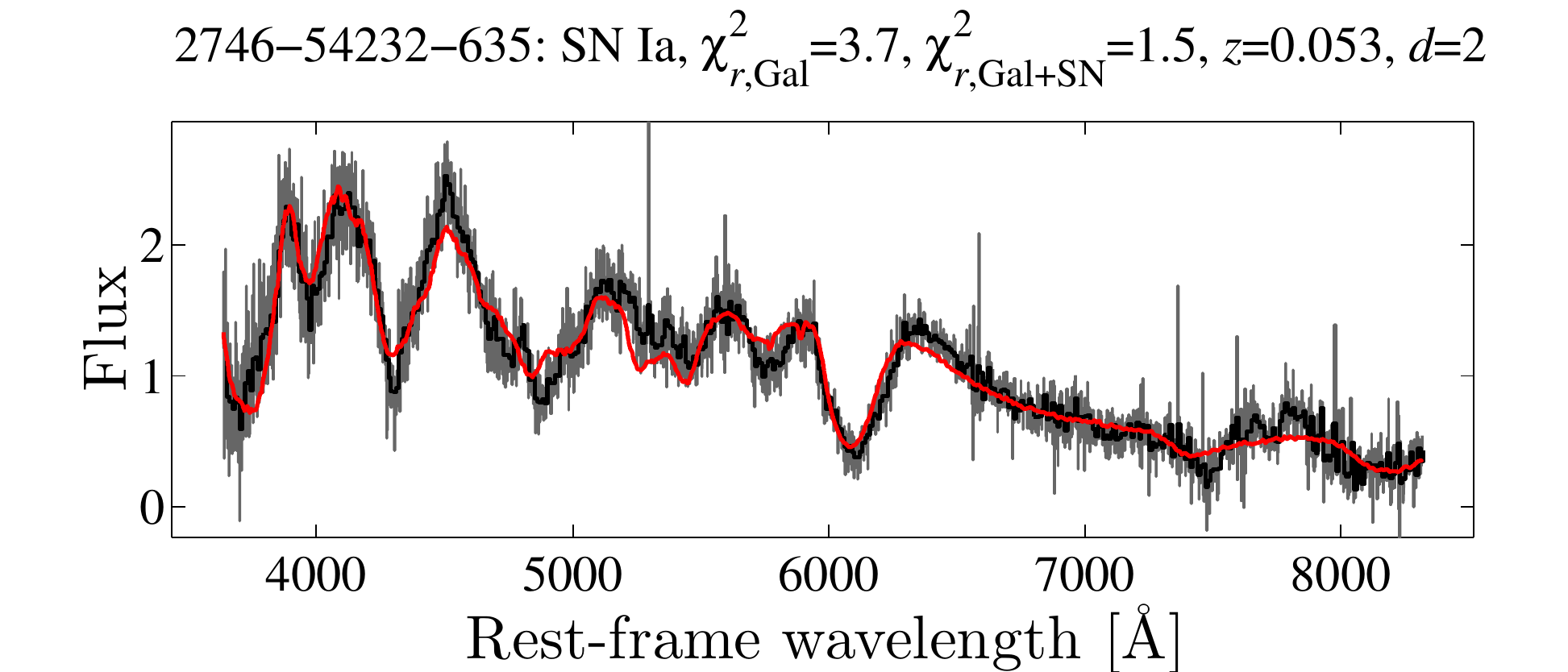} &
   \includegraphics[width=0.475\textwidth]{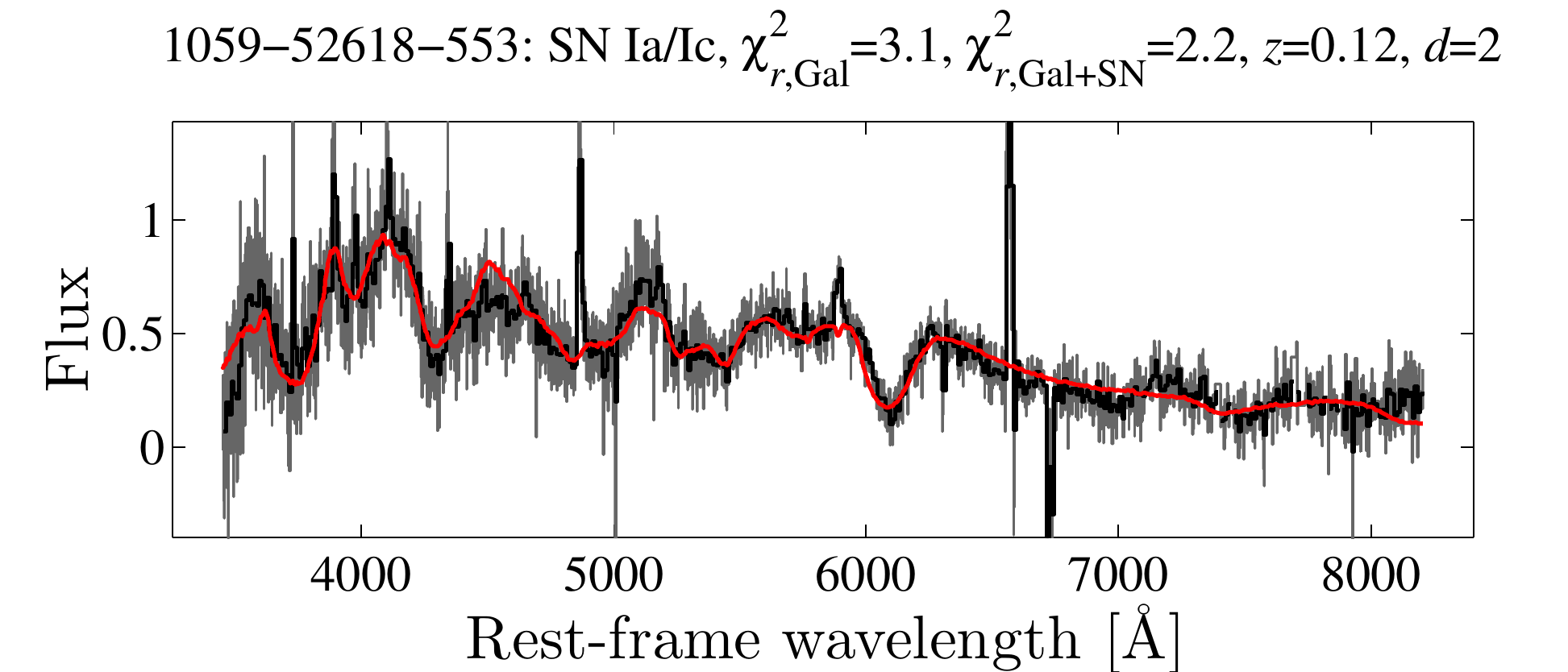} \\ 
   \end{tabular}
  \caption{SNe detected in this work -- continued. SN 1452-53112-120 and SN 1700-53502-359 were reported as SN2004cn and SN2005ca by Connolly (2004) and Subbarao (2005), respectively. SN 1059-52618-553 was previously reported by Tu et al. (2010).}
  \label{fig:sn_all_2} 
 \end{minipage}
\end{figure*}

\begin{figure*}
 \begin{minipage}{\textwidth}
  \vspace{0.2cm}
  \begin{tabular}{cc}
   \includegraphics[width=0.475\textwidth]{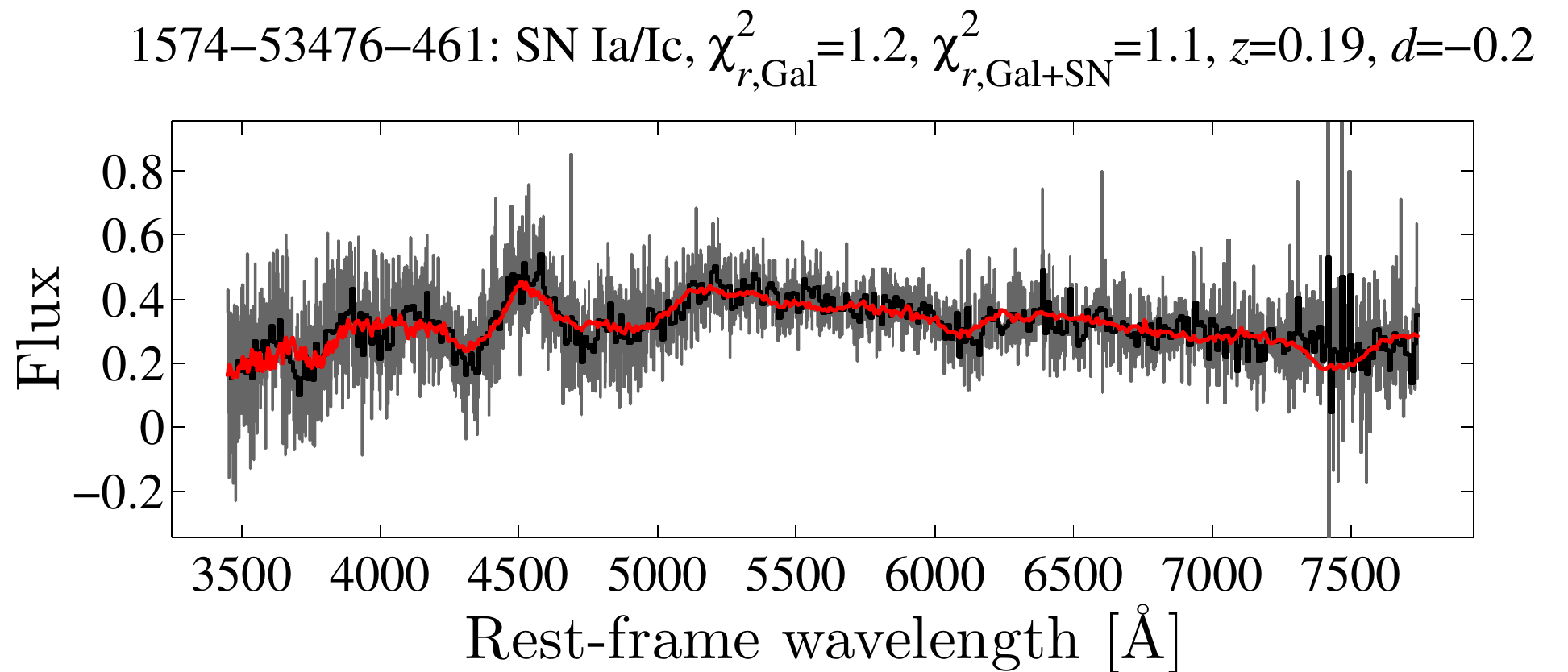} & 
   \includegraphics[width=0.475\textwidth]{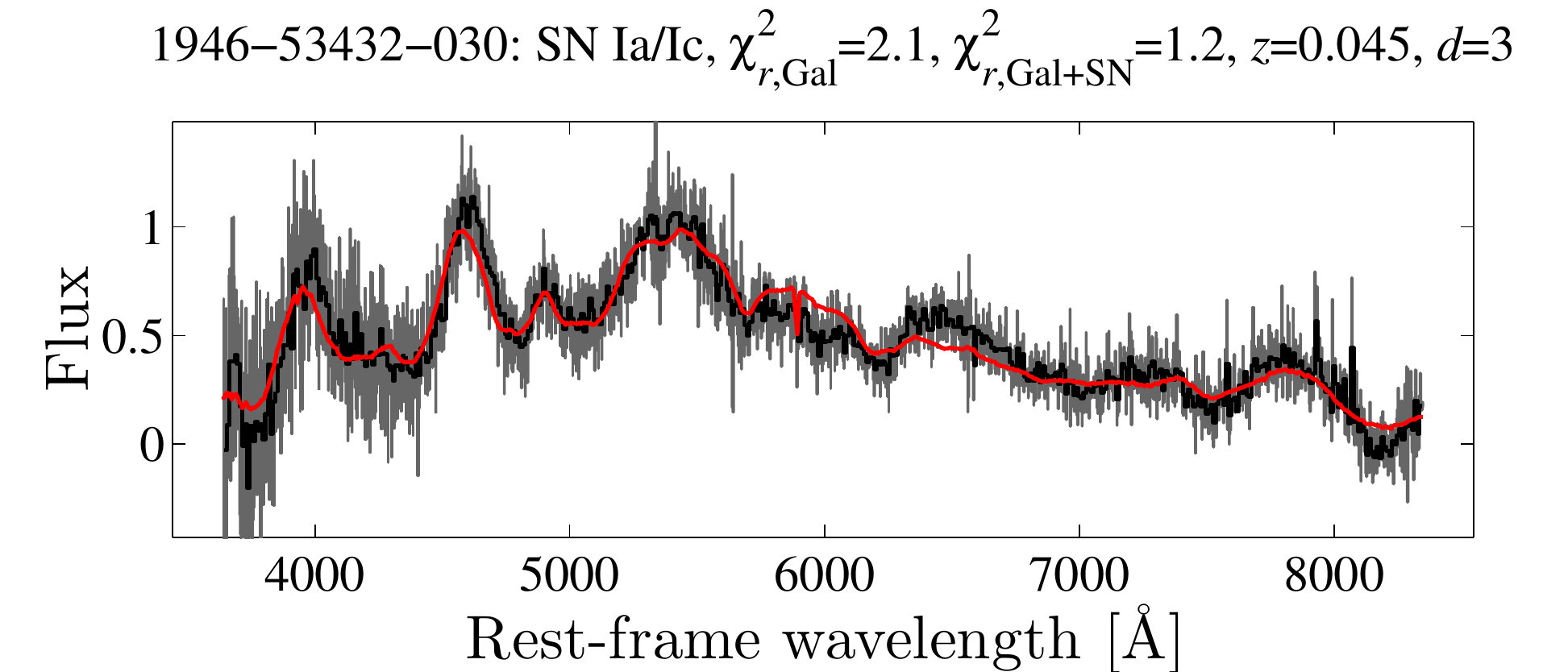} \\
   \includegraphics[width=0.475\textwidth]{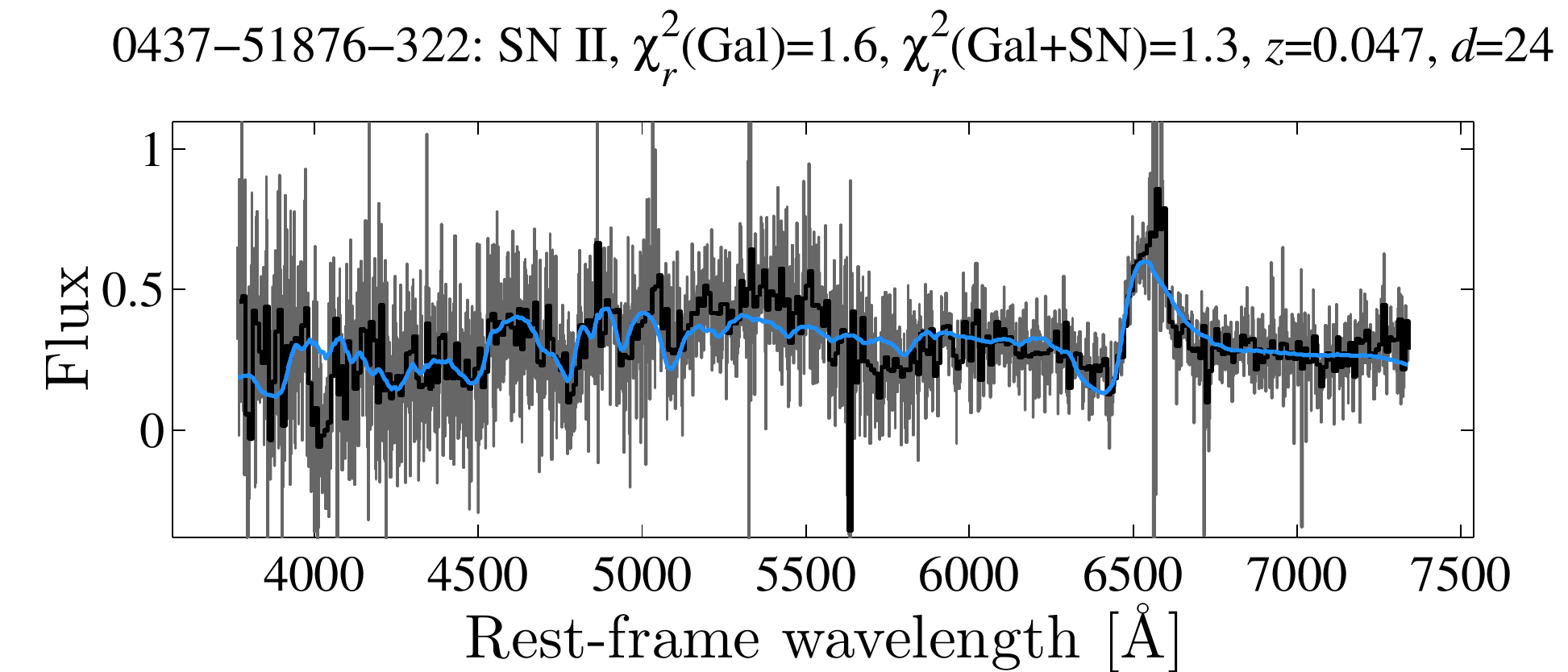} & 
   \includegraphics[width=0.475\textwidth]{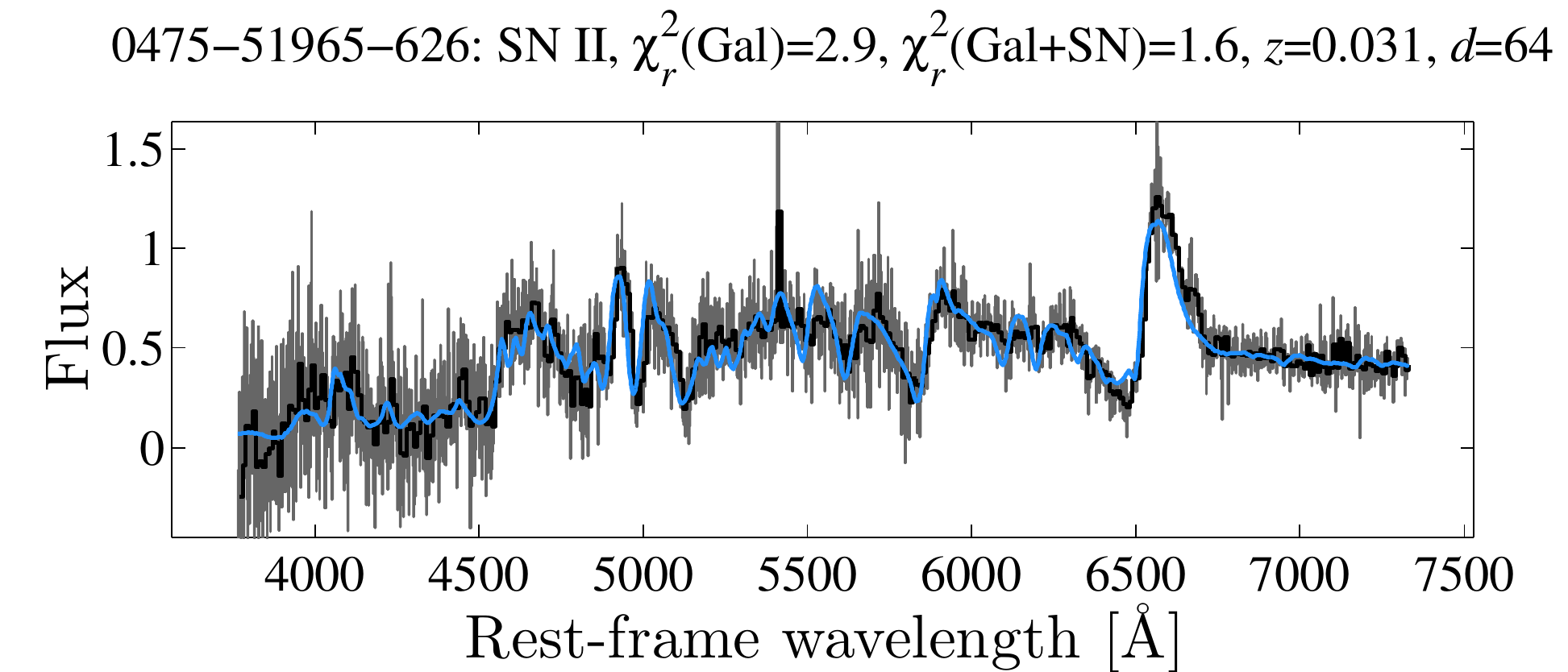} \\ 
   \includegraphics[width=0.475\textwidth]{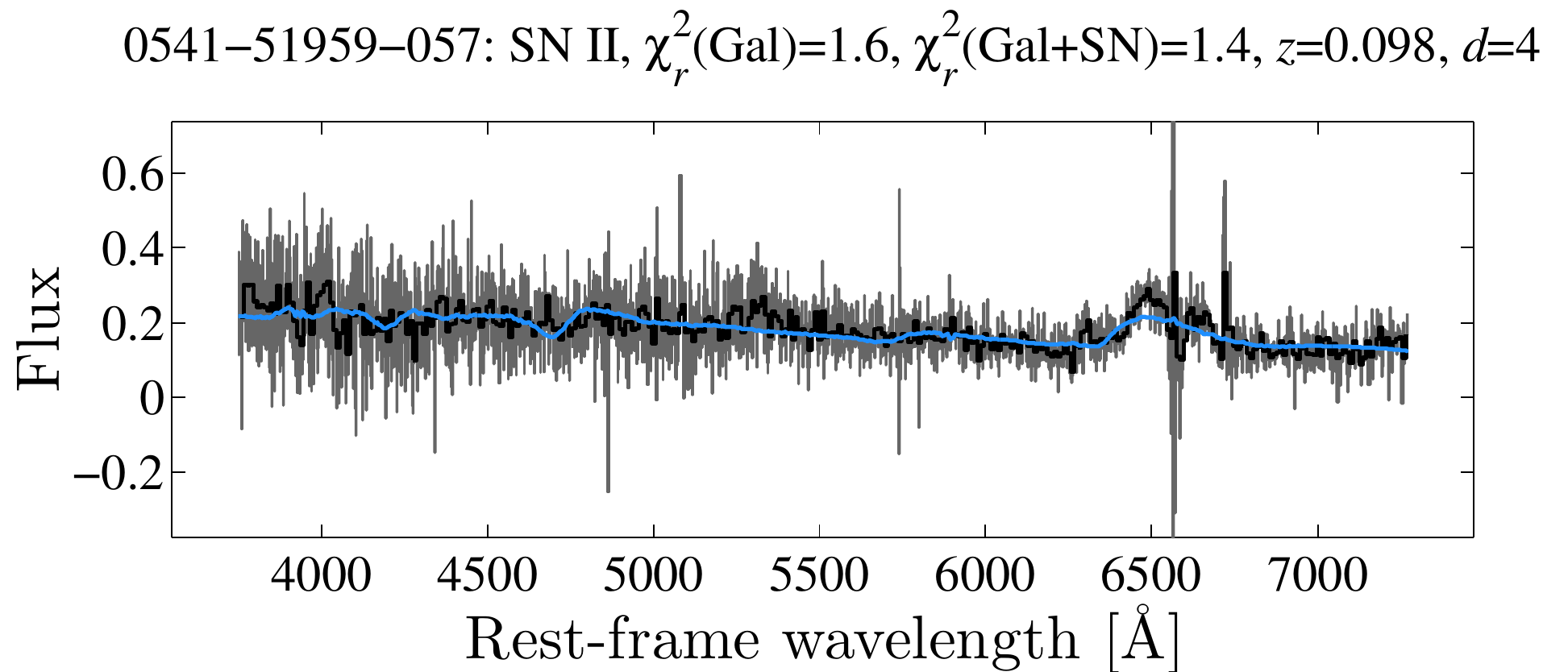} & 
   \includegraphics[width=0.475\textwidth]{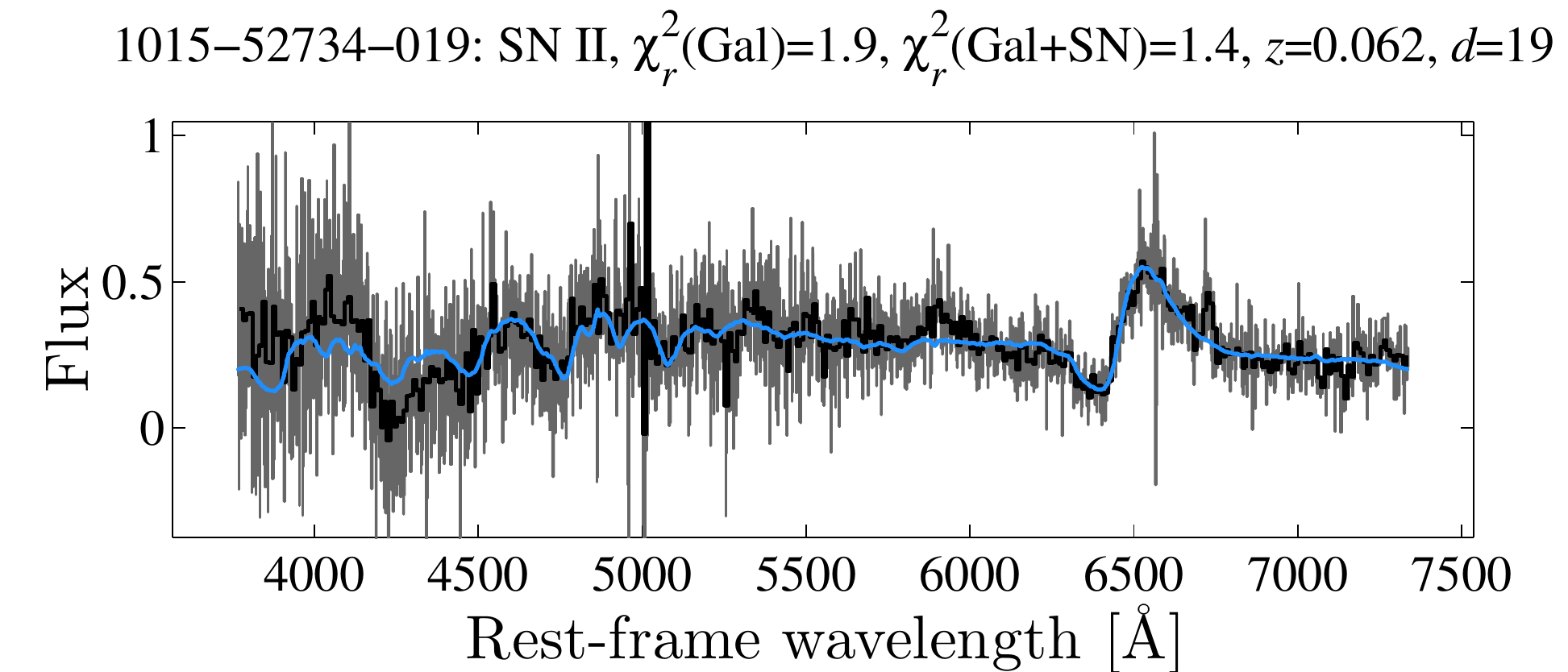} \\
   \includegraphics[width=0.475\textwidth]{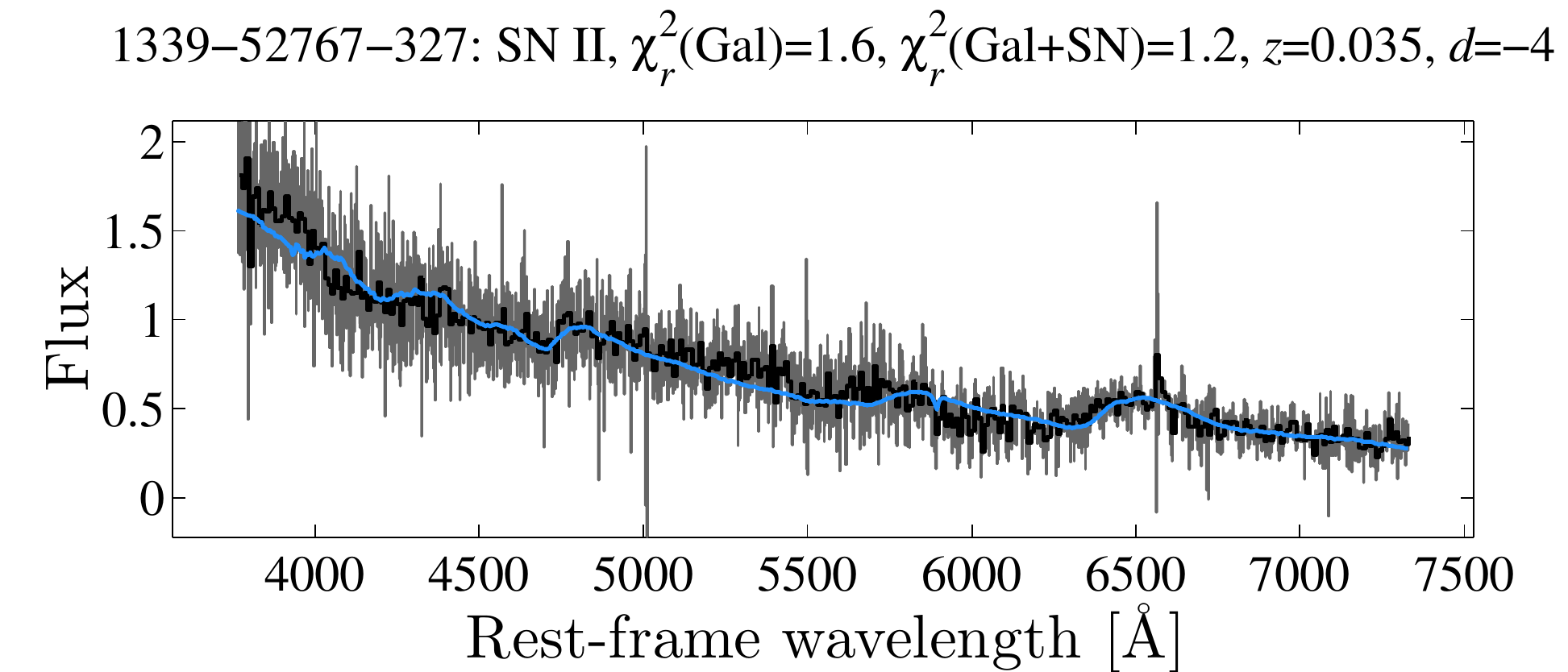} &
   \includegraphics[width=0.475\textwidth]{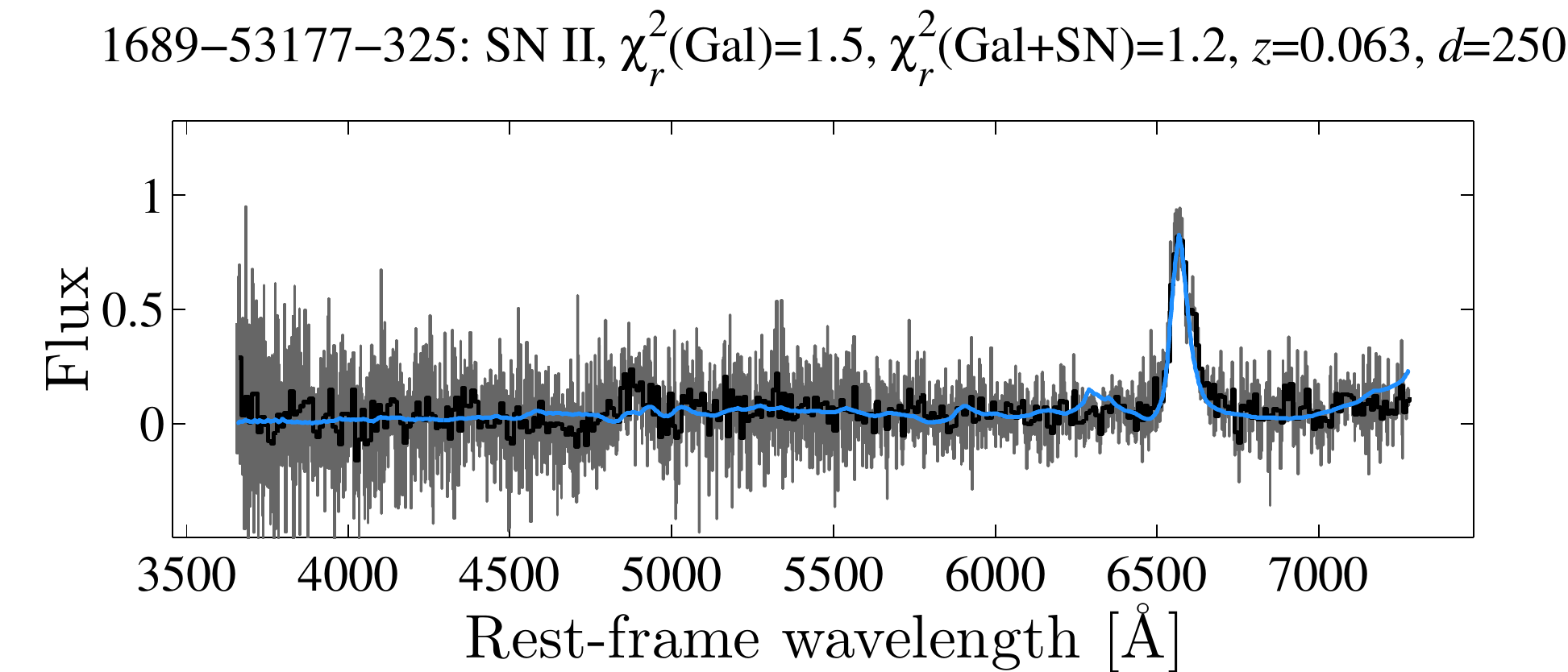} \\   
   \includegraphics[width=0.475\textwidth]{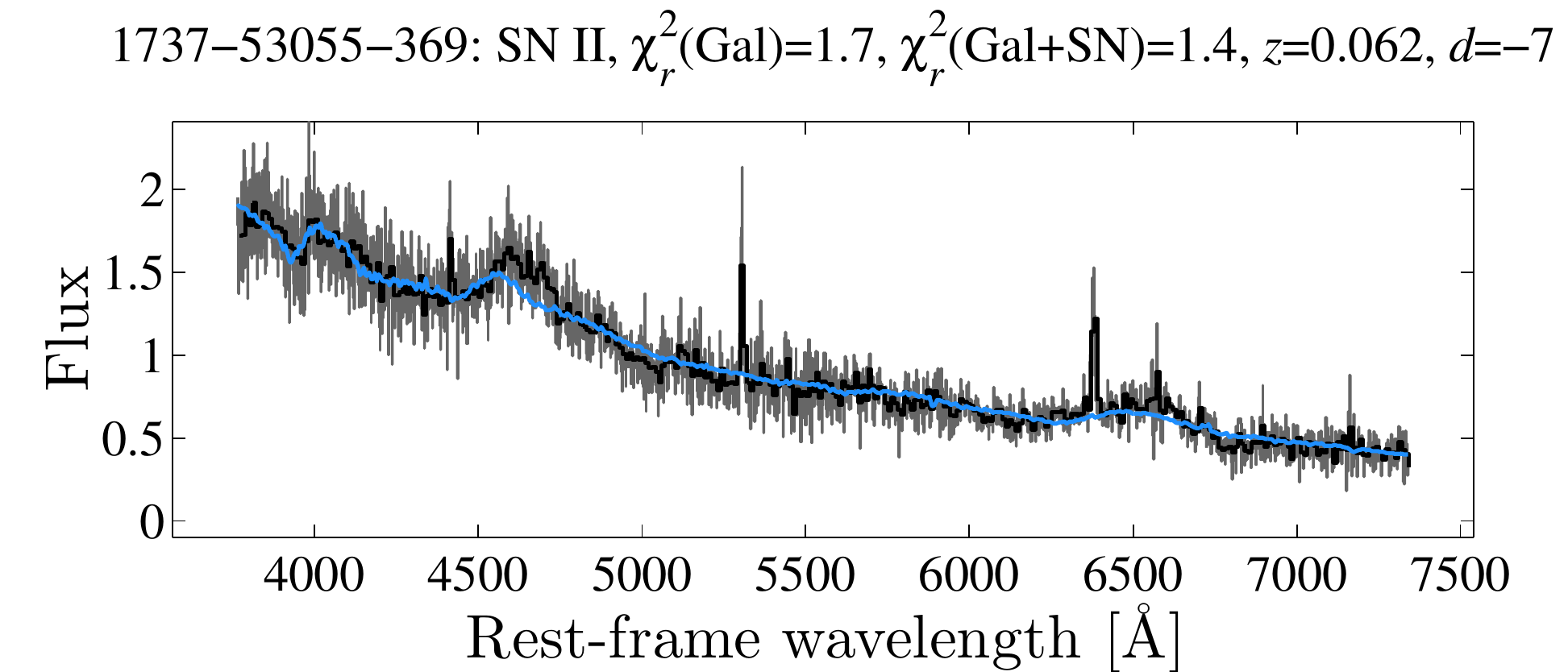} &
   \includegraphics[width=0.475\textwidth]{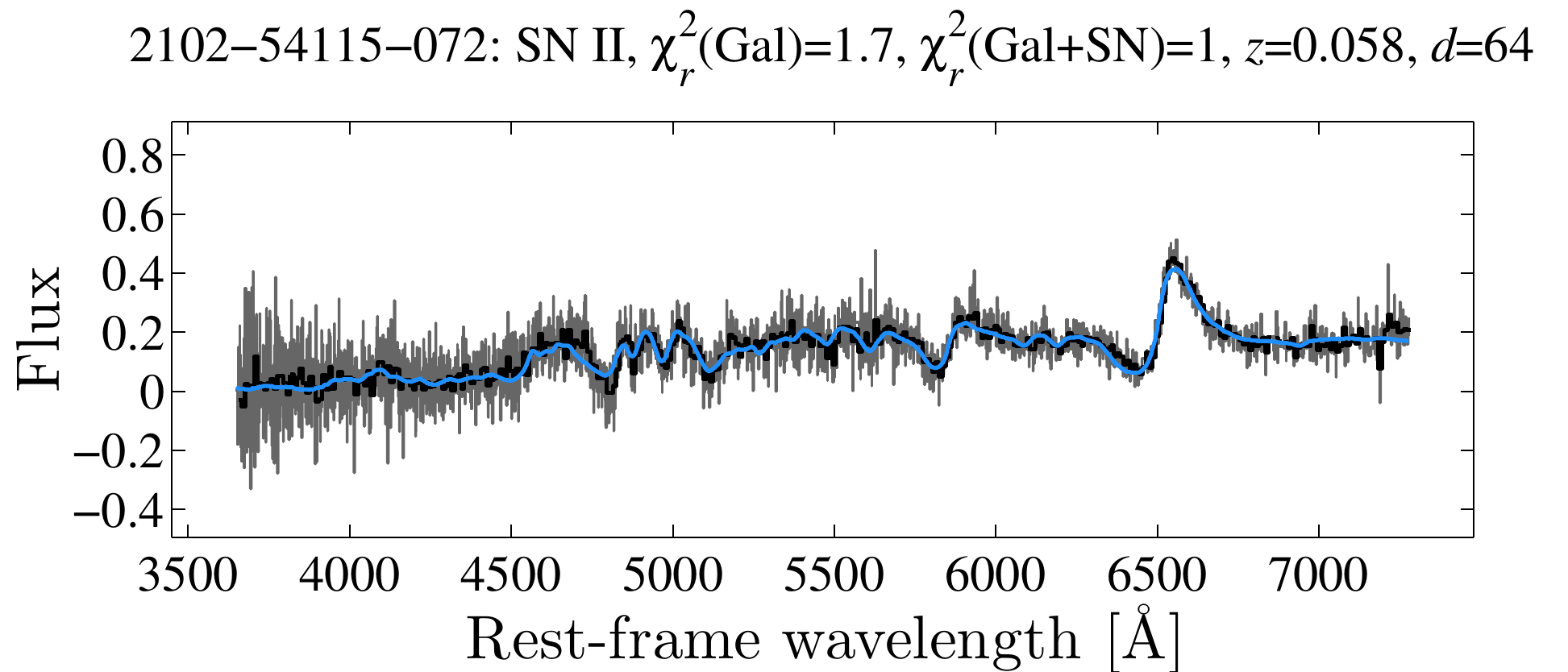} \\ 
   \includegraphics[width=0.475\textwidth]{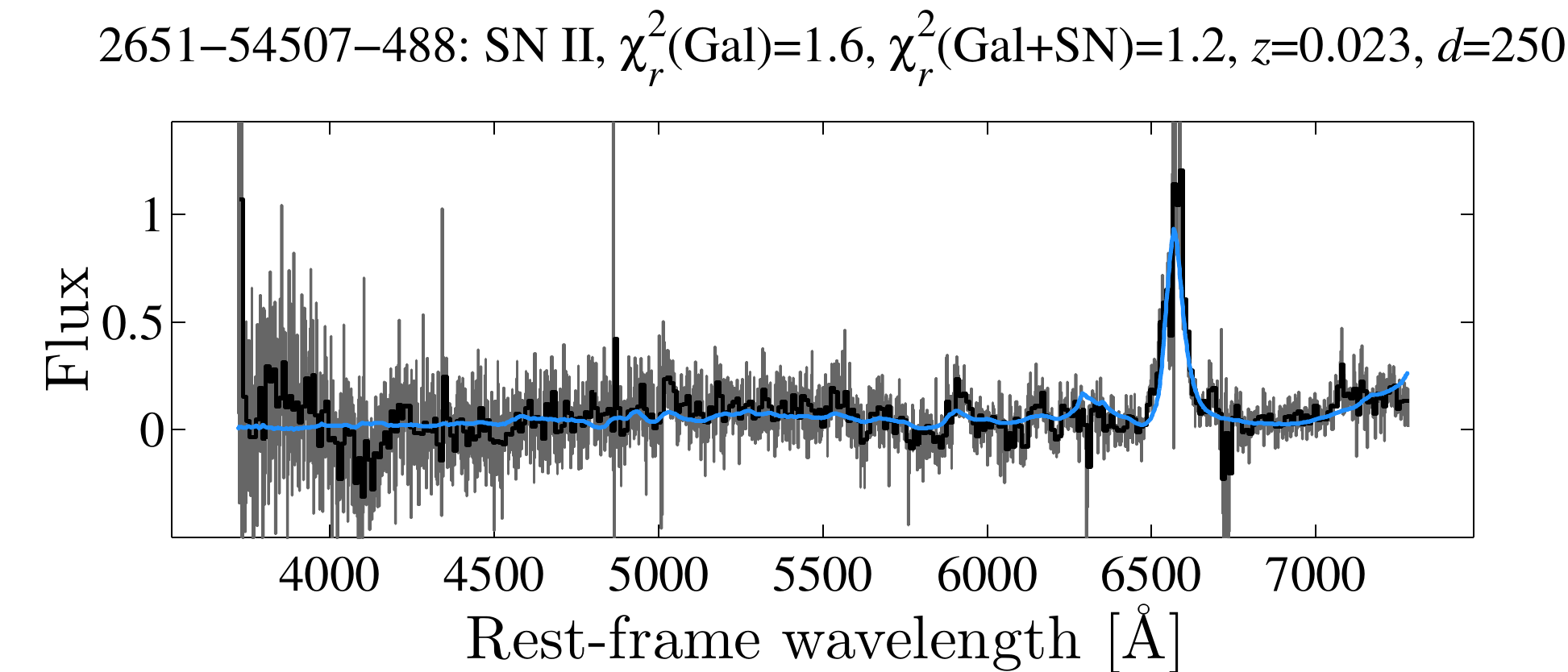} & \\
   \end{tabular}
  \caption{SNe detected in this work -- continued. SN 0475-51965-626 was previously reported by Reines et al. (2013), but is shown here for the first time}
  \label{fig:sn_all_3} 
 \end{minipage}
\end{figure*}